\documentclass[preprint,10pt]{aastex}

\usepackage{graphicx}

\newcommand{\be}{\begin{equation}}
\newcommand{\ee}{\end{equation}}
\newcommand{\nn}{\mbox{} \nonumber \\ \mbox{} }
\newcommand{\ba}{\begin{eqnarray}}
\newcommand{\ea}{\end{eqnarray}}
\newcommand{\om}{\omega}
\newcommand{\Alfven}{ Alfv\'{e}n }
\newcommand{\curl}{{\rm curl}}
\newcommand{\E}{{\bf E}}
\newcommand{\B}{{\bf B}}
\renewcommand{\j}{{\bf j}}
\newcommand{\RHESSI}{{\it RHESSI\,}}

\newcommand\etal{\textit{et al.\ }}
\newcommand\eg{\textit{e.g.\ }}
\newcommand\cf{\textit{cf.\ }}

\newcommand\lo{\mathrel{\raise.3ex\hbox{$<$}\mkern-14mu\lower0.6ex\hbox{$\sim$}}}
\newcommand\go{\mathrel{\raise.3ex\hbox{$>$}\mkern-14mu\lower0.6ex\hbox{$\sim$}}}
\shorttitle{Electromagnetic Blast Waves: Dynamics}
\shortauthors{Lyutikov \& Blandford}
\begin{document}
\date{\today}  
\title{GAMMA RAY BURSTS AS ELECTROMAGNETIC OUTFLOWS}
\author{M. LYUTIKOV$^{1,}$\altaffilmark{2} AND 
R. D. BLANDFORD$^{3,}$\altaffilmark{4}}
\affil{$^1$ Physics Department, McGill University, 3600 rue University
Montreal, QC,\\Canada H3A 2T8, \\
 }
\affil{$^3$ Kavli Institute for Particle Astrophysics and Cosmology, Stanford,
CA 94305}
\altaffiltext{2}{lyutikov@physics.mcgill.ca}
\altaffiltext{4}{rdb3@stanford.edu}
\begin{abstract}
We interpret gamma ray bursts as relativistic, electromagnetic explosions.
Specifically, we propose that they are created 
when a rotating, relativistic, stellar-mass
progenitor loses much of its rotational energy in the form of a
Poynting flux during  an active period lasting $\sim 100$~s.
Initially, a  non-spherically symmetric, 
electromagnetically-dominated bubble expands non-relativistically inside the
star,  most rapidly  along the rotational axis of the progenitor. 
After the bubble  breaks out from the stellar surface
and most of 
the electron-positron pairs
 annihilate, the  bubble  expansion becomes  highly 
relativistic.
After the end of the  source activity most   of the electromagnetic energy
is concentrated in a thin shell inside the contact discontinuity between the
ejecta and the shocked circumstellar material. 
This  
 electromagnetic shell pushes   a relativistic blast wave
into the circumstellar medium. 
Current-driven instabilities develop in this shell at a radius $\sim3\times10^{16}$~cm 
and lead to dissipation of magnetic field and
 acceleration of  pairs which are responsible for the $\gamma$-ray burst.
At larger radii, the energy contained in the electromagnetic shell is mostly  transferred
to the preceding  blast wave.
Particles accelerated at the forward shock may combine with electromagnetic 
field from the electromagnetic shell to produce the afterglow emission.

In this paper, we concentrate on the dynamics of electromagnetic explosions. 
We describe the principles 
that control how  energy is released by the central compact object and interpret 
the expanding
electromagnetic bubble as an electrical circuit. We analyze the electrodynamical
properties of the bubble and the shell, paying special attention to the 
energetics and causal behavior. 
We discuss the implication of the model for the afterglow dynamics and  briefly 
discuss 
 observational 
ramifications of this model of $\gamma$-ray bursts.
\end{abstract}

\section{Introduction}
\label{sec:intro}
In recent years, a ``fireball/internal shock''  model 
 of ``long''
gamma-ray bursts (henceforth GRBs) has been developed \cite[\eg][and
references therein]{mes02,Piran99}. 
\footnote{External shock \citep[\eg][]{derm02} and cannonball \citep{dar03} are
other proposed models.} 
  This associates GRBs with black hole
or neutron star formation 
during the explosion of rapidly rotating, evolved, massive stars - the ``collapsar''
model.
It is proposed that ultra-relativistic jets are formed within the spinning star
and that these jets are subsequently responsible for the $\gamma$-ray emission and the 
afterglow \cite{Woosl02}. This model has been supported by the discovery that GRBs 
occur preferentially in star-forming regions in cosmologically 
distant galaxies \citep[\eg][]{blo01}, that achromatic breaks 
\citep[\eg][]{har01}, indicative of beaming,
have been observed in some afterglows and the observation of 
additional luminous components to the late afterglow \citep[\eg][]{blo02},
which have recently been shown to have a supernova spectrum \citep[\eg][]{hjo03}
in the case of SN 2003dh -- GRB 030329.
The principal phases in this model comprise:
\paragraph{I Energy Release}
A source of power associated with a relativistic stellar mass object, now thought 
to be embedded within a star in the case of the long bursts,
with luminosity $L=10^{50}L_{{\rm s,50}}$~erg s$^{-1}$ operates
for a time $t_{{\rm s}}=100t_{{\rm s,2}}$~s within a region with 
radius $r_{{\rm s}}=10^6r_{{\rm s,6}}$~cm. The energy stored in a combined
rotational, gravitational and internal form is at least $\sim10^{52}L_{50}t_{{\rm s,2}}$
erg. 
The energy release mechanism may involve the release of magnetic energy 
by a torus \citep[\eg][]{woo93,vie98}, a nascent magnetar with initial angular velocity
$\sim10^4$~rad s$^{-1}$
\citep[\eg][]{uso92,dt92,tho94,uso94} or a black hole \citep[\eg][]{pac86}. 
The magnetic field itself may  be produced by strong
dynamo activity \citep[\eg][]{tm01} or shear \citep{klu98}. Interpreting $r_s$ as the 
characteristic size of the light cylinder, the associated strength of the magnetic field 
is $B_s \sim 10^{14}L_{50}^{1/2}r_s^{-1}$~G. 
In an alternative class of models, the energy release involve
the formation of a pair plasma 
by $\go3$~MeV neutrinos \citep[\eg][]{eic89}. Independent of the source,
it is generally supposed that a high effective temperature 
$T_{{\rm s}}=T_{{\rm s,0}}{\rm MeV}\sim(L/4\pi r_{{\rm s}}^2\sigma_{SB})^{1/4}
\sim L_{50}^{1/4}r_{{\rm s,6}}^{-1/2}$~MeV
and entropy per baryon ($S=10^6S_6$~k), optically 
thick ($\tau_T\sim10^{14}L_{50}^{3/4}r_{{\rm s,6}}^{-1}$) 
fireball is produced \citep{cav78}, varying on a timescale
$t_{\rm var} \sim r_{{\rm s}}/c\sim100r_{{\rm s,6}}\ \mu{\rm s}$.
Long bursts are argued to have $L_{50}\sim t_{{\rm s,2}}\sim 
r_{{\rm s,6}}\sim S_6\sim1$ \citep[\eg][]{fra01}. 
\paragraph{II Flow Formation}
As the radiation-dominated fireball expands due to  ``lepto-photonic'' pressure,
the flow is collimated by the surrounding
stellar envelope into two anti-parallel jets with opening angle 
$\theta=0.1\theta_{-1}$.
During the subsequent expansion, the energy is converted into ion bulk motion, 
and becomes matter-dominated at a radius,
$r_{{\rm mat}}\sim10^{10}$~cm, where the fluid Lorentz factor 
saturates with a value $\Gamma_0=10^3\Gamma_{{\rm 0,3}}
\sim S(kT_{{\rm s}}/m_pc^2)\sim10^3$; beyond this radius,
most of the energy resides in the kinetic energy of the protons.
(The stellar photosphere is thought to have a similar radius.)
The photons decouple from the plasma at a photospheric
radius, $r_{{\rm phot}}$ which is also in the vicinity of $r_{{\rm mat}}$
for the envisaged conditions.
\paragraph{III $\gamma$-ray Burst}
Much of the jet power is dissipated through a series of internal 
shocks at a radius $r_\gamma\sim\Gamma_0^2ct_{{\rm var}}\sim 3 \times
10^{12}\Gamma_{0,3}^2$~cm. These shocks are responsible for the 
re-acceleration of relativistic electrons and the production of magnetic 
field and Doppler-shifted, $\gamma$-ray synchrotron emission, up to
$\sim$~GeV energies, that is sufficiently well-collimated by the relativistic 
outflow to escape pair production. This is the Gamma-Ray Burst (GRB).
The constraint that the highest energy $\gamma$-rays be able to escape 
without producing electron-positron pairs, implies that 
$\Gamma_{0,3}\sim0.3$ \citep[\eg][]{lit01}.
\paragraph{IV Afterglow}
When $r>ct_{{\rm s}}/\theta\sim10^{13}t_{{\rm s,-4}}\theta_{-1}^{-1}$,
the debris takes the from of a shell of cold protons driving a blast wave into 
the surrounding medium with density $n=0.1 n_{-1}{\rm cm}^{-3}$. 
An external shock forms and when 
$r\sim r_{{\rm rsh}}\sim 10^{16}L_{50}^{1/2}\theta_{-1}^{-1}\Gamma_{03}^2
n_{-1}$~cm, a reverse shock will also form in the exploding debris. The debris
decelerates after 
$r>r_{{\rm free}}\sim6\times10^{16}L_{50}^{1/2}t_{{\rm s,2}}^{1/2}
n_{-1}^{-1/2}\theta_{-1}^{-1}\Gamma_3^{-1}$~cm
The bipolar blast wave, formed by the shocked circumstellar medium,
which now carries most of the energy of the explosion, will further decelerate
according to $\Gamma\sim\Gamma_0(r/r_{{\rm free}})^{-3/2}$
until it becomes non-relativistic at a radius
$r_{{\rm nr}}\sim\Gamma_0^{2/3}r_{{\rm free}}\sim10^{18}$~cm. 
During this phase, electrons 
are accelerated and magnetic field is amplified at the outer shock,
leading to the formation of the afterglow. 
The non-relativistic blast wave gradually becomes more spherical 
and evolves to resemble a normal, supernova remnant. 

\section{Some Problems with the Fireball Model}
The basic fireball model, which we have just sketched, along with its many 
variations, raises several, important questions.
Included among these, in temporal order of the flow evolution, are: 
\paragraph{1. How is the entropy of the fireball created?} 
In most models, the release of energy is mediated by a strong electromagnetic
field which is invoked to create turbulence in an accretion disk
\cite[\eg][]{wmcf99}, extract the rotational energy of the central black hole 
\cite[\eg][]{kim02} and collimate and confine the  jets
\cite[\eg][]{wmcf99}. As the energy release phase
lasts for $\sim ct_s/r_s\sim10^6$ source
dynamical times\footnote{Longer than we have observed most quasars!},
the magnetic flux must presumably be tied to or trapped by 
a large, conducting mass  and the transients should die away so that 
a quasi-steady, electromagnetic
energy flow will be produced, similar to what happens with pulsars.
The problem is that, if the burst is powered electromagnetically,
how is the large entropy of a fireball created? As we discuss further below, 
there is no natural way to accomplish
this in the vicinity of the source, although there have been suggestions invoking 
magnetic reconnection in an outflowing wind. (Of course, this is not a concern for those models
\cite[\eg][]{swm01} where the energy release is mediated by neutrinos.)
\paragraph{2. How are the hypersonic jets made?}
The fireball model requires that two jets are formed with Mach number
$M\sim2^{1/2}\Gamma_0 \sim 400$ and a ratio
of bulk kinetic energy to internal energy $\go M^2\sim10^5$.
Numerical simulations 
\citep[\eg][]{alo02,mac02} and experience with wind tunnels strongly suggest 
that instability and 
entrainment prevent this from happening inside the star. It is more reasonable
to suppose that the outflow emerges from the stellar surface (radius $R_\ast\equiv
10^{10}R_{\ast,10}$~cm) with a modest
Lorentz factor $\Gamma_\ast$, collimated within a cone with opening angle
$\theta\sim\Gamma_\ast^{-1}$. It will then accelerate linearly $\Gamma \propto r$
 due to radiative  pressure  until either the momentum flux of the radiation field 
falls below that of the ions,  or optical depth to Thomson scattering 
falls below unity. 
For ion jet the raping radius is (Eq. \ref{rT})
$r(\tau=1)= 3\times10^12 {\rm cm} L_{50} \Gamma_{2.5}^{-3} \Delta\Omega_{-2} $.
Thus, if an ion jet starts at $R_\ast \sim 10^{10}$ cm with $\Gamma_\ast \sim $ a few, 
it barely has enough optical depth to accelerate to the required
$\Gamma \sim 300$. (Note that in this case most acceleration happens {\it beyond} the 
photosphere
(for a  ion jet) at $r_{ph}\sim 10^{11} $ cm, see Eq. (\ref{Tbr}).)
\paragraph{3. How can the outflow develop large, parallel, proper velocity gradients 
and avoid producing converging streams?}  
$O(1)$ gradients in the jet proper velocity are invoked in the fireball model in order
to produce internal shocks, and supply the free energy for particle acceleration. The model 
is essentially one dimensional. It is usually supposed that this variation has its origin 
in the source which implies that the GRB be produced within a radius 
$\go\Gamma^2ct_{{\rm min}}$ where $t_{{\rm min}}$ is the minimum variation timescale.  
However, it is also envisaged that the jet be collimated 
and develop $\sim(\Gamma\theta)^2$ causally disconnected streams.  It is difficult to 
understand how this collimation can be achieved without producing angular deflections
$\go\Gamma^{-1}$ which would lead to pair formation by the escaping, high energy $\gamma$-rays.
(This problem is analogous to the one addressed in contemporary cosmology by the theory
of inflation with the important difference that, here,
it must be solved in a continuous flow with a bounding surface.)
\paragraph{4. What determines the baryon loading of the flow?}
Independent of the problem addressed in (2), in the fireball model, the baryon fraction
must be fine-tuned to allow baryons to assume most of the energy of the outflow
and to attain the large outflow Lorentz factors that are necessary;
too small a fraction and the energy will escape before the baryon acceleration is 
complete, too large a fraction and the asymptotic Lorentz factor will be too low to 
allow the highest energy $\gamma$-rays to escape \cite[\eg][]{mes02}.
\paragraph{5. Where are the thermal precursors?}
For ion jets, the escape of thermal radiation at the photosphere should produce a thermal precursor 
with luminosity similar to the main burst \citep{lu00,mr00}. 
These are rarely  seen at the 1\% level \citep{dm02,fro01,gcg03}.
\paragraph{6. How are particles accelerated at relativistic shock fronts?} 
Diffusive shock acceleration, which appears
to operate efficiently at non-relativistic shock fronts, fails
at relativistic shocks, both in the $\gamma$-ray emitting region and at the 
external shock front, because only a minority of the back-scattered particles 
can catch up with the advancing shock front. There are promising, kinematic proposals 
\citep[\eg][]{ach01} for addressing relativistic shock acceleration.
However, they pre-suppose the existence of a subshock in the 
background thermal plasma and it is not clear how this can be maintained. 
(Actually,  thermal and nonthermal particles are not really distinguished in
emission models as it is generally assumed that a single, truncated, power-law distribution
function is transmitted \citep[\eg][]{blm77}.) More fundamentally, it is by no means
certain that relativistic shock discontinuities form at all. It may happen that 
the sharing of momentum between a relativistic outflow and the circumstellar medium
happens gradually rather than abruptly \citep[\eg][]{uso94}.
\paragraph{7. How is the magnetic field amplified?}  
In order to produce a high radiative efficiency and fit the afterglow light curves, 
it is necessary for the post-shock magnetic field strength be amplified by a large factor
over the value it would have due to simple compression. It has been proposed that this 
amplification is due to the Weibel instability \citep[\eg][]{med99}. 
Recent simulations have convincingly shown that  a long coherence range (much larger
than the  ion skin depth) of the  magnetic field
field fluctuation is indeed reached \cite{nish03,fred03}. However,  
the average values of the magnetic energy density, 
$\epsilon_B \sim 10^{-3}$ of the total energy density, is often too low to 
account for observed synchrotron  emission.
\paragraph{8. How is a large degree of $\gamma$-ray polarization created?} 
A very high linear polarization (nominally 80 percent) has been reported in 
RHESSI observations of GRB021206 \citep{coburn03}. 
The observation, if typical,  is  inconsistent with  the internal shock model \citep{lyu03d}
In order to reproduce high polarization the  internal shock model should make a number of
high unlikely assumptions, some of which contradict the very fundamentals of the  model.
There are four assumptions that are made.
(i) the field is confined 
to two dimensional plane, presumably the plane of the shock.
Magnetic field amplification due to Weibel instability at the shock
 indeed produces two dimensional fields \cite{med99,nish03,fred03}, but 
the typical 
 size of resulting  magnetic structures 
with linearly directed currents is 
still microscopic, probably tens or hundreds of ion skin depths (which is of the order
of meters when the  fireball is   $\sim 10^{12}$ cm in size).
 On  larger scales, magnetic field is likely to be  three dimensionally  random.
In addition, the postshock material must be turbulent: in the fireball model
turbulence is {\it needed}  in order to accelerate particle. In order to account 
for large energy fraction in accelerated electrons the turbulent motions 
should have energy density  comparable to the total energy in the shock and thus much 
larger than the energy density in the magnetic field, typically $\epsilon_B \leq 10^{-3}$.
  This turbulence
will easily  destroy  any finely-tuned current structures.
    (ii) the plane of the 
turbulent magnetic field is viewed edge-on in the rest frame (this requires
viewing angle $\sim 1/\Gamma$ in the observer frame); (iii) the  emitting
surface  should be 
quasi-planar; this requires that the  angular size of the  emitting
region be  $\Delta \theta \leq 1/\Gamma$.
(iv) all emitting shells
must have
 the same Lorentz factor to be seen edge on (the burst GRB030329 was multi-peaked).
Assumptions (ii),  (iii) and (iv) are at variance with the
fundamental assumption of the internal shock/fireball model
that  every peak is  interpreted as being due to collisions of shells 
with a range of Lorentz factors. 

\paragraph{9. What determines  the jet opening angle and its structure?}
A number of phenomenological 
jet structures have been proposed (\eg structured,  constant or patchy  jet). 
The fireball  model neither  gives a prediction or expresses a 
 preference for the jet structure.

\paragraph{10. What is the relation between GRBs and X-ray flashes (XRFs)?}
In the fireball model there is no clear relation between GRBs  and XRFs, which can be either
dirty fireballs, less energetic fireballs, or explosions seen ''from the side''.

\paragraph{11. Where are  ``orphan'' afterglows?}
If the GRB and afterglow emission is associated with jets, as described above,
then in the most simplistic interpretation, there will be $\sim\theta^{-2}
\sim100\theta_{-1}^{-2}$ ``orphan'' afterglows expected per GRB. Although the 
current observational constraints are surprisingly poor and the expected number is 
quite model-dependent, it is surprising that no convincing 
examples have been found so far \citep[\eg][]{lev02}.

\section{Electromagnetic model}
\subsection{Overview}
\begin{figure}
\includegraphics[width=1.1\linewidth]{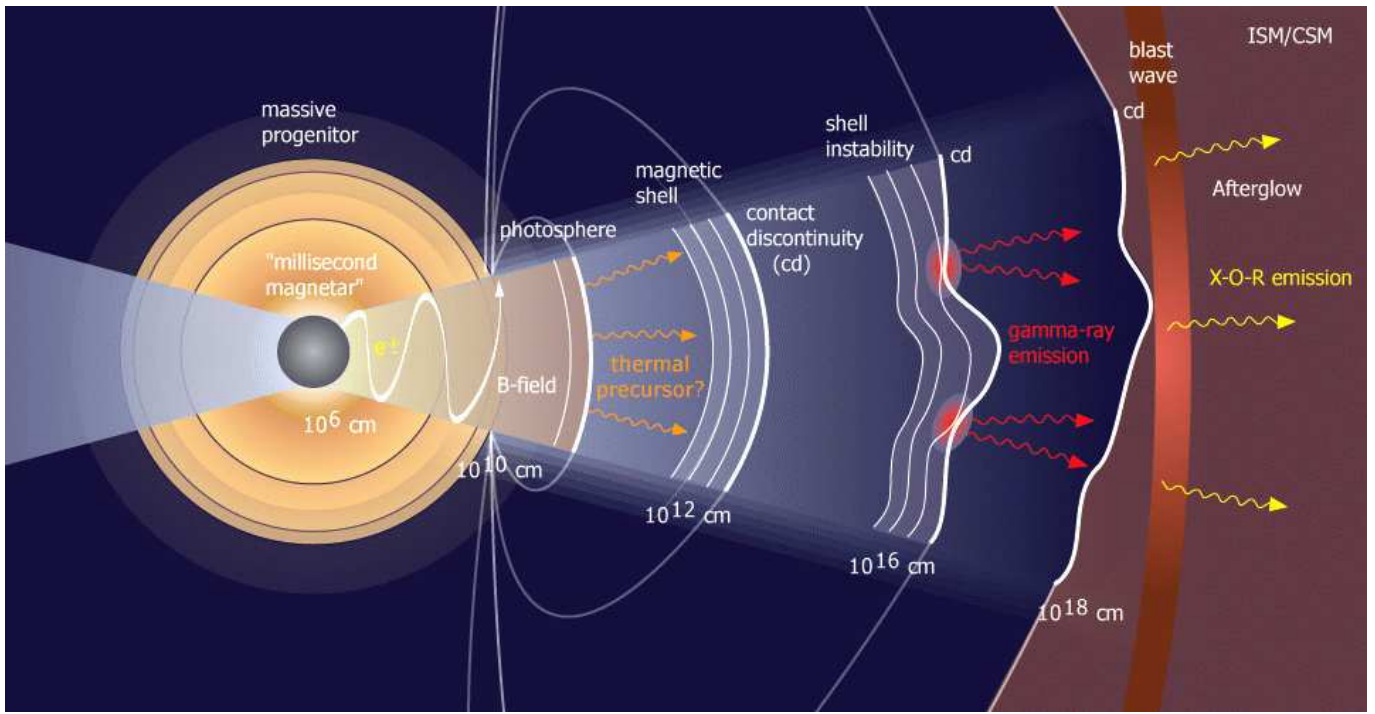}
\caption{
Overall view of the model.}
\label{pict}
\end{figure}

In an attempt to retain the merits of the standard model while addressing 
these questions, we present an alternative, electromagnetic interpretation of GRBs that 
builds upon earlier models of electromagnetic and magnetohydrodynamic explosions 
\citep[\eg][]{blar72,ben78,uso92,uso94,mes92,fer98,klu98,vie98,spr99,whe00,lb00,vla01,bla02,
Cold}.
In our specific version of the electromagnetic model, we give a quite different 
interpretation of the same four phases of a GRB introduced in Section \ref{sec:intro} 
(see also Fig.~\ref{pict}). 
At this point we do not have  detail answers to all the posed questions; 
in some cases we offer only  a plausible explanation. Also,
we specialize to the collapsar model although the principles that 
we describe are easily adapted to other source models that may still be needed for the
majority of GRBs.
\paragraph{I Source Formation (Energy Release)}
The GRB ``prime mover'' is 
a rapidly spinning black hole orbited by a massive disk that 
has just been formed inside an imploding star, or, alternatively, a ``millisecond magnetar''.
For qualitative estimates we may use
 ``millisecond magnetar'' model by  \cite{uso92} 
\citep[see also][]{blar72}, though the numbers will be similar for any relativistic stellar
mass source  rotating with near-critical spin frequency
$\sim 3  L_{50}^{1/2}t_{s2}^{1/2}$~kHz 
  and  
 magnetic field of 
 $B_s\sim10^{14}L_{50}^{1/2}$~G.
 The total rotational energy,
$E \sim I \Omega^2/2 \sim  Lt_s\sim10^{52}$~erg 
is available to power GRB bursts
and the magnetic field is strong enough 
for this energy to be released electromagnetically in a time $t_s\sim10-100$~s.

We suppose that the outflow primarily takes the form of a large scale Poynting flux 
and that the dissipation rate remains low enough that 
the power continues to be dominated by the electromagnetic component rather than the heat of 
a fireball well out into the emission 
region, although there is almost certainly an initial phase in which the electromagnetic 
field is accompanied by a dense pair plasma.

A 
rapidly spinning magnetar  with 
a complicated field structure will form
a relativistic outflow.
  The behavior of such sources remains an unsolved problem, 
even in the simpler case of pulsar winds.
 In this paper 
we adopt a simplifying hypothesis, that the field lines quickly re-arrange 
to become predominantly axisymmetric.
 Thus we hypothesize that the axisymmetric or ``DC'' component of the electromagnetic field
dominates the wave or ``AC'' component
which is either dissipated as heat or diminished through non-dissipative rearrangement.
In this case 
  the 
electromagnetic source acts primarily as a unipolar inductor and drives 
a large quadrupolar current flow, rather like what happens in the
\cite{gol69} model of an axisymmetric pulsar.


\paragraph{II  Bubble Inflation (Flow Formation)}
Initially, the source will inflate an electromagnetic bubble inside the star. 
This magnetized cavity is separated from the outside material by the
(tangential)
contact discontinuity (CD) containing  a surface  Chapman-Ferraro
current. This current  terminates the magnetic field
and completes the circuit that is driven by the source.
On a microphysical level the current is created by the
particle of the surrounding medium completing half a turn in 
the magnetic field of the bubble, so that the 
thickness of the current-currying layer is of the order of  ion gyro-radius.
\footnote{It is expected that the surface current will be unstable \citep[\eg][]{su96,Liang},
so that in reality the motion of particles will be more complicated and 
the penetration depth will related to the scale of  most
unstable modes.}

As we show below, 
the electromagnetic field can be treated as a fluid and behaves similarly to a true 
fluid, with the important difference that the rest frame stress tensor 
is anisotropic. This allows it to self-collimate
\citep[for reviews of stationary flow see, \eg][]{Kon00,Sauty00,HeyN03}. 
\footnote{\cite{vk03a,vk03b} also give  examples of  collimated
MHD outflows applicable to GRBs. }
The poloidal and toroidal components of magnetic field 
are comparable in strength at the light cylinder, but the toroidal field dominates 
beyond this. The velocity of expansion of the bubble is determined by the pressure balance
on the contact discontinuity between magnetic pressure in the bubble and
the ram pressure of the stellar material
\footnote{More precisely, since the expansion is supersonic, 
the pressure balance is between  magnetic pressure and  the pressure of the shocked 
material, which is of the order of the ram pressure at the forward shock.}.
As the magnetic field strength is strongest close to the 
symmetry axis, the bubble will expand 
fastest along the polar direction.
Eventually the bubble will break free of its surroundings
and forming a ``twin exhaust'' along which 
Poynting flux will flow until either the central source slows down or the collimating 
material itself
expands which will both occur naturally on the timescale $t_s\sim100$~s. 
 
Outside the star, the bubble will expand ultrarelativistically and bi-conically.
After it has expanded beyond a radius 
\be
r_{{\rm sh}}\sim ct_s\sim3\times10^{12}t_{s,2}{\rm cm}
\ee 
the electromagnetic energy will be concentrated
within an expanding, electromagnetic shell with thickness $\sim r_{{\rm sh}}$
and with most of the return current completing along its
trailing surface. However, the global dynamics of this shell and its subsequent 
expansion are set in place by the electromagnetic 
conditions at the light cylinder and within the collimation region. 

After  break out, the interaction of the magnetic shell with 
the circumstellar medium proceeds in a similar way, except now the
velocity of expansion is strongly relativistic.
The leading surface of the shell 
is  separated by a 
contact discontinuity (which actually becomes a rotational discontinuity
if the circumstellar medium is magnetized \citep{lyu02}). 
Outside the CD an ultra-relativistic shock front 
may form and propagate into the surrounding circumstellar medium.
The expansion will still be non-spherical.
As long as the outflow is ultra-relativistic, the motion
is virtually ballistic and determined by the balance between the magnetic stress at the 
CD and the ram pressure of the circumstellar medium.

The angular  distribution of magnetic field (and  of the Lorentz factor of the expansion)
depends on the dynamics of the bubble at the non-relativistic stage and the
distribution of the source luminosity. 
The simplest case, which we shall analyze in some
detail and which captures the essential features of the outflow,
 is that the outgoing  current is confined to the poles and the equatorial
plane and closes along the surface of the bubble. 
This produces a toroidal magnetic field that varies inversely with 
cylindrical radius. Accompanying this magnetic field will be a poloidal
electrical field so that there will be a near radial 
Poynting flux, that is carrying energy away from the source
at almost the speed of light. 
In addition to the   outgoing  flux, there is a  much weaker reflected flux
that propagates backward into the flow the information about the 
circumstellar medium. The distribution of reflected current
is determined by the outgoing current and the boundary conditions.

\paragraph{III Shell Expansion ($\gamma$-ray Burst)}                
By the time the shell radius expands to 
\be
r_{{\rm GRB}}\sim ct_s \Gamma^2\sim(Lt_s^2/(nm_pc^2)^{1/4}\sim 3 \times 10^{16} 
L_{50}^{1/4} t_{s2}^{1/2} n^{-1/4}{\rm cm},
\ee
most of the 
electromagnetic Poynting flux from the source will have  caught up with the CD
and been reflected by it, transferring its momentum to the blast wave.  
Simultaneously a strong region of magnetic shear is likely to develop at the  outer part of the
CD \citep{lyu02}. Both of these effects are likely to lead to the rapid development
of current instabilities in the shell that will ultimately result in the acceleration 
of pairs and the emission of Doppler-boosted synchrotron emission in the $\gamma$-ray
band. Although, we defer discussion of the microphysics of particle
acceleration to Paper II, we note here that
expected radius of GRB emission is typically some three orders of magnitude larger 
than in the fireball model.
\paragraph{IV Blast Wave Propagation (Afterglow)}
For $r>>r_{{\rm GRB}}$,
most of the energy of the explosion will reside in the 
blast wave which will eventually settle down to follow a self-similar expansion.
(The structure of the 
 energetically sub-dominant electromagnetic shell will also become self-similar.)
This is the afterglow phase when synchrotron and inverse Compton radiation 
is emitted throughout the electromagnetic spectrum. The initially 
aspheric expansion  will give the appearance of a jet with the ``achromatic break'' 
occurring when the Lorentz factor becomes comparable with the reciprocal
of the observer's inclination angle with respect to the symmetry axis.
When $r> r_{{\rm NR}}\sim(Lt/\rho c^2)^{1/3}\sim2 \times 10^{18}L_{50}^{1/3}t_{s2}^{1/3}
n^{-1/3}$~cm, the blast wave become non-relativistic and will become more 
spherically symmetric, while evolving towards a Sedov solution. 

\subsection{Addressing the Problems of Fireball Models}
\begin{figure}
\includegraphics[width=0.9\linewidth]{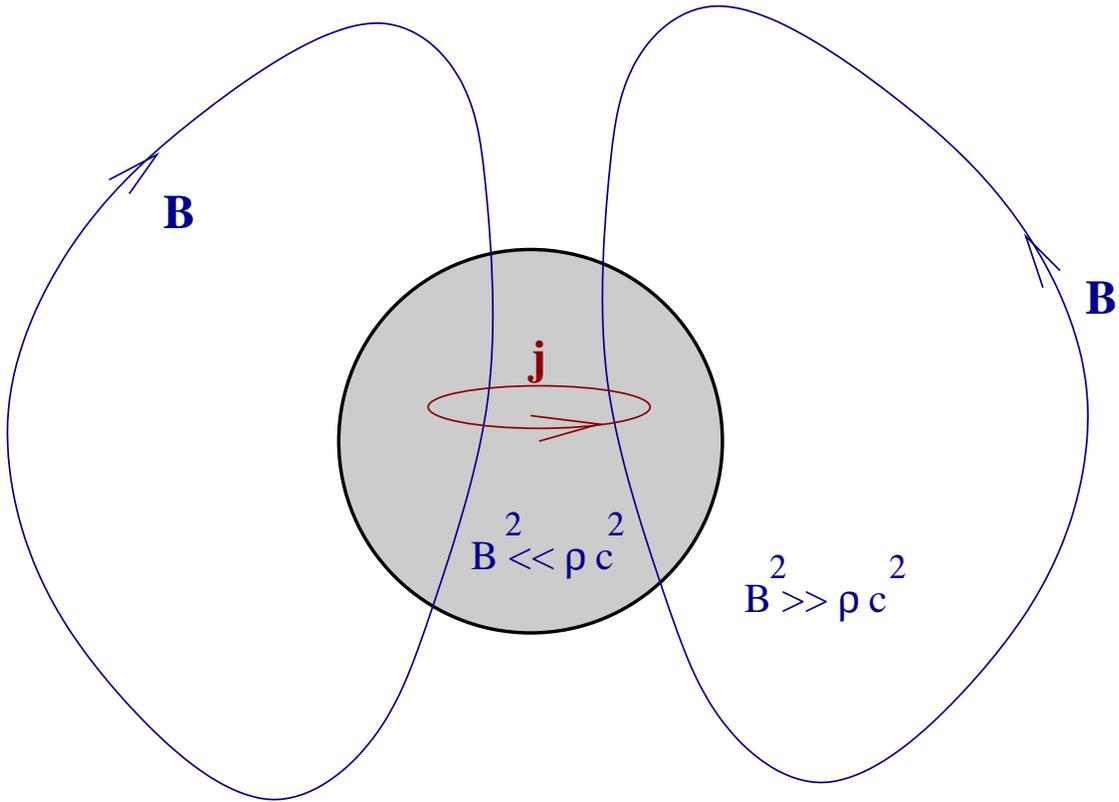}
\caption{Magnetic ''phase separation'' near the central source. The currents 
supporting the strong magnetic field flow in the matter dominated phase inside the neutron star
(or in the accretion disk in case of BH-torus system). 
The outside corona is magnetically dominated (\cf the Sun). }
\label{B-NS}
\end{figure}

Before discussing the dynamical aspects
of our model in more detail, we return to the  problems that 
we identified with the fireball model and outline how they are addressed in the 
electromagnetic model. 
\paragraph{1. How is the entropy of the fireball created?} 
Under the electromagnetic model, entropy production is deferred until late in the evolution 
of the explosion where it occurs naturally as a consequence of 
the development of various instabilities. This also addresses the ``compactness problem''.
\paragraph{2. How are the hypersonic jets made?}
As we discuss further below the effective sound speed is that of light and 
so the jets speeds may be formally subsonic.
Jet are collimated naturally through magnetic hoop stress. Of course inertial and pressure 
confinement by a surrounding stellar envelope can also be important, thought this is 
not necessary.
\paragraph{3. How can the outflow develop large, parallel, proper velocity gradients 
and avoid producing converging streams?} 
No strong constraint need be satisfied because the GRB emission arises at a much greater
radius than in the fireball model. Furthermore, the emitting region is more strongly 
coupled causally.
\paragraph{4. What determines the baryon loading of the flow?}
As the momentum is carried primarily by electromagnetic field, the baryon loading 
can be negligible. However, it clearly cannot be too large. This imposes constraints on the
amount of initial loading
 and entrainment within a stellar envelope. In analogy with the Sun 
one may expect that there are
 two ``phases''  within a source: an internal  matter-dominated one in which
large currents are flowing and an external magnetically-dominated 
(see Fig. \ref{B-NS}). If a flow
is launched from the magnetically-dominated phase the
 matter loading may be expected to be small
(analogous to pulsar wind).
\paragraph{5. Where are the thermal precursors?}
The intensity of the thermal precursor is set by the degree of lepto-photonic
loading which can be arbitrarily small \citep{lu00,dm02}. A small
precursor  seen by  \cite{fro01} indeed had very small luminosity. 
\paragraph{6. How are particles accelerated at relativistic shock fronts?} 
As we discuss further in Paper II, the particle acceleration for the GRB 
does not take place at a relativistic shock front but is instead due 
to magnetic field dissipation in the emission region. 
 Electromagnetic energy is ``high quality'': it can be effectively converted into high
frequency
electromagnetic radiation. For example, in case of Solar flares,
the primary energy output is non-thermal electrons \cite{benz03}.
 The particle acceleration that leads to afterglows
may be shock-related but could also be due to relativistic MHD modes.

\paragraph{7. How is the magnetic field amplified?}  
The electromagnetic field is already present and provides the dominant energy density during
GRB emission. In addition,
 we suppose that, during the afterglow, the magnetic field is supplied by  the 
magnetic shell and may be incorporated into the shocked circumstellar gas through 
interchange instabilities (\eg impulsive  Kruskal-Schwarzschild  instability appendix \ref{IKS})
or due to resistive instabilities of the contact surface. 
\paragraph{8. How is a large degree of $\gamma$-ray polarization created?} 
A strong argument in favor of electromagnetic models 
comes from the recent report
of large polarization in {\RHESSI} observations of the prompt $\gamma$-ray emission
from one GRB \citep{coburn03}. If high polarization is substantiated and found to be generic,
it would imply
that the magnetic field coherence scale is larger
than the size of the visible emitting region, $\sim r/\Gamma$.
Such  fields cannot be generated in a
causally-disconnected,  hydrodynamically-dominated
outflow.  Thus, the large scale  magnetic field should
be present in the outflow from the beginning and is likely to be the
driving mechanism of the explosion \citep{lyu03c}.

\paragraph{9. What determines  the jet opening angle and its structure?}
GRB outflows have large opening angles, but do not have a jet in a proper sense.
Outflows are non-isotropic so an achromatic break is inferred when the viewing
angle is $\sim 1/\Gamma$. The jet internal structure corresponds to a ''structured jet''
with $L_\theta \sim \theta^{-2}$.

\paragraph{10. What is the relation between GRBs and X-ray flashes (XRFs)?}
GRBs are seen from essentially all directions in  the electromagnetic
model.
XRFs are GRBs seen ''from the side''. The typical {\it total}  energy
(inferred from observations of early afterglows) should be similar to GRBs (within an order
of magnitude). 
In the only XRF with a redshift this is indeed the case \citep{sod03}.
In  a flux- or fluence-limited
survey, the bursts viewed at large observer inclination should be systematically closer.

\paragraph{11. Where are the ``orphan'' afterglows?}
 Since in the electromagnetic model GRB outflows have large opening angles
the  incidence of orphan afterglows should be much 
less than in the fireball model. 

There are other appealing features of the 
electromagnetic model. (i) The sources that 
are invoked are very similar to known sources. Pulsar wind nebulae, magnetars and 
extragalactic jets (as well as, perhaps, Galactic jets) are
explained as low entropy outflows associated with spinning,
magnetized, neutron stars, black holes and
relativistic disks. What is novel about the GRB is the combination of high 
field and spin in a stellar object. (ii) Variability may be due to 
the statistical properties of
 dissipation and not the central source activity. 
Magnetic fields are non-linear dissipative  dynamical system which often show 
bursty behavior with 
power law  PDS.
 (iii) ``Standard candle'' - the  narrow distribution
of GRB energies (inferred from prompt emission, \cite{fra01},
from afterglows, \cite{pk02}, and from $K_\alpha$ lines, \cite{laz02}) 
may be related to the  total rotational energy of a critically
rotating relativistic object -  a one parameter (mass) family. 
(iv) Correlations of GRB properties: hard-to-soft temporal evolution 
of GRB spectra and $E_{\rm peak} \propto \sqrt{L}$ correlation  naturally occur
in the model (see section \ref{implic}).

\subsection{GRBs as electromagnetic circuits}

\begin{figure}
\includegraphics[width=0.9\linewidth]{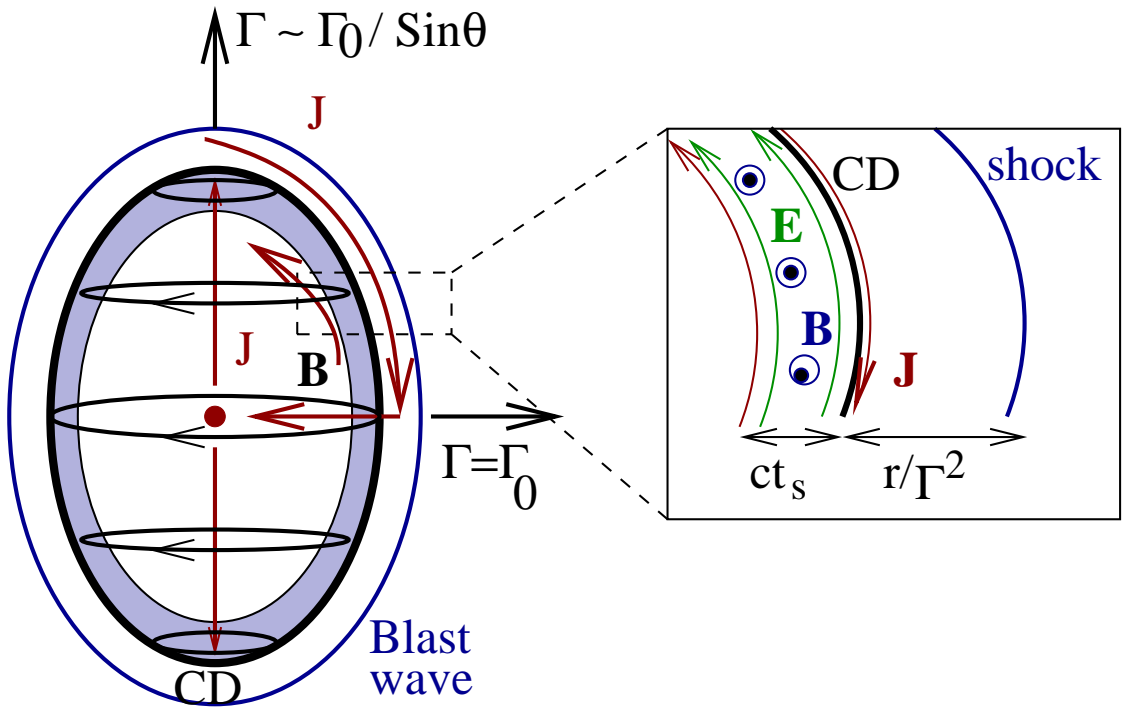}
\caption{Current flow in the electromagnetic bubble.
Current flow mostly along the axis, on the surface of the magnetic shell, along equator
and close-up at the trailing part of the shell. Magnetic shell is preceded by the
forward shock, typically $r/\Gamma^2$ ahead of it.
 Non-sphericity of the shell, which is 
of the order $\sim 1/\Gamma^2$,  is enhanced.}
\label{current}
\end{figure}

The electromagnetic model  exhibits some simple generalities that derive
from treating it as a  circuit.
Suppose that the source is threaded by a
  magnetic flux, $\Phi \sim B_s r_s^2\sim10^{26}$~G cm$^2$,
and has a typical angular velocity $\Omega \sim 10^4$ rad/s$^{-1}$.
 The source will generate  an  EMF, $V$, and an associated current, $I$
\be
\label{eqemf}
{\cal E}\sim\Omega\Phi\sim 3 \times 10^{22}B_{s,14}r_{{\rm s,6}}
\sim \sqrt{ 4 \pi L \over c} \sim 3 \times
10^{22}L_{50}^{1/2}\ {\rm V}.
\ee
There will be an associated current
\be 
\label{eqamp}
I\sim {\cal E} / {\cal Z}_{\rm load} \sim \sqrt{ L c \over 4 \pi}
\sim3 \times 10^{20}B_{s,14}r_{{\rm s,6}}
\sim3 \times 10^{20}L_{50}^{1/2}\ {\rm A},
\label{VI}
\ee
where
$B_{{\rm s}}=10^{14}B_{{\rm s,14}}\ {\rm G}$ is the source
magnetic flux density and ${\cal Z}_{\rm load} \sim  100 \, \Omega$
is the total   impedance  of the source and the emission region,
 which is  of order the impedance of free space
under general electromagnetic and relativistic conditions.
The
source region can be
thought of as a generator capable of sustaining an EMF $\cal E$.
(Under most
conditions the {\it  maximum } energy to  which a particle of charge $Z$ can be
accelerated will be limited by 
$\sim Ze{\cal E}\sim 3 \times 10^{22} $ eV, way above the highest energy of the
observed cosmic rays of $ 3 \times 10^{20}$ eV.)
This implies that the power dissipated in the load
is $ L  \sim{\cal E}^2/{\cal Z}_{{\rm load}}$.
The load consists of external medium, against which $PdV$ work is done
by the expanding shell, and radiation (some of the energy of the shell
is radiated away as a prompt emission).
An equivalent and useful way
to think about this is to say that there is a strong, quadrupolar current
distribution outward along the axes and inward along the equator (or
{\it vice versa}). Our proposal
differs from the conventional interpretation principally through
the assumption that the current flows all the way out
to the expanding blast wave, rather
than completes close to the source (c.f. Fig. \ref{current}).

We now consider, in more detail, the dynamics of the expansion.  We divide the explosion 
into four phases, source formation ($t<<t_s$), bubble inflation ($r_s/c<<t\lo t_s$), 
shell expansion ($t_s\lo t\lo r_{{\rm GRB}}/c$ and blast wave propagation
($r_{{\rm GRB}}/c\lo t$). Each of these phases involves distinct, dynamical behavior 
requiring different approximations to describe.

\section{Source Formation}
\subsection{Nature of the Compact Object}
As explained above, there is now good circumstantial evidence linking 
at least some GRBs with simultaneous supernova explosions -- the collapsar 
model.  In its original and most common form, \citep[\eg][]{woo93}, it is supposed that the 
core collapse of a massive star leads to the formation of a massive black hole
orbited by a dense, thick accretion disk \footnote{The observed coincidence 
between the burst and supernova is contrary to the expectation of the alternative
supranova model \citep{vie98}.} It is also assumed that magnetic flux is 
 generated locally by dynamo action
 \citep[\eg][]{tm01}.
 The actual power 
may derive from the spinning spacetime of the black hole or at the expense of the
binding energy of the orbiting gas. For a rapidly spinning
hole of mass $\sim10$~M$_\odot$, a field of strength $B_s\sim10^{14}L_{50}^{1/2}$~G
suffices in either case. 
The simplest descriptions of these processes  
 comprise solutions of Maxwell's equations in a curved spacetime under the 
force-free approximation. They describe, at least conceptually, a stationary, axisymmetric 
flow of electromagnetic energy and angular momentum away from the surface of the hole and 
the disk. There is an associated current distribution that is quadrupolar so that the sign
of the radial component of the current changes with latitude. 
 
Real source is unlikely to be either 
stationary or axisymmetric.  The  field configuration may well be unstable
and irregularities in the disk will break the symmetry. Our 
fundamental assumption, that underlies most of what follows, 
is that these instabilities do not 
develop to large nonlinear amplitude and completely disrupt the outflow. In other 
words, the  
large scale field is simple and approximately axisymmetric beyond the light cylinder. 
This large scale regularity 
can come about either because the smaller scale magnetic structure is erased
through dissipation or, in the case of a central neutron star,
through a topology-preserving re-arrangement
of the magnetic flux. Our ``DC'' model is therefore rather different from other ``AC''
proposals that GRBs be powered by a rapidly varying Poynting flux 
\citep[\eg][]{lb00,sdd01,sik03}.

In the magnetar model, it is proposed that the neutron star is born with a rotation 
frequency $\sim 3  L_{50}^{1/2}t_{s2}^{1/2}$~kHz and a  dipole field  of strength
$B_s\sim 10^{14}L_{50}^{-1/2}t_{s2}^{-1}$~G so that it can produce the inferred, 
electromagnetic power and energy. 
If the dipole is inclined with respect to the rotation axis, we expect that 
some of  the open magnetic
flux from the northern magnetic pole finds its way into the southern hemisphere beyond 
the light cylinder and {\em vice versa}. 
Adopting our conjecture, provided that the dipole is not too
inclined, the asymptotic electromagnetic configuration
is roughly axisymmetric with field lines ending up in the hemisphere from which they started.
In other words, the corrugations in the current sheet that separates the 
north seeking and south-seeking field lines
 smooth out, with little dissipation, close to the light cylinder.
\footnote{There is a good precedent for this behavior in Ulysses observations
of the quiet solar wind \citep{ulys00} which
reveal that, despite the complexity 
of the measured surface magnetic field, the field in the solar wind
quickly rearranges to form a good approximation to a \citet{par60} spiral.}
Much of what follows is predicated on this hypothesis, that a large scale quadrupolar current 
flow is established and provides a good description of the subsequent evolution of the bubble 
and the shell \citep[\eg][]{bla02}.

In the immediate vicinity of the 
 source the  plasma 
is  separated into two phases:  matter-dominated and magnetically dominated
(c.f., Solar photosphere, pulsar magnetospheres).
 Superstrong  magnetic fields  are  
generated in the dense medium (a  disk or  a differentially rotating  neutron star-like
object). 
 Buoyant magnetic field lines emerge into the tenuous magnetosphere
while remaining 
 anchored in the matter-dominated phase.
Strongly relativistic outflow is generated in the 
magnetically dominated phase. 

Although the magnetosphere is comparatively small,
the electrodynamical conditions at the light cylinder constitute a boundary condition
for the eventual, relativistic outflow, much like what happens with vacuum 
electromagnetic radiation. We can think of these conditions either as establishing
the current distribution or, equivalently, as defining the 
subsequent evolution of the electromagnetic field. As the source will,
typically, remain active for $\sim$  million dynamical times, we suppose that it will be able
to settle down quickly to a quasi-steady state evolving slowly as the hole or neutron star
slows down on a time scale of $\sim$ 100 s.

\subsection{Dissipation at the Source, $r\sim r_s$}
\label{ssec:sourcediss}

In any scenario some fraction of the central source luminosity 
is likely to be dissipated close to the source. In other words, there
is a source impedance ${\cal Z}_s$. 
In the case of electromagnetic extraction of energy from a spinning black hole,
this dissipation occurs beyond the event horizon. For a magnetar, the neutron star
impedance is negligible and ${\cal Z}_s$ is dominated by what happens in the magnetosphere.
In the fireball model, it is implicitly assumed that all of  the energy released is quickly 
converted into heat, forming a high entropy
per baryon, thermal plasma. 
In other words, the load impedance  ${\cal Z}_L$ is located close to the source.
By contrast, in the electromagnetic model this does not 
happen and the energy flows way from the light cylinder mainly 
in the form of an electromagnetic Poynting flux
and the load impedance is located in the emission region.
We argue that this is likely to be the case because, somewhat paradoxically,
it becomes harder to convert electromagnetic energy 
directly to pair plasma the stronger the magnetic field becomes. The reason is that 
the plasma surrounding the source is 
 highly conductive. 
In such plasma there is usually plenty of charge available to screen the component
of the  electric 
field along the magnetic fields, so that the first electromagnetic invariant
is close to zero: $\E\cdot\B \sim 0$ (the perpendicular component
of the electric field just defines  a plasma velocity as long as $E<B$, see below). 

In strongly relativistic plasmas, a possible source of (inertial) resistivity 
 is related to the break down of the $\E\cdot\B \sim 0$ approximation in cases
when the real charge  density falls below some critical value.
In the case of rotating magnetic field, this  minimum charge
density needed to short out the component of electric field along the magnetic
field is  known as the ``Goldreich-Julian''
density
\be
n_{GJ} = {{\bf \Omega} \cdot \B \over 2 \pi e c} \sim 10^{17} \mbox{cm$^{-3}$}
\label{GJ}
\ee
The minimum energy density associated with this amount of plasma is  
$ n_{GJ}mc^2 $.  It is convenient to introduce a parameter
$\sigma$ as the ratio of the magnetic 
energy density $u_B=\B^2/8\pi$ to the total plasma energy density $u_p$ including photons
 ($u_p$ has contributions
from rest mass of ions and pairs plus internal energy)
\be
\label{sigmadef}
\sigma={u_B\over u_{p}}
\label{sig}
\ee
If the  energy density $u_p$ is dominated
by leptons with density given by Eq. (\ref{GJ}), then the ratio (\ref{sig}) 
is roughly the ratio of the light cylinder radius
to the electron gyro radius which can be as large as $\sim10^{18}$.

A conservative flow of plasma may 
 not be able to satisfy the constraint that
the local charge density  always exceeds  $n_{GJ}$.
If this happens, a gap  will develop where the 
 field-aligned electric fields  are nonzero $\E\cdot\B \neq 0$.
However, the  maximum potential drop that is available for dissipation
will be  limited by various mechanisms of pair production. Typically,
after an electron has passed through  a potential difference
 $\Delta V\sim 10^9-10^{12}$ V
it will  produce an electron-positron
 pair either through the  emission of curvature photon
or via  inverse Compton scattering. This will be followed by an electromagnetic cascade and
the   newly born
pairs will create a charge density that would shut-off the accelerating
electric field.
The typical potential difference  required for the  creation
is orders of magnitude smaller than the total available EMF 
 ${\cal E}  \sim 10^{22}$ V.
In this case the magnetization parameter may be estimated as
\be
\sigma  \sim { {\cal E} \over \Delta V} \sim 10^{10}
\ee

Another possible way through which an electromagnetically-dominated flow
can create entropy directly and reduce $\sigma$ appreciably is through the development of 
an electromagnetic turbulent cascade
operating down to wavelengths, $\lambda_{{\rm min}}$,
so small that electromagnetic energy can be dissipated directly in particle acceleration.
This is analogous to the viscous dissipation that terminates
a fluid turbulence spectrum. If this really can operate then it is hard
to see how more than a few percent of the electromagnetic energy will be
dissipated in this fashion.  More specifically, for a differential cascade
 a typical cascade time is 
the interaction time multiplied by the logarithm of the outer and inner scale \citep{Zakharov}.
If the typical interaction time is a light crossing time, then 
we can estimate 
 $\sigma\sim
\ln(r_s/\lambda_{{\rm min}})\sim100$ (where 
$\lambda_{{\rm min}}$ of the order of Larmor radius), so that
 the flow will remain
electromagnetically dominated.

Another
 source of dissipation is magnetic reconnection. This is surely important if 
 the outflow 
retains an AC component, contrary to our hypothesis, or if  current sheets develop 
as a result of the nonlinear evolution of electromagnetic instabilities.
Driven relativistic reconnection may proceed at velocities approaching the speed
of light \citep{lu03}, 
 however the initial  development of dissipative current sheets 
  occurs  on a timescale intermediate between very long resistive and very short
dynamic (light travel) 
timescale and  may be too slow
 to dissipate much of the magnetic energy density \citep[]{lyu03}.

Thus, 
an abundance of pairs can be created 
without changing the electromagnetic dominance and the more pairs are present, the 
harder it becomes to create the small ``gaps'' that would be necessary to 
replenish them. 
 Put another way, the pair density required to supply the electrical 
current and space charge scales linearly with the field strength, while the electromagnetic 
energy density scales as its square. The stronger the field, the more likely it is to persist
into the outflow. 
It is because GRBs are so powerful that the dissipation in the source is probably low.

\section{Bubble Inflation, $r\leq c t_s$}
\label{ssec:bubble}

In the previous section
we have argued that the  flow formation occurs on a scale of a light cylinder, 
$r\sim r_s \sim 10^6$ cm and that the 
dissipative processes are not likely  to drain all the potential EMF, so the 
flow   is likely to remain magnetically-dominated.
 The velocity with which the plasma leave the neighborhood of the light cylinder
is determined by the details of the acceleration and magnetic field structure at the source.
For the purpose of this paper we leave the question of the detail
structure of the  central source open, assuming that the 
formation of the flow  occurs in  a way similar to  pulsar magnetospheres \citep{Michel69,gj70}, 
with an important difference -- in GRBs pressure effects may play an important role.  
Beyond the light cylinder the magnetized flow generated by the central source
will expand due to magnetic and pressure forces
\citep[see also][]{whe00,mabk03}. It will  become super-fast-magnetosonic
and thus causally disconnected from the source \citep{gj70}. 
For subsequent evolution the conditions close to source  
may be considered as 
boundary conditions at which the rate of energy and magnetic flux injection is some
given function. 

The asymptotic structure (at $r\gg r_s$) of axisymmetric, magnetized outflows is a
challenging problem 
that has a considerable literature \citep[\eg][and references
therein]{HeyN03}.
Heyvaerts and Norman argue that a perfect, non-relativistic MHD flow
with five conserved 
constants of the motion evolves either to a state where the current is
confined
to the axis and a finite number of thin current sheets or that the
current closes and that  the
outflow energy flux becomes purely mechanical (for relativistic flows
the latter happens at distances much larger than astrophysically relevant scales).
 In the context of a
quadrupolar current 
distribution, the first option is equivalent to stating that the current
becomes 
concentrated on the axis, surface and equator. They also argue
that the second 
possibility is ultimately favored but that finite flows may not, in
practice, achieve 
this state.   
\footnote{An alternative 
way to justify this particular electromagnetic configuration is to note
that 
it minimizes the magnetic energy associated with a given quantity of
magnetic flux.}.

As the wind expands, it starts to interact with the surrounding medium, so that its
properties are determined both by initial conditions and interaction with the
 surrounding medium. 
The wind is  slowed down by  interaction with the stellar material, so that 
initial expansion is non-relativistic, and 
 later, after breakout, it becomes strongly relativistic. 
The physical processes governing these two phases are quite distinct. 
We consider them in turn.

\subsection{Non-relativistic expansion}
\label{ssec:nrexpansion}
Consider a newly-formed, compact object inside a star or other dense gas
distribution
generating an EMF ${\cal E} (t)$ and driving a quadrupolar current flow
$I(t)$ as described above. This will inflate an electromagnetic bubble
expanding 
at a rate controlled by the 
external gas density.  Let the bubble radius be $R(\theta,t)$ where
$\theta$ 
is the polar angle measured from the symmetry axis defined by the spin 
of the compact object.  For the moment,
suppose that the bubble expands non-relativistically. 

We expect magnetic flux (integrated over the meridional plane) to 
cross the light cylinder and be supplied to the bubble at 
a rate $\dot\Phi\sim\mu_0Ic/2\pi$. Similarly, Poynting flux will 
be supplied to the bubble at a rate $\dot U_{{\rm EM}}\sim {\cal E}I$
\footnote{If there is outward and return current at intermediate
latitude, 
then $\dot U_{{\rm EM}}=\int d{\cal E}I$, where $I$ is the total enclosed
current.}.
We can also compute the magnetic flux $\Phi={\cal L} I$ and the energy
stored 
within the bubble $U_{{\rm EM}}={\cal L} I^2/2$, using the self inductance ${\cal L}$.
If, as we discuss 
further below, the magnetic field in the bubble is predominantly toroidal
between cylindrical radii $\varpi_{{\rm min}}$ and $\varpi_{{\rm
max}}=R\sin\theta$, this
is given by
\be
\label{selfinduct}
{\cal L} \sim{\mu_0\over2\pi}\int dz\ln\left({\varpi_{{\max}}\over \varpi_{{\rm
min}}}\right).
\ee
We therefore see that, if the bubble expands 
homologously and sub-relativistically, the rate of supply 
of both flux and energy exceeds the rate at which 
the flux and energy can be stored by a factor 
$\sim[\ln(\varpi_{{\rm max}}/\varpi_{{\rm min}})(dz/dt)/c]^{-1}$
\citep[\cf][]{rg74}. Therefore,
too much flux and energy is generated  by the source when the expansion speed is less
than 
$\sim0.1c$.  

The fate of this surplus flux and energy depends upon the amount of
dissipation within the
bubble. 
If, despite strong driving,  the 
 resistance in the  electrical circuit is sufficiently low 
(much less than $\sim100\Omega$), then the  electromagnetic
energy would be either reflected back to the source changing its properties
(this can
and does
happen with solid conductors, \eg waveguides, but is unlikely to happen in a
plasma environment), or 
there will be ``inductance breakdown'', so that ${\cal L}$ will decrease below the estimate
(\ref{selfinduct}). This will be achieved by destroying axial symmetry of the flow by MHD instabilities
and 
creation of smaller scales current, much along the lines of what has been suggested in  the Crab
Nebula by \cite{beg99}.
On the other hand, 
 if the resistance in the electrical circuit is sufficiently
high, 
magnetic flux and electromagnetic energy will flow toward those parts of
the current flow 
where the resistance is located. Magnetic energy then can be dissipated
at a very high rate (this is similar to the case of  driven magnetic reconnection, which
in relativistic plasma may proceed at the speed approaching a speed of light \citep{lu03}).
 The total electrical resistance
needed can be estimated from the ratio of the potential difference
at the light cylinder to the current as $\sim60\Omega$ times a
logarithmic factor 
and we will estimate the total resistance  as $100\Omega$. In this paper, we assume that
the 
current flows out to the boundary of the bubble and the Poynting flux is
transformed into
heat {\it until the expansion speeds exceeds $\sim0.1c$}. Thereafter, it can
be accommodated 
non-dissipatively by the expanding, electromagnetic bubble.

However, if the resistance in the outer part of the bubble {\it  always }
exceeds $\sim100\Omega$, then the current will
complete closer to the compact object and the outer parts of the bubble
will comprise 
hot radiation/pair-dominated plasma.  This is the implicit assumption
underlying fireball
models of GRBs that invoke electromagnetic sources.

As mentioned above, ideal MHD flows are naturally collimating. 
At the non-relativistic stage of expansion there is another 
collimating effect due to resistive dissipation, which,
as we have argued, must be  present in the flow.
 Most
of the dissipation
is likely to occur near the axis where the current density is highest and
the susceptibility to known instability is the greatest.
In this case a lateral 
flow of energy will set in carrying the poloidal field lines with it towards the axis
(dissipation of magnetic energy on the axis will result in a loss of  magnetic pressure, which
resists the inflow of plasma towards the axis, and will be communicated
to the bulk of the flow by a
rarefaction wave propagating away from the axis).  
This, in turn, will leads to the
pile-up of magnetic field near the axis and to faster radial expansion
(the toothpaste tube effect).

Next  we consider the structure of the axial current, whose radius
$\varpi_{{\rm min}}$ we have already introduced.  If the energy that is
dissipated were 
to be radiated immediately and to escape freely, then the field
structure 
would evolve to become force-free everywhere. This would imply that 
$\varpi_{{\rm min}}\sim r_s$, the radius of the light cylinder.  However,
this is surely 
not the case. The optical depth of the plasma will be far too large and
the 
radiation will be trapped and thermalize. This implies that pressure is
likely
to become quickly important in the core. For $r>>r_s$, the net poloidal
field is 
quite small.  The simplest structure to consider is that of a Bennett
pinch.
In a Bennett pinch, the internal energy associated with the plasma is 
$3\mu_0I^2/16\pi$ per unit length so that 
the mean value of $\sigma$ is $4\ln(\varpi_{{\rm max}}/\varpi_{{\rm min}})/3$.  A
substantial 
plasma energy density is needed to oppose the magnetic stress
independent of the 
choice of $\varpi_{{\rm  min}}$. However Bennett pinches are notoriously
unstable
and so this is also unlikely to be a complete description of the
current.
The instabilities that would  develop within the core lead to 
a less ordered and dynamic magnetic field with pair plasma contributing
to the overall stress 
tensor to an extent controlled by the balance between its rate of
creation and annihilation
which we discuss  below. Furthermore, we suppose that these
instabilities lead to dissipation
and that the effective impedance in the circuit automatically adjusts to
match that required to
account for the electromagnetic energy and flux supplied by the source
at the light cylinder
and the rate of expansion of the bubble.

In summary, we adopt a particularly simple  distribution of currents generated by the central source:
along the axis, the surface of the magnetic bubble and closing in the equator. 
More general
current
distribution 
do  not change the picture qualitatively, but will lead to
quantitative
changes in our conclusions.

\subsection{Dynamics of magnetic bubble inside a star}
 \label{inside1}
 
Electromagnetic bubbles are crucially different from expanding fireballs in another way. They are 
naturally self-collimating and will create bipolar outflows even if the 
stellar mass distribution is spherically symmetric.

The dynamics of such non-spherically expanding bubble may be described using the
method of \cite{komp60} which was initially developed for propagation
of non-spherical  shocks \citep[for astrophysical applications see][]{Icke88,bks95}.
 Consider  a small section  of non-spherical 
 non-relativistically
expanding 
CD with radius  $R(t, \theta)$.
 The CD   expands under the pressure of magnetic field 
so that  the normal magnetic stress at the bubble surface is balanced
by the ram pressure of the surrounding medium.
At the  spherical polar
angle $\theta$ the CD  propagates  at an angle 
\be
\tan \alpha = - { \partial \ln R \over \partial \theta}
\ee
to the radius vector.
Balancing the pressure inside the bubble $B^2/(8 \pi) = I^2 /( 2 \pi c^2 R^2)$
with the pressure of the shocked plasma  we find
\be
\label{bubbleexp}
\left({\partial R\over\partial t}\right)^2_\theta= \kappa {I^2(t)\over2\pi R^2\sin^2\theta
\rho(R,\theta)}\left[1+\left({\partial\ln R\over\partial\theta}\right)^2_t\right]
\label{inside}
\ee
where $\kappa $ is a coefficient of the order of unity
which relates the pressure at the CD to the pressure at the forward shock.

Equation (\ref{inside}) shows that non-spherical expansion inside the star is due both
to the anisotropic driving by magnetic fields  {\it and} 
collimating effects of the stellar 
material (the term in parenthesis, which  under certain conditions
tends to 
amplify  non-sphericity).
Note that this use of the Kompaneets approximation assumes that the shock  and the CD
are located close enough so that there is no lateral (in $\theta$ direction)
 redistribution of pressure in the shocked material. This is a 
good approximation for
accelerating shocks and is used here only to illustrate qualitative behavior of the
solutions.

The rate of expansion of the bubble inside the star
depends upon the dynamics of the stellar envelope and the time evolution of the 
current $I(t)$. 
For a given dependence $\rho(R, \theta)$ and $I(t)$ Eq. (\ref{inside})  determines the 
velocity of the CD. Generally solutions will be strongly elongated along the
axis. A simple analytical solution 
 for $I,\, \rho \sim const$ is  
\be
 R (t,\theta) = \left( { 2 \over \pi} {I^2 \over \rho c^2 } \right)^{1/4}{  \sqrt{t} \over 
\sin \theta}.
\ee
(current is related to the luminosity by Eq. (\ref{VI}).
Qualitatively, the bubble and the forward shock will cross
the iron core ($r_c\sim 2.5 \times 10^8$ cm) in $t \sim r_c \sqrt{\rho} / B(r_c)
\sim .3 $ sec, short enough to produce an ample supply of $^{56}Ni$ \cite{Woosl02}.
 If we define $M(R)$  as the stellar mass external to radius $R$, then we can  also
estimate the  breakout  time  
\be
\label{eqexp}
t_{{\rm breakout }}(\theta) \sim { 1}  \theta_{-1}^2
\left({M\over M_\odot}\right)^{1/2}\left({R\over R_\odot}\right)^{1/2}
L_{50}^{-1/2}~{\rm s}
\ee
The electromagnetic bubble can be confined equatorially
by the star for the duration of the burst  $t_{{\rm breakout }}(\pi/2) \sim 100 $s 
and will expand non-relativistically as
we have assumed. However the expansion along the axis proceeds
on a short timescale and  breakout should occur
early in the burst. 
Furthermore, at the time of breakout, the axial expansion 
speed of the bubble will become relativistic.  

Non-relativistic expansion  lasts for several seconds along the axis. This time is much
shorter than the burst duration. Recall that  when the velocity of expansion is $\leq 0.1$,
 a considerable fraction of the magnetic  energy is indeed dissipated. But since the flow
quickly becomes relativistic, the relative fraction of dissipated energy is small, so that
the flow magnetization remains large, $\sigma \geq 100$.
An important advantage of the electromagnetic models is that as long
as $\sigma \geq 1$, the asymptotic evolution of the flow is mostly
independent of $\sigma$. Most of the dissipation described above will result in creation
of lepto-photonic plasma, which decouples after photosphere, so that the remaining
flow remains strongly magnetized. For $\sigma \geq 1$ the energy associated with the 
thermalized component is not important for the  flow dynamics.

In summary, our contention is that, initially, when the bubble expands non-relativistically, 
the dissipation 
is concentrated along the pole and the current returns to the source along the surface of the 
bubble. The polar  current pinch is supported against collapse at its
core by a combination of plasma pressure and, possibly, by  dynamical, disorganized magnetic
field. After breakout, the expansion becomes relativistic and the resistance falls so 
that the electromagnetic energy that is still being supplied by the source is mostly absorbed 
by the inflating bubble and by doing work against the surroundings.

\subsection{Early optically thick expansion: mini-fireball}
\label{early}

In the previous subsection we have argued that  energy release and initial 
non-relativistic stage of expansion  are necessarily
accompanied by partial dissipation of the magnetic energy, but the flow
is likely to remain magnetically dominated with $\sigma \gg 1$.
For any reasonable $\sigma \leq 10^{10}$ the lepto-photonic component
will be optically thick near the central source
 and consequently in a  thermodynamic quasi-equilibrium.
In this subsection and in appendix \ref{warm}
 we   discuss the 
 optically thick, quasi-spherical  expansion of a relativistically hot,
magnetically-dominated flow  after 
the flow became weakly 
relativistic and no further dissipation of magnetic energy is 
happening in the flow,
but at the early enough stages 
so that photons  remain trapped
 \citep[see also][for a more extensive analysis]{vk03a,vk03b,fo03}.

Under the  electromagnetic hypothesis, most of the energy released by 
the source comes out in 
the  form of Poynting flux. However, as we argued above,  there must be some dissipation
that would lead to creation of  a 
lepto-photonic component. In addition, some ions may be present
in the flow. The luminosity of the source then can be written as 
\be
L= \int d \Omega \Gamma^2 r^2 \beta  c \left(b^{2}/2+ w \right) 
\label{L}
\ee
where $w$ is plasma enthalpy, $b$ is  a toroidal magnetic field in the plasma rest frame,
$ \Omega$ is a solid angle. The luminosity (\ref{L})
includes contributions from  the rest mass energy density
of ions and pairs, their kinetic pressure and  the trapped photon gas. 
At  early stages the  plasma enthalpy is strongly dominated by 
lepto-photonic plasma with a temperature 
\be
T \sim \left( { L \over a \Delta \Omega r^2 \beta  c \Gamma^2 (1+\sigma)} \right)^{1/4}
\ee
where $\Delta \Omega$ is a typical opening  solid angle.
(The magnetization parameter in this case is $\sigma = b^{2}/2 /w$).
For any reasonable values of $\sigma$ close to the source this temperature is high enough, so that
pairs are freely produced.
At  breakout, $r_{breakout} \sim 10^{10}$ cm,  plasma is moving weakly relativistically,
$\Gamma  \sim 1$, so that 
the temperature
may still be  high enough for plasma to be  pair-dominated
\be
T_{breakout} \sim 100\, {\rm keV}  L_{50}^{1/4} \Omega_{-2}^{-1}  \sigma_{2}^{-1/2} r_{breakout,10}^{-1/2}
\label{Tbr}
\ee

At this point the flow will accelerate to relativistic velocities.
Initially, the expansion is mostly
 pressure-driven, even in the strongly magnetized case.
This results in  dynamics qualitatively similar to the unmagnetized case.
During outflow, the wind plasma accelerates $\Gamma \sim r$ while 
its density, pressure and temperature   decrease
$n\sim r^{-3}$, $p \sim r^{-4}$, $T \sim r^{-1}$.   During the pressure-driven  expansion the
flow becomes superfast magnetosonic $\Gamma^2 >  \sigma$, while the
magnetization parameter remains approximately constant  (appendix \ref{warm}).
(Note that the  magnetization parameter  is equal to the ratio of the  Poynting to the
 particle
fluxes. Since it remains approximately constant for quasi-spherical expansion, 
there  is little transfer of energy between the  magnetic field and the  plasma at this stage
for the assumed quasi-spherical geometry of the outflow.)

When the  temperature falls below $\sim 10-20 keV$, most of the  pairs annihilate. 
This occurs at 
 $r_{ph} \sim   10^{11} $~cm. 
This suddenly reduces the  optical depth to Thomson scattering below unity.
(Under certain conditions
photons may  remain trapped
in the flow  \citep[Section \ref{tauT} below, also][]{lu00}. In this case, thermal
 driving  by photon pressure continues, until the  thermal photons
escape.) As a result
the lepto-photonic part of the flow decouples from the  magnetic
field
and $\sigma$  increases by roughly seven  orders of magnitude
to $\sigma \sim 10^{9}$. 
By this time most of the thermal energy has been spent on accelerating  the flow
  to $\Gamma \sim 10 $.
Beyond the  photosphere, thermal photons  propagate freely. The
 thermal radiation from the lepto-photonic component
has a rest-frame temperature $T_0 \sim 10-20 keV$ times  a boost
due to the bulk motion.
This  thermal radiation, which should peak around $\sim 100 $ keV  may  put constraints on the
initial $\sigma$ \citep{lu00,dm02}. There are indications that the
thermal precursor has been observed \cite{fro01},
 with 
 intensity $\sim 1\%$ of the total burst intensity. This can be used to estimate 
the magnetization parameter below photosphere as 
$\sigma \sim 100$, so that beyond the photosphere $\sigma \sim 10^9$.

\subsection{Thomson and  pair production depths }
\label{tauT}
Next,  we consider conditions on the  optical depth to Thomson scattering and pair 
production in electromagnetic models beyond the  photosphere when the ejecta  
  plasma is cold, collisionless (so that
the  number of pairs
is conserved) and   strongly magnetized ($\sigma \sim 10^9$).
Under certain conditions, determined below,  it 
may remain optically thick to Thomson scattering
and  to pair production. 

Under the approximation above, the
 total energy is  carried by  the Poynting  and 
particle fluxes (the latter   includes a
contribution from the pair rest mass and
proton rest mass). If we neglect  possible effects of pair production due to
the  dissipation
of  magnetic field, the
magnetization parameter becomes
\be
\sigma = { {b}^2 \over 4 \pi  c^2 m_\pm n_\pm' ( 1+ \xi)}
\label{ss1}
\ee
where $\xi =  m_i n_i' / m_\pm n_\pm'$ is the ratio of ion to pair mass fluxes. 
(In a standard fireballs, $\sigma =0$, and pair are dynamically unimportant
 $\xi \sim  m_p/m_e$;  for magnetized, 
pair-loaded   flows , $\sigma \gg 1$,  $\xi \leq 1$. )
The total energy flux then can be written as
\be
L =  \Gamma^2  {b}^2  r^2 c { 1+ \sigma \over \sigma} \Delta \Omega
\ee
and we  find that 
\be 
n'_\pm = { L \over  \Delta \Omega r^2 m c^3 \Gamma^2 (1+ \sigma)  (1 + \xi) }
\ee
Introducing  the 
 compactness parameter 
\be
l_c ={  L \sigma_T r'  \over   \Delta \Omega  \Gamma^2  r^2 m  c^3 } 
\label{lc}
\ee
where $r' \sim r/\Gamma$  is the typical size of emission region in the rest frame, 
 we 
find that the
optical depth to Thomson scattering is 
\be 
\tau_T = { L  \sigma_T r' \over  \Delta \Omega r^2 m c^3 
\Gamma^2 (1+\sigma) (1 + \xi) } = {l_c \over  (1+ \sigma) (1 + \xi) }
\label{lcc}
\ee
In order to escape, we require that
photons must have 
$\tau_T \leq 1$. This places a lower limit on the radius:
\be
r > \sqrt{ L  \sigma_T r'  \over  \Delta \Omega m c^3 \Gamma^2 (1+\sigma) (1 + \xi) }
\ee
which under the  assumption $ r'  \sim r/\Gamma$ gives
\be
r>  =  { L  \sigma_T  \over  \Delta \Omega   m c^3 \Gamma^3 (1+\sigma) (1 + \xi) }
\label{rT}
\ee
Thus, for large  magnetization, $\sigma > 1$, even for pair dominated flow, $\xi \ll 1$ 
the ejecta become optically thin to Thomson scattering at
\be
r >  10^{8} (1+\sigma)_9^{-1} \Delta \Omega_{-2}^{-1} \Gamma_2^{_3} \, {\rm cm} 
\ee 
(For an  electron-ion plasma $\xi = m_p/m_e$, so that for a given $\sigma$ the flow
becomes optically thin at radii which are three orders of magnitude 
smaller than for a pair plasma.)

Next, we 
 estimate the radius at which the  optical depth to pair production falls below unity. 
Assume that most photons emerge with energy
 $E_{br}\sim 200$ keV. The 
optical depth to pair production then  becomes
\be
\tau_{\gamma -\gamma} = n'_\gamma(E_{br}) \sigma_{\gamma-\gamma} r ' =
{ \epsilon_\gamma  L_\Omega \over  r^2 \Gamma c E_{br}}  {11 \over 180}
 \sigma_T {r \over \Gamma} \Delta \Omega
\label{taug}
\ee
where  $\epsilon_\gamma$ is the fraction of the  luminosity 
emitted in $\gamma$-rays,
$\sigma_{\gamma-\gamma} = (11/180) \sigma_T $ is the cross-section for pair production
by particles with power law spectrum \citep{Sven87}, and  primes denote quantities measured
in the outflowing frame.

Using the  definition of the compactness parameter (\ref{lc}), we can write the
 optical depth to pair production
(\ref{taug}) as
\be
\tau_\gamma = {11 \over 180} \epsilon_\gamma  \Gamma l_c {mc^2 \over E_{br}}
\label{taug0}
\ee
This can be called  photon compactness.
Since $mc^2 /E_{br} \sim 2$, 
\be
\tau_\gamma \sim 0.1  \Gamma l_c 
\label{taug1}
\ee
The 
condition $\tau_{\gamma -\gamma} < 1$ gives
\be
r_{\gamma-\gamma} > { 11 \over 180}  {mc^2 \over E_{br} } \epsilon_\gamma 
  {  L_\Omega   \sigma_T  \over  m c^3  \Gamma^2 }
= 3 \times   10^{16} \,  L_{50}\, \Gamma_2^{-2} \, {\rm cm}
\label{taug3}
\ee

The ratio of $\tau_T$ to $\tau_\gamma$ may then be used to find
the ratio of photons to electrons
\be
{n_\gamma ' \over n_\pm} \sim 0.1 
{\Gamma \sigma (1 +\xi)  } \gg 1
\ee
Thus,
 there are  many  more photons than electrons and  
the flows will  first become optically thin to Thomson scattering, but will
remain optically thick to pair production  up to much larger distances. 
For strong magnetization, $\sigma \gg 1$ the flow will become Thomson thin  right after
 breakout, while 
remaining  optically thick to pair production up  to  $r\sim 10^{16} $ cm (Eq. (\ref{taug})).
Note that in the electromagnetic
 model   pair production may play a more 
important role than in the hydrodynamic models since it can  regulate the component of 
 electric fields along magnetic field
 and control acceleration of pairs.

\subsection{Magnetic acceleration and collimation in the  relativistic regime}
\label{Collim}

The behavior of magnetized winds depends both on the conditions at the source
(\eg the  lateral  distribution 
of the energy and magnetic  fluxes)  and on the subsequent dynamics.
 Since for $t < t_s$ the  expansion is quasi-stationary,
we can assume that, at this stage, the progenitor produces a steady relativistic wind.
There is an extensive literature on acceleration
and magnetic collimation of outflows from pulsars
\citep{Michel71,ben84,sl90,beg92,fca95,bts99,lyub02} and from disks \citep{bp82,lws87,cl94,cam95,fo03}. (Unfortunately, even in the cleanest case
of pulsar winds, the   dynamics of relativistic MHD outflows 
remains  an unsolved problem).

Magnetic  outflows can be naturally self-collimating through the action of 
 magnetic hoop stresses, $\propto B_\phi^2/4 \pi $.
Generally, this   magnetic hoop stress is   counterbalanced by the gas
pressure and magnetic field gradients, so that the flow evolution
(collimation or de-collimation) depends on this balance 
 \citep[\eg][]{Heyvaerts,Chiueh,vk03a,fo03}.
Collimation also affects the  magnetization of the flow. For radially expanding  flow,
$ \sigma$ remains constant since both the  magnetic energy density and  the plasma rest mass 
energy density scale as $\propto r^{-2}$, so there is no transfer of energy between them
(in this case magnetic hoop stresses are exactly compensated by the gradient
of magnetic energy density).
Collimation (or decollimation) leads to transfer of energy from magnetic field 
to plasma (or vice versa). 

An important property of ultra-relativistic outflows is that their motion is 
virtually ballistic, so that any collimation should be achieved  close to the source
where the flow is only 
 mildly relativistic. 
 This is a well-known  problem
for stationary pulsar outflows and in AGNs \cite[\eg][]{beg92,bts99}.
The reason for the  weak collimation is that far from the source
 the gas pressure and the poloidal 
magnetic fields are unimportant and the 
magnetic stress $\propto B_\phi^2$ is almost exactly balanced by the 
oppositely directed electric stress  $\propto \beta^2  B_\phi^2$.
Thus, the resulting stress is quite small
  $\propto B^2/\Gamma^2$ \citep[\eg][]{bogo01}.
 Increasing the  magnetic field
strength does not help either, since in the strongly magnetized outflow
the rest frame  energy density is $ \propto B^2$,  the resulting
lateral acceleration is independent of the magnetic field
strength.
A simple way to see this is to consider the MHD force balance equation
in the $\theta$ direction (see Eq. (\ref{x501}))
\be
\partial_t[(w+b^2)\gamma^2 \beta_\theta]+{1\over{r\sin\theta}}
\partial_{\theta}[\sin\theta(p+b^2/2)]-\cot\theta{{p-b^2}\over r}=0
\label{later}
\ee
Eq. (\ref{later}) shows that typically on a flow expansion time
$\beta_\theta \sim 1/\gamma^2\theta$. Thus for angles $\theta \geq 1/\gamma$
the lateral dynamics is frozen-out for ultra-relativistic flow.
and   the geometry of the  outflow
is likely to be  determined in the acceleration region near the light
cylinder.

One  particular stationary outflow  configuration  contains 
a   quadrupolar current flow concentrated close  to the polar axis 
and the equator.
This current distribution minimizes the total energy given a total toroidal magnetic flux
and has been advocated in relativistic stationary winds  \citep{HeyN03}.
In a  stationary 
wind,  the return current flows at infinity
from the pole to the  equator; in case of electromagnetic explosions, the 
return current flows along the surface of the bubble - the contact discontinuity. 
The magnetic field in the bubble is inversely proportional to  the cylindrical
radius.  In what follows,  we  adopt this particular current structure.
Eventually, the problem of flow acceleration near the source (near the light
cylinder of the progenitor) and collimation is likely to be solved by
numerical simulations. At this moment there are  no strongly relativistic numerical
simulations that trace the flow evolution from the subsonic accelerations
region, through the special points of the flow to asymptotic infinity.

There are two important exceptions to the application of these principles.
First,
as we discussed in Section \ref{ssec:nrexpansion},
at the non-relativistic stage
 plasma pressure, poloidal magnetic field and tangled component
of the magnetic field may be important in  providing  support against hoop stress
inside the current regions. Secondly,
 at late times,  $t  \geq  t_s$, the flow need not  
 be in equilibrium since it 
changes on dynamical (light travel) time scale and, in addition,
energy may flow towards the axis where it is dissipated 
and radiated as $\gamma$-ray emission.

The minimum size of the  core region is the   magnetic Debye radius
\be
r_D=\sqrt{ I \over  2 \pi n e c}
\label{rD}
\ee
(for a given density $n$  and  total current $I$ the relative drift of current-carrying
particles in this case is of the order of the speed of light). 
Note that it is orders of magnitude larger than the light cylinder radius $\sim r_s$.
The corresponding minimal 
 angular size of the core region can be estimated if we  use definition (\ref{ss1})
to eliminate the  density
\be
\theta_0 = {r_D  \over R} =
\left( { m^2 c^5 \sigma^2 \Gamma^2 \over L e^2}  \right)^{1/4}
\approx 10^{-3} L_{50}^{-1/4}  \sigma_9^{1/2}\Gamma_2^{1/2}
\label{theta0}
\ee

 A type of collimation that we envision in case of electromagnetic explosions
is somewhat different from the conventional "jet" model of AGNs and pulsars.
We expect that large Poynting fluxes
 associated with explosive release of $\sim 10^{51}$
ergs are sufficient to drive a relativistic outflow over a large solid angle,
so that during  the  relativistic stage the resulting cavity is almost spherical. But the
Lorentz factor $\Gamma$ of the CD may be a strong function of the polar angle.

\subsection{Quasistationary force-free outflow, $r\leq c t_s$}
\label{GSE}

At small radii, $r_{ph} \leq r \leq  c t_s$, the interaction with the circumstellar
matter is not important, so that the  flow may be considered as stationary. 
At this stage the 
 poloidal magnetic field is small,  flow is ultra-relativistic and moving ballistically. 
We can drive general equations governing the behavior of such flows.
Although the flow may still be supersonic, so that matter inertia
is important, in the asymptotic regions most important forces normal to the field
are electromagnetic \citep{nit95,HeyN03}.

Consider a simplified model problem of
{\it steady state}  force-free wind carrying only toroidal magnetic field.
The time-independent force-free flows are described by  the stationary 
Maxwell equations
with time derivatives set to zero:
\ba &&
\curl \E = 0
\label{curlE}
\nn &&
\curl \B =    {\bf j}
\ea
(factor $4 \pi$ has been absorbed into definition of current density).
 From Eq. (\ref{curlE}), it follows that the 
 electric field
is a gradient of a potential,
${\bf E} = -\nabla \Phi$, and
\ba &&
 {B  \partial_\theta \sin \theta B \over \sin \theta}= \partial_\theta \Phi \Delta
\Phi
\nn &&
 {B \partial_r r B \over r } =
\partial_r \Phi \Delta \Phi
\label{N}
\ea
Eliminating $ \Delta \Phi $ from eqns (\ref{N}) we find
\be
{\partial_r r B \over r} \partial_\theta \Phi -
{\partial_\theta \sin \theta B \over \sin \theta} \partial_r \Phi=0
\ee
Which means  that    electric potential is
a function of total  current enclosed within a magnetic loop
at polar angle $\theta$:
\be
\Phi \equiv \Phi (I)
\ee
The equation for the current $I(r,\theta)$ then becomes
\be
-{ 4 I  \over \sin^2 \theta  r^2 } + \Delta I  {\Phi'}^2 +
(\nabla I )^2 \Phi' \Phi^{\prime \prime} =0
\label{ZX}
\ee
This equation resembles closely the relativistic Grad-Shafranov equation
\citep[\eg][]{bes97}. The main difference is that under assumption of
zero poloidal field the conventional  flux function loses its meaning.
In the   standard theory of the  Grad-Shafranov equation,  there is a poloidal field which
is related to the derivative of the flux function.
We replace  the flux function
with  the current enclosed by a given field loop and treat this  as an independent
variable. Equation (\ref{ZX}) is considerably simpler than the
full relativistic Grad-Shafranov equation. It has one free function -
distribution of electric potential that in the steady state
determines uniquely the current
distribution.
In the  steady state the plasma experiences radial and lateral  drift
with velocities
\be 
\beta = {E_\theta \over B} =-  \Phi'  \sin \theta \partial_\theta
\ln\sqrt{I}
, \hskip .3 truein
\lambda = -{E_r \over B}=   \Phi' r  \sin\theta \partial_r \ln \sqrt{I}
\label{bl}
\ee
For radial motion $\lambda=0,\, I=I(\theta)$, equation (\ref{ZX})
gives
\be
\partial_\theta \left( I(\theta) \sqrt{ 1- \beta^2} \right) =0
\ee
which can be integrated
\be
I(\theta) \sqrt{ 1- \beta^2} =I_0 \sqrt{ 1- \beta_0^2}
\ee
where $I_0$ is the axial current and $\beta_0$ is the ratio of the axial charge
density per unit length to axial current.
Note that condition $\beta < 1 $ requires $I_0  \neq  0$. 
Thus, subluminal expansion along conical surfaces requires presence of a line
current \citep[see also][]{HeyN03}. 
Further solutions of this equation can be written down but will not be considered here.

\section{Electromagnetic Formalism}
\label{sec:emform}

\subsection{Relativistic
Force-free Electro-dynamics}

 The conventional method for handling relativistic, magnetized flows
is to use the relativistic extension of regular, non-relativistic
magnetohydrodynamics. Relativistic MHD (RMHD) 
 is  considerably more
complicated. 
In the non-relativistic case  the displacement current is neglected,
 $ \partial_t\E \rightarrow 0$, and a charge neutrality is assumed,
 $\rho_e \rightarrow 0$.
 Both of these assumptions
may be violated in relativistic plasmas. In addition, when calculating 
dynamic properties one needs to take into account the   inertia of 
 the electromagnetic field.  
However, there is a simpler extension, the  relativistic
force-free approximation (RFF),
which is appropriate when the plasma is sufficiently tenuous and subsonically moving, so
 that its
inertia can be ignored. 
On the other hand, 
 the  local microscopic  plasma time scale (\eg plasma frequency)
 typically is much shorter
than the global  dynamical time scales
and 
there
is plenty of charge available to screen the component of
electric field along the magnetic field.
In this case we can {\it neglect the inertia of the plasma particles but have to include their
electromagnetic interaction}. This is relativistic force-free (RFF) approximation.
 (Inertial contributions
 may be included as  perturbations to magnetic forces.)

There are two equivalent ways of deriving RFF equations.
First, they can be derived from the
two Maxwell
equations (factor $4 \pi$ has been absorbed into definitions of currents and charge 
densities):
\ba
\label{twomax}
{\partial{\bf E}\over\partial t}&=&\nabla\times{\bf B}-  {\bf j}\\
{\partial{\bf B}\over\partial t}&=&-\nabla\times{\bf E}
\ea
with the current density perpendicular to the local magnetic field
 determined by the force-free condition,
\be
\label{forcefree}
\rho{\bf E}+{\bf j}\times{\bf B}=0
\ee
This immediately  implies that the invariant
${\bf E}\cdot{\bf B}=0$ and its temporal derivative
can be set to zero; in addition,  electromagnetic energy
is conserved, ${\bf E}\cdot{\bf j}=0$. This  allows one to relate   the 
current to the electro-magnetic fields
\be
{\bf j}={({\bf E}\times{\bf B})\nabla\cdot{\bf E}+
({\bf B}\cdot\nabla\times{\bf B}-{\bf E}\cdot
\nabla\times{\bf E}){\bf B}\over   B^2}
\label{FF}
\ee
This may be considered as the Ohm's law for relativistic force-free electrodynamics.

For consistency of RFF, one also needs to assume that 
the second electromagnetic invariant is positive, $B^2-E^2>0$.
This implies that  there
is a  reference frame where the   electric field vanishes
 Equivalently, the electromagnetic stress-energy tensor can be
diagonalized. Note, that there is no mathematical constraint that would 
ensure that $B^2 - E^2$ remains positive. Thus, even if initially 
$B>E$ everywhere,  both numerical  and analytical
solution (\eg Appendix \ref{resicol})  may have regions where this condition becomes
violated. The implies that the RFF is no longer valid, so that  either plasma inertia or
dissipation should be taken into account (Section \ref{applicability}).

  The conditions ${\bf E}\cdot{\bf B}=0$
and $B^2>E^2$  allow us to
define an electromagnetic velocity
${\bf v}={\bf E}\times{\bf B}/B^2$ perpendicular
to the magnetic field. The velocity along the field
is not defined. This degeneracy  comes  from a neglect of plasma dynamics
associated with non-force-free motion of plasma along the field
(however strong the magnetic field is, it does not influence the plasma motion along
the field).
Note, that under the  force-free approximation all plasma species 
drift across the magnetic field with the same velocity. Thus,
the current perpendicular to the magnetic   
field is exclusively due to the plasma charge density.

RFF  equations (\ref{twomax}-\ref{FF}) represents  a simple
evolutionary dynamical system \citep{uch97,kom02}, which can be solved numerically
\citep[\eg][]{kom01}.
When one includes the constraints ${\bf E}\cdot{\bf B}=
\nabla\cdot{\bf B}=0$, there are four independent
electromagnetic variables to evolve and four characteristics
along which information is propagated.  In the linear
approximation, these correspond to forward and backward
propagating fast and intermediate wave modes
with phase speeds $c$ and  $c\hat{\bf k}\cdot\hat{\bf B}$ respectively
($\hat{\bf k}$ and $\hat{\bf B}$ are corresponding unit vectors).

RFF dynamics can be developed in a manner
that is quite analogous to regular hydrodynamics, with the anisotropic Maxwell
stress tensor taking the place of the regular pressure and the electromagnetic
energy density playing the role of inertia \citep[\cf][]{uch97,kom02}. 
There is an important difference,
though, in that the existence of a luminal fast mode means that
electromagnetic ``flows'' do not become truly ``supersonic'' (see
Section \ref{applicability}).

We would like to stress an important difference between non-relativistic and
relativistic force-free fields: the relativistic force-free theory is {\it dynamic}.
In  laboratory (non-relativistic)
 plasma the notion of force-free fields is often related
to the stationary configuration attained asymptotically  by a  system
(subject to some boundary conditions and some constraints, \eg conservation
of helicity). This equilibrium is attained on time scales of the order of the
\Alfven crossing times. In  strongly magnetized
relativistic plasma the \Alfven speed may become of the order of the
speed of light $c$, so that
 crossing time becomes of the order of the light travel time. But if plasma
is moving relativistically its state is changing on the same time scale.

\subsection{Applicability of force-free approximation}
\label{applicability}

Force-free electrodynamics assumes that the inertia of plasma is negligible.
This approximation is bound to break down for very large effective plasma
 velocities, when electric field becomes too close in value to magnetic field,
$E \rightarrow B$. The condition that inertia is negligible is equivalent to the
condition that the  effective plasma four-velocity,
 $u  \sim E/(B\sqrt{1-(E/B)^2})$, is smaller that the
\Alfven four-velocity in the  plasma.  In relativistic plasma
\be
u_A^2 = {\sigma }
\ee
This puts an upper limit on the value of electric fields consistent with
force-free approximation
\be
{B^2 -E^2 \over E^2} \gg {1 \over \sigma}
\ee

A second constraint comes from the fact that in RFF the charge number  density $n_e$
 may be comparable
to the total plasma number density $n$ and the current density may reach
$j \sim n e c$.
Force-free electrodynamics assumes that there are enough charge carriers
in the plasma to assure that the condition $\E \cdot \B=0$ is satisfied. 
This condition can be violated if the plasma density is too small and/or if   
fields
change on  very small scales (\eg due to development of electromagnetic 
turbulent cascade
which will bring energy to smaller and smaller scales). In this case plasma becomes charge-starved.
 To estimate charge-starved condition, 
 assume that   a typical fluctuating amplitude
of  magnetic field at the scale $l'$ is $ b_{l'}$ (both quantities measured in the plasma rest frame). 
 The corresponding
 current $ j'_{l'} \sim c  b_{l'} / 4 \pi$ is limited by the number density
of real charges $n'$: $ j'_{l'} \leq 2 e n' c$, which gives  \citep{tb98}
\be 
n' \sim { \delta b_{l'}  \over 2 \pi e  l' }
\label{n}
\ee
If this condition is violated,  the electromagnetic perturbations behave
more like electromagnetic waves than MHD  waves. In particular,
for smaller densities the condition $\E \cdot \B=0$ is not satisfied, so that
electric fields can accelerate particles, which  leads to dissipation of 
electromagnetic energy. This can be an important acceleration mechanism.

Thirdly, non-ideal effects (such as resistivity) may lead to violation of the ideal RFF approximation. 
It is possible to formulate equations of
resistive RFF by  introducing a macroscopic  resistivity \citep{lyu03}. 
In resistive RFF, the  dynamics is still controlled by the  electromagnetic field, but now
the currents that support these fields may become dissipative.
Resistive effects in magnetically-dominated 
plasma are somewhat different from the  non-relativistic
 analog.
In a force-free plasma,   conduction currents only flow
along  the magnetic  field in the plasma rest frame, so that  only the component of
current along the field is  subject to resistive dissipation. 
Ohm's law in resistive RFF becomes \citep{lyu03}
\be
(\j \cdot \B) + \partial_t\left( \eta \sqrt{ 1- \left( (\E \times \B)/B^2 \right)^2}
(\j \cdot \B) \right) =
({\bf B}\cdot\nabla\times{\bf B}-{\bf E}\cdot
\nabla\times{\bf E}){\bf B}
\label{Ohm-1}
\ee
In spite of the differences between resistive RFF and non-relativistic plasmas, 
effects similar to familiar resistive instabilities, like
tearing mode, should develop in RFF  \citep{lyu03}.
This has important implications for the  numerical modeling of RFF fields: 
similar to fluid dynamics, where sound waves turn into shocks, RFF fields tend to form
small scale structures in which resistive effects becomes important.

\subsection{Relativistic force-free versus MHD}

It is instructive to contrast this electromagnetic approach with the
relativistic MHD (RMHD)  formalism that is currently used by most investigators
\citep[\eg][]{phi82,cam86,tak90,tak00,par02}.
In the RMHD formulation dynamic  equations  include inertial terms  and
 pressure gradients.
This means that a fluid velocity field must be tracked, along with 
enthalpy density and  pressure.
 The electric field
is  supposed to be related to current and magnetic fields through an Ohm's law.
Under relativistic MHD,
the constitutive relation, Eq.~(\ref{FF}) must be augmented with inertial
terms. The equations are still evolutionary, though more complex - there are
seven independent variables to evolve, with seven independent wave modes
(two fast, two intermediate, two slow and one adiabatic) to follow.

The introduction of these extra complications, when the inertia
of the plasma is relatively small, can be questioned
on several grounds. Firstly, it is assumed that the electric field
vanishes in the center of momentum frame of the plasma. This is not
guaranteed by  plasma physics in a tenuous, relativistic
plasma. Secondly, it is usually
assumed that the plasma slides
without friction along the magnetic field. In other words, there is no
dissipation. This is unlikely to be the case
in the face of instabilities and radiative drag
\citep{bes93}. Thirdly,
it is generally supposed that plasma is conserved. This is untrue in GRBs
 where pair creation may be important until far out in the outflow.
 Finally, there
is the common assumption that the particle pressure tensor is isotropic.
In practice this is rarely the case in observed plasmas.
In contrast to that, the  magnetic flux is a  much better conserved
quantity.

The
relativistic force-free equations may be derived from
 relativistic MHD formulation in the limit of negligible inertia.
This offers an advantage that the system of equations may be set in form
of conservation laws and can be easily generalized to general relativistic form. 
The basic equations of RMHD include  the relativistic Ohm's law
and the  
relativistic dynamics, Maxwell's and mass conservation equations \cite{lich67,uch97,kom01}:
\ba &&
T^{\mu\nu}_{;\nu} =0,
\label{x1}
 \\ &&
F^{\ast\, \mu\nu}_{;\nu} =0,
\label{x2} 
\\ &&
(\rho u^\mu)_{;mu}=0
\label{x3}
\ea
where the stress energy tensor is a sum of contributions from matter and 
fields:
\ba &&
T^{\mu \nu } = T^{\mu \nu }_{(m)}+ T^{\mu \nu }_{(em)}
\nn &&
T^{\mu \nu }_{(m)}=w  u^\mu  u^\nu +p
g^{ij}
\nn && 
T^{mu\nu}_{(em)} = F^\mu_\alpha F^{\nu \alpha} -
{1 \over 4} F_{\alpha \beta} F^{\alpha \beta} g^{\mu \nu} = 
b^2 u^\mu u^\nu +{b^2 \over 2}
g^{\mu \nu} -b^\mu \,b^\nu 
\ea
where $w $ is  the
 plasma proper enthalpy,
 $\rho$ is proper plasma density and $p$ is pressure,
$b_\alpha = {1\over 2} \eta_{\alpha \beta \mu \nu} u^\beta F^{\mu \nu}$ are the four-vector of magnetic
field, Levi-Civita tensor and electro-magnetic field tensor, $u^\alpha=(\gamma, \gamma {\bf \beta})$
are the plasma four-velocity, $F_{\mu \nu}$ is electromagnetic field tensor.
An additional constraint is that the first electromagnetic invariant is zero:
\be
F_{\mu \nu}\, ^\ast F^{\mu \nu} =0
\ee 
where $^\ast F^{\mu \nu}$ is a dual electromagnetic tensor. 

It has been also implicitly assumed in the derivation of 
these equations that the second  electromagnetic invariant $F_{\mu \nu}\, F^{\mu \nu} \neq 0$, so that 
the electromagnetic stress energy tensor can be
diagonalized.
This assumption is important since we are interested in the limit
when matter contributions to the stress energy tensor vanish;
the possibility of diagonalization of the electromagnetic stress energy tensor
distinguishes 
MHD and vacuum electromagnetic fields  where such diagonalization
is not possible.

When we set the matter
 contributions to the stress energy tensor to zero,
$w, p \rightarrow 0$, the  RMHD equations become equivalent to the  RFF equations 
with the exception that the velocity along the field is not defined anymore.
  This can also  be checked by
combining RFF equations in the form of conservation laws
\ba &&
{1\over 2} \partial _t ( E^2+ B^2) +
\nabla   \cdot  \left( {\bf E} \times {\bf B} \right)  =0
\nn &&
\partial _t {\bf E} \times {\bf B}= \nabla \times {\bf B}\times {\bf B}+
\nabla \times {\bf E}\times {\bf E}+(\nabla \cdot {\bf E}){\bf E}.
\label{ER1}
\ea

\section{Relativistic expansion of the magnetic shell, $r> c t_s$ }
\label{relatexpan}

After  breakout, the electromagnetic flow is accelerated by pressure and magnetic forces and becomes
super-fast-magnetosonic. At 
the  photosphere the lepto-photonic component decouples
from the magnetic field, so that
$\sigma$ increases to $\sigma \sim 10^8 -10^{10}$,
so that the wind becomes sub-fastmagnetosonic again.
Beyond this  point the
outflow is strongly  magnetically dominated and can be described by RFF equations.
The wind is then further accelerated by magnetic pressure, while its expansion
is determined by the interaction with external medium.
In this   section we study the   dynamics of  such magnetically-dominated explosions.

There are two   principal differences
 between this work and other (mostly stationary MHD-type)
approaches. First, 
 the flow is  commonly  assumed  to expand freely, without interaction with
the circumstellar medium, so that 
the flow dynamics is determined  by the  internal 
structure of the flow.   MHD flows usually cross fast magnetosonic critical surface
after which they become  causally disconnected from their source \citep{gj70}.
 A  condition
that the flow remains regular at the critical surface is what determines the global
properties of the wind. Unlike MHD,
 force-free flows  are sub-fastmagnetosonic so no conditions
at the fast critical surface appear. In this case it is the {\it interaction with
boundaries 
 that determines the properties of the flow} (similar to subsonic hydrodynamic
flows).
 In the case of freely expanding
 magnetized wind the terminal Lorentz
factor $\Gamma_\infty$ depends  very sensitively on details of geometry:
for radially expanding MHD flows
$\Gamma_\infty \sim \sqrt{\sigma}$ \cite{Michel69,gj70}.
Thus, the distinctive feature between MHD and force-free flows is whether
the wind becomes fast supersonic (MHD regime, $\Gamma > \sqrt{\sigma}$) or not
(force-free regime, $\Gamma < \sqrt{\sigma}$). 
In case of GRB, beyond the  photosphere $ \sigma \sim 10^{9}$, so it the
interaction of the force-free wind with the  external medium (and not the
internal wind dynamics) that  is likely to
 limit acceleration to
$\Gamma\sim 10^{4}$  (see below).

Secondly, the flows under consideration are strongly non-stationary, so that their
dynamics
 is  qualitatively different
from stationary solutions and is more reminiscent of the
strong explosion problem \citep{Sedov}. Relativistic effects 
introduce an important difference, though. Primarily, the differences lie in the
 causal  structure of the shell.
  It takes a
very long time for a wave to be reflected by the relativistically expanding
surface of the shell  (contact discontinuity)
 and return to the origin.  This is generally true of ultra-relativistic flows
so that  stationary solutions, which take a long time to be established,
can be quite misleading. 

As we have discussed  above the long time scales for establishing a ''feed-back'' on the source
have an important consequence for energy dissipation in the flow: 
there is no necessity to destroy magnetic flux through ohmic
dissipation, until the wave can actually propagate back to the source.
Stated another way, there {\em need} be little resistance in
the electrical circuit. The effective load can
consist of the performance of work on the expanding blast wave. This is
where most of the power that is generated by the central magnetic
rotator ends up.
Thus, 
until the outflow   becomes non-relativistic,
the distinction between the
inertial load   and a different,
dissipative load is quite unimportant for the behavior of the
central source.
 As long as the expansion
remains ultra-relativistic, it is a very good approximation
to impose a Sommerfeld radiation condition on the solution at the source 
 and to ignore the reflected wave. 

The general problem of relativistic force-free expansion is quite complicated: we need to solve
for the  time-dependent,  three-dimensional evolution of electric and magnetic fields
 with a non-linear coupling through  the 
force-free Ohm's law (\ref{FF}). 
(note that even 
normal modes in such plasma experience strong refractive phenomena).
To make the problem tractable we make several simplifying assumptions. First, 
we limit our approach to distances much larger than the
source scale $r_s$ and,  since at large distances magnetic field is dominated by the toroidal fields,
we neglect poloidal magnetic fields altogether (thus we also neglect the angular momentum loss by the source). 
Secondly, 
instead of treating the flow acceleration and collimation at the source,
we assume that the central source can be represented as  an inner boundary condition
through which  a toroidal magnetic flux is injected with a given total luminosity and
a given polar angle dependence  $L(\theta)$. 

We assume also that at large distances, $r\geq r_{ph}$, the outflow may be  described by ideal RFF equations.
In this case the expansion of the relativistic shell will be subsonic, and virtually
ballistic (radial).
The energy input of the central source then can be represented as fast modes propagating
from the source to the contact discontinuity. 
Motion of the CD 
is determined by the pressure balance between the Poynting flux from the source and
the ram pressure of the ISM and thus depends on the
source luminosity $L_\Omega(t')$ per unit solid angle $d \Omega$
at the retarded time $t'$ such that $R(t) = t -t'$.
Typically $ R(t) / c \sim \Gamma^2 t_{\rm source}$.

A particular form of the forward-propagating wave train (its temporal variations,
 amplitude and lateral distribution
of the energy flux) depends on the history of the source activity and 
 detailed properties of the
source.
Since the
 central source is expected to vary considerably during its activity,
 the pressure on the inner boundary of the CD
will be fluctuating as the new pulses from the source start to contribute
to the pressure. This will be reflected in the "jitter" of the CD motion.
At later times multiple reflections
from the contact discontinuity and the center become important as well.
 As we have discussed in Section \ref{Collim} 
the lateral distribution of the   magnetic field  
and   energy fluxes in the  outgoing waves is likely to consist
of a line current. Then, 
a very important observational consequence follows. For outgoing waves
corresponding to the  line current  
the distribution of Poynting flux  is $ L \sim 1/\sin^2 \theta$,
implying that the central source releases an equal amount of energy per
decade of $\theta$.  This would later translate into a  similar form
of the forward shock.

The form of the ingoing reflected wave depends both on the 
 outgoing wave and the form of the contact discontinuity.
Since in the strongly relativistic regime the wave will be reflecting from
receding CD the amplitude of the reflected wave will be $\sim \Gamma^2$ times smaller.

The relativistic stage of the shell may be separated into two stages:
''coasting'' and ''self-similar'' depending on whether or not most of the
fast waves emitted by the central source have caught up with the CD. 
\begin{itemize}
\item  {\it ``Coasting'' electromagnetic
 shell ($ct_{{\rm s}}<r<r_{{\rm sh}}\equiv
(L_\Omega t_{{\rm s}}^2/\rho c)^{1/4}$)}
(in the observer frame this phase lasts also $\sim100$~s).
At $r>r_{ph}$ the shell becomes
a relativistically expanding shell of thickness $\sim ct_{{\rm s}}\sim3\times10^{12}t_{{\rm s,2}}\ {\rm cm}$.
The shell still contains toroidal magnetic field but the current now
detaches from the source and completes along the shell's inner surface.
At this stage the  CD  is constantly re-energized by the
fast-magnetosonic waves propagating from the central source.
The  average motion of the CD is determined by the
average luminosity,  $\Gamma \sim (L_\Omega/\rho c^3)^{1/4}  r^{-1/2}$ (in a constant
density medium) or  $\Gamma \sim {\rm const}$ (in a $\rho \sim r^{-2}$ wind).
In addition, there is a "jitter" of the CD, in response to the
"jitter" in the source luminosity.
The internal structure of the magnetic shell is a messy
 mixture of the outgoing waves from the source and the ingoing
waves reflected from the CD, similar to  a pre-Sedov phase in
hydrodynamical explosions.
  Unlike the case of a
hydrodynamic  blast wave with energy supply, no internal discontinuities form
inside  magnetic shell.
\item {\it Self-similar  electromagnetic shell} ($r_{{\rm sh}} <r<r_{{\rm NR}}\equiv
(L_\Omega t_{{\rm s}}/\rho c^2)^{1/3}$).
After one dynamical time scale, $t_d \sim \Gamma^2 t_s \sim 3 \times 10^{16}$ cm,
 all the regions of the shell come
into  causal contact -- most of the
 waves reflected from the CD have propagated throughout the shell.
As the expanding shell performs
 work on the surrounding medium its total energy
decreases;
the amount of energy that remains in the shell during  the self-similar
stage is small, $\sim E_\Omega /\Gamma^2$.
 Most of the energy  is still
concentrated in a thin
shell with $\Delta R  \sim R/ \Gamma^2$ near the surface of the shell
which is moving according to $\Gamma \sim \sqrt{E_\Omega / \rho c^2}\, r^{-3/2}$
(in a constant density medium), or $\Gamma \sim r^{-1/2} $
 (in a $\rho \sim r^{-2}$ wind). The energy of the shell decreases only weakly
with radius, $dE/dt \sim 1/r $ in constant density and
 $dE/dt \sim 1/r^3$ in a wind, so that the surface of the shell keeps moving
relativistically as long as the preceding shock wave is moving relativistically,
until $r \sim (E_\Omega /\rho c^2)^{1/3} \sim 10^{18}$ cm - the shock never
becomes completely free of the shell.
Interestingly, the   structure of the {\it magnetic}
 shell (in particular the  distribution of energy)
 resembles at this stage the structure of the {\it hydrodynamical}
 relativistic blast wave  \citep{blm76} (Section  \ref{self}, m=3).
\end{itemize}

\subsection{RFF equations for axisymmetric asymptotic flows}

In the asymptotic domain $r \gg r_s$ the ejecta will be dominated by
toroidal magnetic field and, since we assumed that all small scales field
variations are smoothed out, it will be axially symmetric.
In this case 
the only
nonzero components of the fields are $B=B_\phi, E_r, E_\theta$ and  the RFF
equations
 (\ref{twomax}) and (\ref{FF}) give
\ba &&
\partial_t B = -{1\over r} \partial_r( r E_\theta) + {1\over r}  \partial_\theta E_r
\nn &&
\partial_t E_ \theta =  -{1\over r} \partial_r( r B) +
{ E_r  \over B} \left(
{1\over r \sin \theta} \partial_\theta (  \sin \theta E_\theta) +
{1 \over r^2} \partial_r (r^2 E_r) \right)
\nn &&
\partial_t E_r = {1\over r \sin \theta} \partial_\theta (  \sin \theta B)-
{ E_\theta \over B} \left(
 {1\over r \sin \theta} \partial_\theta (  \sin \theta E_\theta)
+{1 \over r^2} \partial_r (r^2 E_r) \right)
\label{ff1}
\ea

The charge  density associated with a given solution  is given by
\be
\rho= {1 \over r^2} { \partial_r ( r^2  E_r )} +
{1\over r\sin\theta}{\partial\over\partial\theta}E\sin\theta
\label{rr}
\ee
and the
 currents are
\ba &&
j_r = \rho  { E_{\theta} \over B} =
{ E_{\theta} \over B} \left(
{1 \over r^2} { \partial_r ( r^2  E_r )} +
{1\over r\sin\theta}{\partial\over\partial\theta}E\sin\theta \right)
\nn &&
j_\theta =  - \rho  { E_r \over B} =
  - {E_r \over  B} \left(
{1 \over r^2} { \partial_r ( r^2  E_r )} +
{1\over r\sin\theta}{\partial\over\partial\theta}E\sin\theta \right)
\label{jj}
\ea
After some rearrangement the
 system (\ref{ff1}) gives
\ba &&
{1\over 2} \partial _t \left( E_{\theta}^2+ E_r^2 + B^2 \right)+
{1 \over r^2} \partial _r r^2 E_{\theta} B - {1 \over r \sin \theta}
\partial _{\theta} \sin \theta E_r  B =0
\nn &&
\partial _t E_{\theta} B +
{1 \over r^2} \partial _r r^2 \left(B^2+ E_{\theta}^2 \right) -
{1\over r \sin \theta}
\partial _{\theta} \sin \theta E_r E_{\theta} - { E_r \over r^2} \partial _r r^2
E_r=0
\nn &&
\partial _t  E_r B + { E_r \over r} \partial _r r E_{\theta}  +
{ E_{\theta} \over r^2} \partial _r r^2 E_r +
{ E_{\theta} \over r \sin \theta} \partial _{\theta} \sin \theta E_{\theta}-
{B \over r \sin \theta} \partial _{\theta} \sin \theta B -
{E_r \over r} \partial _{\theta} E_r =0
\label{Wqq}
\ea
To connect to  RMHD description we note that the plasma drift velocity
is $(E_\theta/B) {\bf \hat{e}}_r -( E_r/B) {\bf \hat{e}}_\theta$. Setting 
 $ B =  \gamma b, \, E_{\theta} = \beta \gamma b , \,
E_r = \lambda \gamma b$ and the system (\ref{Wqq}) gives
 eqns (\ref{x62}).

Equations (\ref{ff1})  are the main equations that determine dynamics
of the expanding magnetic  shell.

\subsection{Boundary condition on the contact discontinuity}

To find the internal structure of the expanding magnetized shell the
dynamical equations should be supplemented by boundary conditions on the
contact discontinuity separating the  relativistically expanding
magnetic shell and external medium. The first boundary condition requires
that the pressure of the magnetic field  of the
ejecta should be balanced by the  pressure of the external medium.
The pressure of the external medium is   the ram pressure
if the forward shock has not formed, or the kinetic pressure of the shocked
material if the forward shock has  formed. The two cases differ  only  by
a numerical coefficient $\kappa$   (appendix \ref{shocked}).

 As the shell propagates into the
surrounding medium it reflects particles with typical Lorentz factor $\sim
 2 \Gamma^2$.
In the frame of the shell  there must be a pressure equilibrium
on the contact discontinuity.  The magnetic pressure inside the shell,
$B^2 /( 8 \pi \Gamma^2)$ should balance the ram pressure of the incoming
 external medium , $2 \kappa \Gamma^2 \rho_{\rm ext} c^2  $, (here $\rho_{\rm ext}$
is the
density  of the external medium and $ \kappa$ is a coefficient of the order of unity
relating the ram pressure before the shock and  kinetic pressure on the CD):
\be
2 \kappa \Gamma^2 \beta^2  \rho_{\rm ext} c^2 ={ B^2 \over 8 \pi \Gamma^2}
\label{b1}
\ee
where $B$ is the shell
magnetic field measured in the lab frame.
Thus, the  magnetic field in the plasma  rest frame should scale as 
$b \sim \Gamma$,  while magnetic field in the lab frame scales as
 $B \sim \Gamma b \sim \Gamma^2$.

Alternatively,   a force balance on the CD can be considered as a Lorentz force
of  magnetic field and the surface current acting against the ram pressure.
Consider, for example,   magnetic field occupying  a half space $x>0$
and moving with velocity $\beta_x {\bf  \hat e}_x+ \beta_z{\bf  \hat e}_z$, 
so that the fields are given by
 ${\bf B} = B_0 \Theta(x - \beta_x t) {\bf  \hat e}_y,
\, {\bf E} = - \beta_x B_0 \Theta(x - \beta t) {\bf \hat e}_z+
\beta_z  B_0 \Theta(x - \beta t) {\bf \hat e}_x$
where $\Theta$ is the Heaviside function.   Then  from Maxwell equations
the surface current ${\bf g}$ is
\be
{\bf g} =  { B_0 \over 4 \pi \Gamma^2}  \delta(x - \beta t) {\bf e}_z ,
\label{g}
\ee 
where $\Gamma^2 1/(1-(\beta_x^2 + \beta_z^2))$, so that
the Lorentz force on the surface is
 ${\bf g}\times {\bf B}= B_0^2/4 \pi \Gamma^2 $.

The second boundary condition comes from the requirement that  the flow
 should smoothly
connect to the CD, which implies that in the frame of the CD
 $E_\theta$ on the CD equals zero, or, alternatively, 
 the component of the 
 electromagnetic velocity normal to the surface
 of the CD should equal the velocity of the CD.
 If at a given point the CD has a velocity $v_{CD}= \{ \cos \alpha_{CD}, 
\sin \alpha_{CD}, 0 \}$ in spherical coordinates $(r, \theta,\phi)$, then
\be
{E_\theta \cos \alpha_{CD} - E_r \sin \alpha_{CD} \over B} = v_{CD} 
\label{b2}
\ee
where 
$\tan \alpha_{CD} = \partial _ \theta \ln R(\theta,t)\sim  {\cal{O}} (1/\Gamma^2) $
and $v_{CD} = \sqrt{1-1/\Gamma^2}$.
Note, that in order to calculate    the angle 
 $\alpha_{CD}$ between the radial direction and normal to the CD
 one needs to know the whole history of the shell expansion since it 
 depends on the position of the CD at a given time $R(\theta,t)$.
 On the other hand the velocity of the CD  $ v_{CD}$ depends only on the source
 luminosity at a retarded time.

The third boundary condition, for the radial electric field $E_r(R)$, determines the 
 surface charge density on the  shell.

\subsection{Motion of the contact discontinuity }

Next we determine an average motion of the CD given an average source
luminosity.
At the coasting phase,
 equating the source luminosity $L_\Omega(t')$ with
  the Poynting flux at the CD $L_\Omega(t')= B^2 R(t)^2 \beta/4 \pi $
and using boundary condition (\ref{b1}) we find
\be
L_\Omega(t') =  4 \kappa \rho c^3 \Gamma^4 R(t)^2 \beta^3
\ee
If during its activity period the source produces an approximately constant
luminosity $L_\Omega $, then
 for  a constant density medium $\rho(r)=\rho_0$
\be
\Gamma_{CD} =  (L_\Omega / 4  \kappa \rho_0 c^3)^{1/4} R^{-1/2},
\ee
while for a wind environment $\rho(r)=\rho_0(r_0/r)^2$
\be
\Gamma_{CD} =  (L_\Omega / 4  \kappa  \rho_0 r_0^2 c^3)^{1/4} = {\rm const}
\ee
If the distribution of luminosity is as argued above, $L_\Omega = L_0 \theta^{-2}$,
then 
$\Gamma_{CD} \propto 1/\sqrt{\sin \theta}$ (see Section \ref{axial}).

The work that the shell makes on the circumstellar medium per unit time and unit
solid angle is (see Eq. (\ref{g}))
\be
p dV/dt  = \left. {\bf g} \times \B \right|_{CD} R^2 \beta
=  4 \kappa  \rho c^3 \Gamma^2 R(t)^2 \beta
\ee
Thus, in  case of constant density the most efficient energy transfer to the
forward shock occurs at the end of the coasting phase, $p dV/dt 
\propto R$, $R \leq 2 \Gamma^2 t_s c$,  while  for the wind environment
 $p dV/dt  \sim {\rm const}$.

By the end of  coasting phase most of the waves emitted by the central source
will have reflected from the CD, reached the center and reflected
again. At this stage the  structure of  the shell will become self-similar.
 At the self-similar stage $\Gamma \sim R^{-3/2}$
for constant density and $\Gamma \sim R^{-1/2}$ for the wind environment.
By the end of the  coasting phase, a large fraction of the energy of the shell
will have been transferred to the forward shock.
The transfer of the remaining small fraction (of the order of $E_{tot} / \Gamma_c$,
where $\Gamma_c \sim 100$ is the Lorentz factor at the end of the coasting stage)
 to the forward shock
becomes very inefficient ($ p dV/dt d\Omega \sim 1/r$ for  both $\rho={\rm const}$
and $\rho \propto 1/r^2$), so that the energy
inside the shell decreases only logarithmically with time until the
motion  becomes
non-relativistic.

\section{Internal structure of the magnetic shell}
\label{internal}

To determine the internal structure of the magnetic shell one needs to solve 
Eqns. (\ref{ff1}) together with boundary conditions  (\ref{b1}-\ref{b2}).
This   is a 
  quite complicated system  and should generally be solved numerically. 
In this section we first consider an idealized case of purely radial  motion
which allow exact solutions,
and then derive equations and obtain 
approximate solutions for our preferred  case, when the
outgoing wave consists of a  line current.

\subsection{Radially expanding  magnetic shell}

To illustrate the behavior of the solutions we consider in details  a particular case of  radial
propagation of both outgoing and reflected waves. In this case  exact
solutions can be found. For radial propagation $E_r=0$ and the 
 relevant equations become
\ba &&
{\partial B\over\partial t}={-1\over r}{\partial\over\partial r}rE
\label{p3} \\ &&
{\partial E\over\partial t}=-{1\over r}{\partial\over\partial r}rB
\label{p4} \\ && 
\partial_\theta \left( \sin^2 \theta (B^2-E^2) \right) =0
\label{EQ}
\ea
Eqns (\ref{p3}-\ref{p4})
 can be  reduced to a 1-D wave equation, which solutions can be expressed
as a superposition of  simple wave \citep[]{LLIV}. Taking into account that
in the force-free plasma the fast magnetosonic waves propagating across magnetic
field have velocity $c$, the simple waves are given by
$r=\pm t + f(v)$, where $f(v)$  is some function determined by the boundary or initial
conditions.
Allowing for angle-dependent velocity of expansion
the
general solution of (\ref{EQ}) becomes
\footnote{
It is instructive to derive these equations using the RMHD formulation.
Under assumption of radial motion Eqns (\ref{x62}) become
\ba &&
\beta \gamma \partial _ t b + b \partial_r \gamma =0
\nn &&
\beta \gamma{ \partial_r ( r b) \over r} +  b \partial_t \gamma =0
\ea
Which using substitution $ r b \rightarrow  \exp\{b_1 \},\,\,
\gamma \rightarrow \cosh y,\,\, \beta \rightarrow \tanh y$
reduces to
\ba &&
\partial _ t b_1 + \partial_r y =0
\nn &&
\partial _r b_1 + \partial_t  y =0
\ea
with a  general solution
$b_1 = c_1(\theta) (\tilde{f}_1(t-r)+\tilde{f}_2(t+r)) + c_2 (\theta)$,
$y=  c_1(\theta) (\tilde{f}_1(t-r)+\tilde{f}_2(t+r)) + c_3 (\theta)$, which
reduces to (\ref{3}).
}
\begin{eqnarray} &&
B=r^{-1}\left[f_1(t-r) z(\theta) + { f_2(t+r) \over z(\theta)} \right]g(\theta)
\\ &&
E=r^{-1}\left[f_1(t-r) z(\theta) - { f_2(t+r) \over z(\theta)} \right]g(\theta)
\\ &&
(B^2-E^2)\sin^2\theta=h(r,t)
\label{3}
\end{eqnarray}
The two terms in each expression are fast modes propagating outward and inward.
The form of the functions $f_1$ and $f_2$ is determined by the boundary conditions
at the surface of the shell.
\footnote{In case of a dominant toroidal magnetic field, these equations can be
generalized for MHD plasma, when inertial and pressure effects are important.
For propagation perpendicular to magnetic field lines (and only in this case)
 simple waves in MHD may be reduced to hydrodynamic simple wave with
a special form for the equation of state, including magnetic field. Then the
one-dimensional  relativistic simple waves become
$r=t (v  \pm u_f)/(1-v u_f) + f(v)$  where $u_f$ is the fast magnetosonic speed
(\cite{LLIV}). }

For  given  fields (\ref{3})
the  radial velocity of the plasma is defined as
\be
\beta = { f_1 z(\theta)^2 - f_2 \over f_1 z(\theta)^2 +  f_2}
\ee
while the charge density $\rho = {\rm div }  \E$ and current
${\bf j} = \curl \B - \partial_t \E$ become
\ba &&
\rho ={ \partial_\theta (\sin \theta z(\theta) g(\theta)) \over \sin \theta r^2} f_1 -{  \partial_
\theta (\sin \theta  g(\theta) / z(\theta) ) \over \sin \theta r^2} f_2
\nn &&
j_r= { \partial_\theta (\sin \theta z(\theta) g(\theta)) \over \sin \theta r^2} f_1 + {  \partial_
\theta (\sin \theta  g(\theta) / z(\theta) ) \over \sin \theta r^2} f_2
\label{jj1}
\ea
In the plasma rest frame electric field is 0 while magnetic field is
$
b = { 2 \sqrt{f_1 f_2}  g(\theta) / r}
$. (Note that it  depends both  on the amplitude of  the forward and backward
waves.)
 
The boundary conditions  (\ref{b1}-\ref{b2}) become 
\ba &&
4 F_1 F_2 = { (F_1  z(\theta) +F_2/ z(\theta))^2 \over \Gamma^2}
\label{1}
\\ &&
(F_1  z(\theta)  + F_2/z(\theta) )^2 = \alpha R^2 \sin^2 \theta \Gamma^4 \beta^2
\label{2}
\ea
 and $w = 16 \pi  \kappa  \rho_{\rm ext} c^2$ and $F=f(r=R(t))$. 
We can solve the system (\ref{1}-\ref{2}):
\ba &&
F_1 = { \beta \over 1-  \beta} {  R \sqrt{\alpha} \over 2 z(\theta) g(\theta)}
\approx {  R \sqrt{\alpha} \Gamma(\theta)^2 \over z(\theta) g(\theta)}
\label{30}
\\ &&
F_2= { \beta \over 1+ \beta} {  R \sqrt{\alpha}  z(\theta)  \over 2  g(\theta)}
\approx {  R \sqrt{\alpha}  z(\theta)  \over 4  g(\theta)}
\label{31}
\ea
where approximations assumes $\Gamma \gg1$.
In addition, Eq. (\ref{3}) gives
\be
g(\theta) = {1 \over \sin \theta}
\label{32}
\ee
Equations (\ref{31}-\ref{32}) define possible angular dependence of the solution
since 
both $F_1$ and $F_2$ are angle independent by definition. 

Since  for strongly relativistic motion $R\sim t$. 
from  Eqns. (\ref{30}-\ref{31}) it follow that there are two 
possible choices. First,  $w = {\rm const}=  w_0$, $z(\theta) = g(\theta) =1/\sin \theta $,
 $\Gamma \propto  g(\theta) =\Gamma_0(t)/\sin \theta $ with the  fields  given by 
\ba &&
B={ f_1(t-r) \over r \sin ^2 \theta} + {f_2(t+r) \over r}
\nn &&
E = { f_1(t-r) \over r \sin ^2 \theta} -  {f_2(t+r) \over r}
\label{bala}
\ea
The second possible choice is 
 $W \propto g(\theta)^2= w_0/\sin^2 \theta $, $z(\theta) = {\rm const}$, 
 $\Gamma(\theta)= {\rm const}= \Gamma_0(t)$.
\ba &&
B={f_1(t-r)+f_2(t+r) \over r \sin \theta}
\nn &&
E_\theta={f_1(t-r)-f_2(t+r) \over r \sin \theta}
\label{bala1}
\ea
In both cases  the functions $f_1$ are and $f_2$ are found from boundary conditions
\ba &&
F_1 = 
{  t \sqrt{w_0} \Gamma_0(t)^2 }
\label{30a}
\\ &&
F_2= 
 {  t \sqrt{w_0}   \over 4  }
\label{31a}
\ea

Note, that the amplitude of the reflected wave $f_2$ is
$\Gamma_0^2$ times smaller
than that of the forward propagating waves $f_1$ due to  the Doppler
shift during reflection off  the relativistically moving surface of the shell.
Eqs. (\ref{30a}-\ref{31a}) describe the jitter of  the CD as a function of the
source history of activity through the  dependence of $f_1$ on the retarded time
(this is true only at the coasting stage, when no waves reflected from the
CD have reflected from  the origin and reached the CD for the second time).

In the ultra-relativistic case $\Gamma \gg 1$ a  further simplifications is possible.
Since for relativistic motion $ R(t) \sim t$ the function $f_2$ can be found
explicitly
\be
f_2(t+r) = { (t+r)  \sqrt{w_0}    \over 8}
\ee
while $f_1$ can be found for any given expansion velocity of the shell
$\Gamma_0(t)$  and $R(t)$.

\subsubsection{Self-similar solutions}
\label{self}

A particularly simple and analytically tractable case of the radial expansion
of the magnetic shell is when the
Lorentz factor of the contact discontinuity is a power-law function
of time $\Gamma^2 = \Gamma_{in}^2 (t/t_{in})^{-m}$. This includes a point-like
explosion and a power-law luminosity variation of the central source.
We will call these self-similar solutions.
In this case $ R(t)= t\left(1-1/(2 (m+1) \Gamma^2) \right)$, so that 
at the boundary 
\ba &&
F_1 \sim  F_1( t/(2 (m+1) \Gamma^2))
\nn &&
 F_2  \sim  F_2 (2t)
\ea  
We can then find functions $f_1$ and $f_2$
\ba &&
f_1(z) = \sqrt{w_0} \Gamma_{in}^2 
\left( 2 (m+1) \Gamma_{in}^2 z \right)^{(1-m)/(1+m)}=
 \sqrt{w_0} \Gamma^2 t^{ 2 m/(1+m)} \left( 2 (m+1) \Gamma^2 z \right)^{(1-m)/(1+m)}
\nn &&
f_2(z) = \sqrt{w_0}  z/8
\label{f2}
\ea
Equations  (\ref{f2}) determine the amplitudes of the outgoing and
the ingoing waves for self-similar solution.

\begin{figure}
\plotone{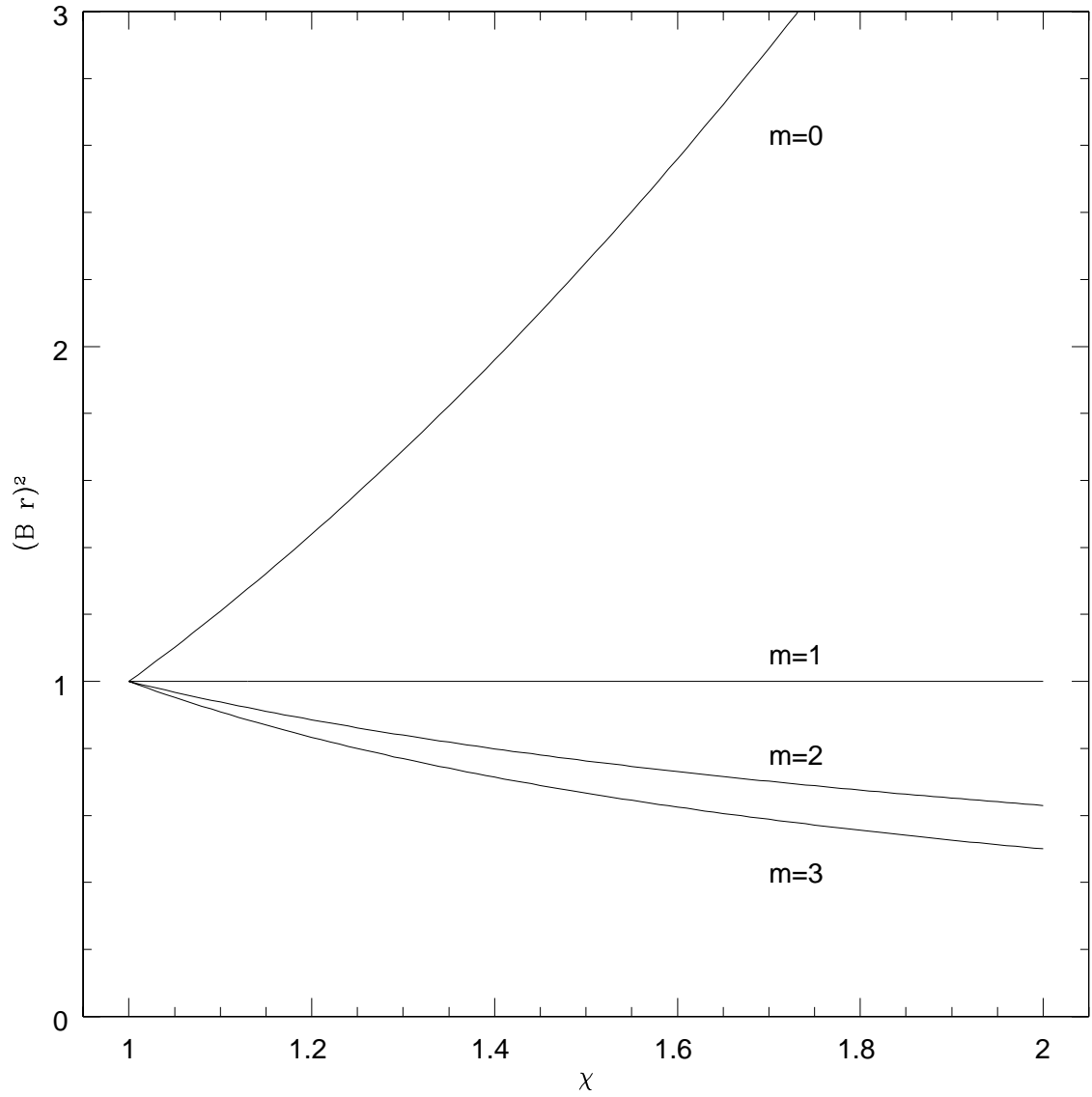}
\caption{
Energy density of magnetic field  per range of radii $B^2 r^2 dr$ as a function
of the  self-similar variable $\chi$. }
\label{Bsqrd}
\end{figure}

We can find the structure of fields in the shell for the self-similar expansion.
Keeping only the leading terms  in $\Gamma^2$ we find
\be
B \propto {f_1(t-r) \over r  } =
\sqrt{w_0} \Gamma^2 { t\over r }  \left[ 2 (m+1) \Gamma^2 (1-r/t) \right]
^{(1-m)/(1+m)}  \equiv \sqrt{w_0} \Gamma^2 { t\over r } \chi^{(1-m)/(1+m)}
\label{BB}
\ee
where $\chi =  2 (m+1) \Gamma^2 (1-r/t)$ \citep[c.f.][]{blm76} (see Fig. \ref{Bsqrd}).
For a given time $t$ the magnetic field depends on a slowly varying
function $1/r$ multiplied by  fast varying (on a scale $R/\Gamma^2$)
variable $\chi ^{(1-m)/(1+m)}$.
For example, for a constant energy source in constant density medium, $m=1, $
$ B \propto  \sqrt{w_0} \Gamma_{in}^2/r$ (times some angular dependence),
 which is  constant in time inside a 
shell.

Result (\ref{BB})  implies  that in the self-similar case 
 the energy density of the magnetic field near the surface of the CD 
has the  same distribution
as the
kinetic energy density in the case of relativistic shock wave 
\cite{blm76}.
This is a  surprising result,
since the two systems are completely different (electromagnetic shell and
hydrodynamical shock wave) and solutions come from different equations
(Maxwell and Euler).
Generally, 
for  $1<m<3$ the magnetic field in the self-similar solutions
 piles up near the
contact discontinuity,
 concentrated in a layer of thickness $R/\Gamma^2$. In this case the
above solutions may be obtained by the Blandford-McKee method
(which assumes that the most of the energy is concentrated near the
outer boundary of the cavity)
from the RMHD equations (see appendix \ref{BMR}).
To continue our analogy with the hydrodynamic case, we note that the 
above force-free  solutions may be interpreted in terms of RMHD if 
we define a  Lorentz factor 
\be
\gamma^2 = {B^2 \over B^2 -E^2} = {f_1^2 +f_2^2 + 2 f_1 f_2 \over 4 f_1 f_2}
\sim {f_1 \over 4 f_2}
=  {t \over r} \Gamma^2 {2  \over (r+t) } 
\chi ^{(1-m)/(1+m)}
\ee
Note that near the  CD, $t \sim r$ we get
\be 
\gamma^2_{near \, CD} = 
\Gamma^2 \chi^{( 1-m)/(1+m)},
\ee
again
similar to the case of  hydrodynamical blast wave \citep{blm76}.

Two kinds of self-similar solutions are most interesting. The first is the 
$m=1$ case corresponding to a constant luminosity source and expansion
into  a constant density medium.
 In this case 
$
f_1(t-r) =  \sqrt{w_0} \Gamma_{in}^2 t_{in} = {\rm const}
$
and  the fields are given by
\ba &&
B= {  \sqrt{w_0} \over r \sin \theta } \left( { \Gamma_{in}^2 t_{in}  }  +
{t+r \over 8}  \right)
\, \hskip .3 truein
E_\theta = { \sqrt{w_0}  \over r  \sin \theta } \left(  { \Gamma_{in}^2 t_{in} } 
- {t+r \over 8}  \right)
\nn &&
B= {  \sqrt{w_0} \over r } \left( { \Gamma_{in}^2 t_{in}  \over   \sin^2 \theta }  +
{t+r \over 8}  \right)
\, \hskip .3 truein
E_\theta = { \sqrt{w_0}  \over r}  \left( { \Gamma_{in}^2 t_{in}  \over   \sin^2 \theta } -
{t+r \over 8}  \right)
\ea

The second is the $m=3$ case corresponds to the late self-similar stage 
when most of the energy of magnetic shell has been transferred to the forward shock.
This case becomes applicable when all the waves emitted by the central source
have been  reflected from the CD and reached again the center establishing a causal contact. 
In this case $f= \sqrt{w_0} \Gamma_{in}^2 t_{in}^3 / (t^{3/2} \sqrt{\chi})$, and 
 the fields are given by
\ba &
B= {  \sqrt{w_0} \over r \sin \theta} \left( { \Gamma_{in}^2 t_{in}^3  \over t^{3/2} \sqrt{\chi} }
+
{t+r \over 8}  \right)
, \hskip .3 truein  &
E_\theta = {  \sqrt{w_0} \over r  \theta } \left( { \Gamma_{in}^2 t_{in}^3  \over t^{3/2} \sqrt{\chi} } 
-{t+r \over 8}  \right)
\nn &
B= {  \sqrt{w_0} \over r } \left( { \Gamma_{in}^2 t_{in}^3  \over t^{3/2} \sqrt{\chi} \sin^2  \theta }  +
{t+r \over 8}  \right)
, & 
E_\theta = { \sqrt{w_0}  \over r } 
\left(  { \Gamma_{in}^2 t_{in}^3  \over  t^{3/2} \sqrt{\chi} \sin^2  \theta }
- {t+r \over 8}  \right)
\ea

Generally,
self-similar solutions have a very limited application since at the coasting stage
they are limited by finite activity period of the source, limiting their application
to a narrow layer near the CD, while at the self-similar stage
(when the whole shell comes into a causal contact) the total energy remaining
in the shell is small -
 most of the energy has been transfered  to the forward shock.

It is instructive to  relate this simple solution to the properties
 of characteristics.
   In force-free electro-dynamics,
there are just two
characteristics, a fast mode and an intermediate (\Alfven) mode. The former
is simply an electromagnetic wave with  electric
vector  parallel to $\vec k\times\vec B$. It propagates
with phase and group velocity $c$ in all frames
\citep{ucb97,kom02}.
The intermediate mode
is best described in the shell  frame, where the background 
electric vector vanishes.
It has its electrical perturbation perpendicular to $\vec B'$ and
magnetic perturbation  parallel to $\vec k'\times\vec B'$. The phase
velocity is $\hat{\vec k'}\cdot\hat{\vec B'}$ and the group velocity
is equal in
magnitude to $c$ and directed along $\vec B'$. When we transform to a general
frame, the group velocity, $\vec V_g$,
which is what is important for the transmittal
of information, has a magnitude of $c$ and a component
$\vec E\times\vec B/B^2$ resolved perpendicular to $\vec B$,
with the remaining component, of magnitude $(1-E^2/B^2)^{1/2}c$,
directed along $\pm\vec B$. The magnetic perturbation is along
$\vec k\times\vec V_g$.

The backward-propagating, reflected wave
is a fast mode but
has a small amplitude when $\Gamma \gg 1$.  (In this approximation,
the intermediate model's group velocity is purely toroidal and carries
no radial information.) However, the fast mode
is able to propagate radially inward and, in effect, create
an electromagnetic pressure wave which decelerates the electromagnetic
velocity of the outflow and reduces it to match that of the contact
discontinuity. The magnetic field is toroidal and carries information
about the current flow. Put another way, if we were to change the
properties of the load, e.g., by encountering a sudden increase in the ambient
gas density, which would cause sudden jumps in $\Gamma$,
then the boundary conditions on the interior flow would change, along with
changes in the
amplitude of the reflected wave. This allows the interior solution to adjust.

In this section we have considered, mostly for didactic purposes,
two exact solutions for the radial expansion
of the shell. Both of them are not physically realizable:
solution (\ref{bala}) requires a large (formally divergent) axial current
 and a similarly  large distributed return current, while  the solution (\ref{bala1}) requires
an unreasonable density distribution. Despite of these limitations these
solutions 
 illustrate the principal point of the shell dynamics: that the shell motion
is a result of the electromagnetic waves reflecting from it, that in some  cases it  is
possible  to define electromagnetic velocity of the flow so that electromagnetic waves
behave in a manner similar to hydrodynamics, and that the self-similar structures of
an electromagnetic bubble may resemble  self-similar structure of a fluid shock.

\subsection{Structure of the shell for axial outgoing current}
\label{axial}

Next we consider the most relevant case, when the outgoing wave
consists of an axial current (and axial charge) only. Since generally the reflected wave
is non-radial, this case is considerably more complicated, so we are forced to make
several simplifying assumptions. First,
 we assume that
 the outgoing wave carries only a line current and no distribute
current or charge density. The reasons behind this assumption are discussed in
Section \ref{Collim}.  In this case the outgoing wave
is radially propagating.
Secondly, as we discussed 
in the previous section, the reflected waves are typically $\Gamma^2$ weaker.
Thus, we can expand the electromagnetic fields in the shell assuming that
the reflected wave is weak. 
Expanding the field 
 in terms of small amplitude of the reflected wave $\epsilon \sim 1/\Gamma^2 \ll 1$,
\ba &&
B = {2 I_0 \over r  \sin \theta } f(t-r)+\epsilon B ^{(1)}
\nn &&
E_\theta = {2 I_0 \over r  \sin \theta } f(t-r) + \epsilon E_\theta^{(1)}
\nn &&
E_r = \epsilon E_e,
\ea
we find from  eqns. (\ref{ff1})
\ba &&
\partial_t ( B ^{(1)}  r) + \partial_r (E_\theta^{(1)} r) - {\partial_ \theta E_
r}=0
\label{p1} \\ && 
\partial_r(  B ^{(1)}  r) + \partial_t ( E_\theta^{(1)} r) =0
\label{p2} \\ &&
{ \partial_ \theta ( \sin \theta( E_\theta^{(1)} - B ^{(1)} )) \over \sin \theta
} + {
\left( \partial_r + \partial_t \right) ( E_r r^2) \over r} =0
\label{Q}
\ea
These
equations simplify if we introduce
\be
B^{(1)}= { b ^{(1)} \over r}, \hskip .3 truein 
E_\theta^{(1)}= { e_\theta^{(1)}\over r}, \hskip .3 truein 
E_r = {e_r \over r^2}
\ee
Then
Eqns. (\ref{p1}-\ref{p2}) 
 may be combined to give
\be
\partial_r^2(e_\theta^{(1)} ) - \partial_t ^2 (e_\theta^{(1)} ) - 
\partial_r \partial_\theta {e_r \over r^2} =0,
\ee
which shows that generally  for $ E_r \neq 0$ the inward propagating wave is not radial.

We can simplify the system (\ref{p1}-\ref{Q})  if we introduce $y =( e_\theta^{(1)} -  b ^{(1)})$:
\ba && 
r^2 \left( \partial_t -  \partial_r \right) y + 
\partial_\theta e_r =0
\nn &&
{ \partial_\theta \sin \theta y \over \sin \theta } +
\left( \partial_t +  \partial_r \right) y=0
\label{Sa}
\ea
which can be further reduced  to a single  PDE for one variable
$y $:
\be
\left(
\partial_t+ \partial_r \right) r^2 \left(
\partial_t -  \partial_r \right) y + \partial_\theta \left( {  \partial_\theta (\sin \theta y) 
\over \sin \theta } \right)=0
\label{y}
\ee
This non-linear PDE equation describes the dynamics of the reflected wave.

Equation (\ref{y}) allows a separation of angular variable. Assuming that
$y=f(r,t)g(\theta)$, equation  for $g(\theta)$ becomes
\be
\partial_\theta \left( {  \partial_\theta (\sin \theta g) 
\over \sin \theta } \right) +  l(l+1) g=0
\ee
which has solutions $f=P^1_{l}(\cos \theta)$ where $P^1_{l}$  are generalized
Legendre polynomials (for $l=0$ the solutions is $f= C_1/ \sin \theta + C_2 \cot \theta $).

Equation for $f(r,t)$ may, in turn, be simplified if we introduce 
retarded and advanced time variables
$\xi_1 = t+r$ and $\xi_2 = t-r$:
\be
\partial_{\xi_1} \left( (\xi_1-\xi_2)^2 \partial_{\xi_2} f \right) - 
 l(l+1) f =0 
\label{Y}
\ee

If a solution $f(\xi_1 , \xi_2) $ to the Eq. (\ref{Y}) is found, then the corresponding fields
are found from relations $b^{(1)}, e^{(1)} \propto P^1_{l}(\cos \theta)$,
$e_r \propto \partial_\theta(\sin \theta P^1_{l})/ \sin \theta$ 
and 
\ba &&
2 \partial_{\xi_1}b^{(1)}= -\left( \partial_{\xi_1}  + \partial_{\xi_2}  \right) f
\nn &&
2 \partial_{\xi_1}e^{(1)}= \left( \partial_{\xi_1}  - \partial_{\xi_2}   \right) f
\nn &&
2 \partial_{\xi_1} e_r = -f
\ea

Solutions of the Eq. (\ref{Y}) cannot be found analytically, except for 
$l=0$,
 in which case the general solution can be written as 
a sum of two solutions $y_{l=0}=y_{l=0;1}+y_{l=0;2}$:
\ba &&
y_{l=0;1} ={C_1 \over \sin \theta}  
\left(H_1(\xi_1) + \int { H_2(\xi_2) \over (\xi_1-\xi_2)^2} d \xi_2 \right)
, \hskip .3 truein
e_{r,1} = \left(C + \ln \sqrt{\cot \theta/2} \right)  H_2(\xi_2) 
\nn &&
 b^{(1)}_1= {1 \over 2} \left( G_2(\xi_2) - H_1(\xi_1) + {H_2(\xi_2) \over  \xi_1-\xi_2} - 
\int { H_2(\xi_2) \over (\xi_1-\xi_2)^2} d \xi_2  \right)
\nn &&
e^{(1)}_1=  {1 \over 2}  \left( G_2(\xi_2)+  H_1(\xi_1) + {H_2(\xi_2) \over  \xi_1-\xi_2}  + 
\int { H_2(\xi_2) \over (\xi_1-\xi_2)^2} d \xi_2  \right) 
\ea
and 
\ba &&
y_{l=0;2} ={C _2 \cot  \theta}
\left(H_1(\xi_1) + \int { H_2(\xi_2) \over (\xi_1-\xi_2)^2} d \xi_2  \right)
\nn &&
e_{r,2} =  {1 \over 2}  { \int H_1(\xi_1) d \xi_1  }
, \hskip .3 truein
b^{(1)}_2 =  - {H_1(\xi_1) \over 2} \cot  \theta
, \hskip .3 truein
e^{(1)}_2= {H_1(\xi_1) \over 2} \cot  \theta  
\ea
where $C$, 
$C_1$ and $C_2$ are constants and $H_1(\xi_1)$, $H_2(\xi_2) $ 
and $G_2(\xi_2)$ are arbitrary functions ($G_2(\xi_2)$ is  
just a correction to the amplitude of the outgoing wave and can be set to zero).

Solutions of the Eq. (\ref{y}) 
should be matched to boundary conditions.
Eq. (\ref{b1}) implies immediately
\be
\Gamma = {\Gamma_0(t) \over \sqrt{ \sin \theta}}
\label{bg}
\ee
In order to use condition  (\ref{b2}) we need to know 
how the angle $\alpha_{CD}$ between the radial direction and the normal
to the CD  depends on $\theta$ and how it evolves with time. 
This can be done only in the self-similar regime,
 when the motion of the CD is a simple given function
$\Gamma_0(t)= \Gamma_{in} (t/t_{in})^{-m} 
$ where 
$\Gamma_{in}$ is the initial Lorentz factor at time $t_{in}$.
In this case we find
\be
\tan \alpha_{CD} =
 - { \cos \theta \over 2(m+1) \Gamma_0^2 \left(1- \sin \theta /( 2(m+1) \Gamma_0^2) \right)}
\ee
Thus, 
\be
\alpha_{CD} \approx { \cos \theta \over 2(m+1) \Gamma_0^2 }
\ee
which is only weakly 
dependent on $\theta$ for $\theta \ll 1$.

The angle $\alpha_{CD }  \sim  1/\Gamma^2 \ll 1 $ is the angle 
at which the waves emitted by the central source fall onto the CD ($\alpha_{CD}$ is measured
in the  laboratory frame). Note, that the   
reflected angle, on the other hand, is of the order of unity
$\sim \alpha_{CD}  \Gamma^2$. 

Expanding the boundary conditions for $ \alpha_{CD} \sim O(1/\Gamma^2) $ we find
\be
r \left( B^{(1)} - E^{(1)} 
+E_r \alpha_{CD} \right) \Gamma^2 \sin \theta = I_0 F_1(t-R)
\label{b21}
\ee

Recall next, that
$ B^{(1)} - E^{(1)} \propto P^1_l (\cos \theta) $ and
$ E_r \propto \partial_\theta( \sin \theta P^1_l )/ \sin \theta$.
This implies that in order to satisfy Eq (\ref{b21})
 the reflected wave  should consist of  a large number of 
spherical harmonics $ P^1_l$ with temporal and radial dependence
given  by  solutions of Eq.
(\ref{Y}). 

At this point 
 we stop our analytical approach since the  resulting equations are clearly more easily
solved by numerical simulations. 
In this section we showed how the dynamics of time-dependent force-free outflows should be calculated,gave simple analytically tractable examples and finally wrote down equations
describing a particular type of electromagnetic explosion, when the outgoing current
is confined to a small region around the axis. Though we did not solve for the structure of the 
reflected wave, its properties have
little effect on the form of the CD since typically the  amplitude of the reflected wave is 
$\Gamma^2$ smaller than of the forward wave.
 For example, the basic property of the expanding shell, dependence of 
Lorentz factor on the angle Eq (\ref{bg}),  can be obtained without solving for internal structure.

The upshot of this section is that our preferred  current distribution --  with current concentrated
near the pole, along the CD  and in the equatorial region -- will produce a nonspherical
relativistically expanding shell. The internal structure of the shell is determined by  a combination
of outward and inward propagating waves and is in many ways similar to hydrodynamic subsonic flow. 
Our solutions are formally divergent on the axis, $\Gamma_{CD} \propto 1/\sqrt{\sin \theta}$, 
$L_\Omega =L_0/\sin^2 \theta$ but the
total energy flux is only  weakly (logarithmically) divergent $L_{tot}=  \int L_\Omega d\Omega 
\approx 2 \pi L_0 \ln \sin \theta $.
The outflow then should have a current carrying core. The core may, in fact, be cylindrically 
collimated \citep[\eg][]{HeyN03}, but if we assume that its typical polar angle is 
$\theta_0 \sim 10^{-3} \ll 1$ (see (\ref{theta0})), then, qualitatively, 
\be
L_\Omega \approx {L_0 \over \theta^2+\theta_0^2}, 
\hskip .3 truein
L_{tot} \approx 2 \pi L_0 \ln 1/ \theta_0
\ee
The  energy released per solid angle and the total energy of the explosion will have a similar
dependence
\be
E_\Omega= {E_0   \over \theta^2+\theta_0^2},
\hskip .3 truein
E_{tot} \approx 2 \pi E_0 \ln 1/\theta_0
\label{Etot}
\ee
This will play an important role in the interpretation of afterglows (see Section \ref{after}).

\section{$\gamma$-ray emission}
\label{emission}

\subsection{Dissipation of magnetic energy}

One of the principal implications of the electromagnetic hypothesis
is that the  conventional model of particle acceleration  -
 acceleration at internal shocks - cannot work in
this model.
In the RFF limit, the fast speed is the speed of light so that
fast shocks do not  form. If we add a limited quantity of plasma and 
use RMHD, the shocks are weak and not likely to be efficient 
particle accelerators. There are no intermediate shocks in the RFF 
limit, though rotational discontinuities can be present.

We therefore propose that the $\gamma$-ray-emitting electrons are accelerated
by current instabilities during the magnetic shell phase.  
Current-driven instabilities play a major role in a variety of laboratory
experiments, \eg  TOKAMAK discharges like sawtooth oscillations and
major disruptions, \citep[\eg][]{kad75}, Z-pinch collapse,
 \citep[\eg][]{rud97}, in the  Earth's magneto-tail
\citep[\eg][]{gca78} and on the Sun \citep[\eg][]{asch02}.
The development of
current instabilities usually results in enhanced  or anomalous
plasma resistivity  which can lead to an efficient 
dissipation of the  magnetic field.
The magnetic energy is  converted into heat, plasma bulk motion
and, most importantly, into   high energy particles, which, in turn, 
are responsible for the
  production of the
prompt $\gamma$-ray emission.  The conversion of magnetic energy
into particles may be very efficient.
Recent RHESSI observations of the Sun indicate that, in reconnection
regions, most magnetic energy goes into non-thermal electrons \citep[]{benz03}.
We propose, that in case of GRBs it is the 
dissipation of magnetic energy
that is responsible for 
particle acceleration. 

Why is  magnetic field dissipation is negligible close
to the source and  become important at large distances?
We argue that this is because the 
particle acceleration is  suppressed 
near the central engine  by efficient pair production which screens out
 the electric field that led to the particle acceleration.
Eventually,  the optical depth to pair production becomes small
so that large electric fields may develop (see Eq. \ref{taug3}).
This estimate is consistent with the assumption that particle acceleration and emission 
generation takes
place at $r \geq 10^{16}$ cm. (If particle acceleration 
takes place at $\tau_{\gamma-\gamma} \geq 1$
all the IC photons that will necessarily  accompany such acceleration
will make pairs. This will increase the  density of pairs and will shut off acceleration. 
In this regime most dissipated  energy 
  goes to pair production, not synchrotron emission.)

Magnetic dissipation is complicated. After 40 years of intensive
research it is not clear at the moment what is the correct {\it macroscopic}
model of magnetic  reconnection, Sweet-Parker or Petschek \citep{bis00,pf00}. 
The dissipation of magnetic energy in relativistic, magnetically-dominated
plasma is likely to be complicated  as well. Only the first steps have been made
in that direction
 \citep[\eg][]{lu03,hos02,llm02,lyu03}. A fundamental problem is that 
in case of magnetic reconnection 
most of the observed properties depend sensitively on the {\it kinetic}
 details of the model. This can be contrasted with the 
shock acceleration model, where  a basic kinetic property - spectrum of
accelerated particles -  can derive using simple macroscopic 
quantities - shock jump conditions \citep[\eg][]{blandEich}.

Overall, the observed radiative  properties of 
magnetic dissipation and internal
 shocks should be similar. Emission generation (synchrotron) is likely 
to be the same in reconnection models, so that
most of the well detailed radiation models will still hold
(modulo, perhaps, the assumption of the  equipartition magnetic field).
On the other hand, the acceleration mechanism is very different, but, unfortunately, 
  hardly distinguishable observationally -
both internal shocks
and reconnection regions represent  transient internal dissipative
regions, which heat the plasma and  accelerate particles.
This problem is also complicated by the fact that, in the emission region, the plasma
should be  near equipartition (since it emits most efficiently then). Unlike with
 the fireball model, 
in the electromagnetic model, 
 equipartition is reached  by dissipation of the magnetically dominant plasma.

Particle acceleration  by dissipative magnetic fields
 may proceed in a number of ways. The best studied non-relativistic
example is particle acceleration in reconnection regions either 
  by inductive 
electric fields, resistive electric fields inside the 
 current sheets \cite[\eg][]{cl02} or formation of shocks in the downstream
of reconnection regions \citep[\eg][]{bf94}. Investigation of the particle
acceleration in the relativistic regime in only beginning \cite[\eg][]{llm02}.
Relativistic reconnection may also produce power-law spectra of accelerated particles 
\citep{hos02}.
For example, in the relativistic Sweet-Parker reconnection model
\citep{lu03}, 
if one balances linear acceleration inside the reconnection layer by the
resistive electric field, $d_t  {\cal E} \sim e E c$ with the rate
of particle escape (proportional to relativistic gyro-frequency),
$d_t \ln N({\cal E}) \sim \om_B (mc^2 / {\cal E})$, one finds
\be
N({\cal E})  \sim {\cal E}^{-\beta_{in}}
\ee
where ${\cal E}$ is the energy of a particle, $N({\cal E})$ is the particle number
and $\beta_{in} \sim E / B$ is the inflow velocity. For relativistic reconnection
the inflow velocity can be relativistic \citep{lu03}. This simple estimate
is confirmed by numerical simulations \citep{llm02}.  
The fact that reconnection models can produce spectra
which are prohibitively hard for shock acceleration may serve as a distinctive property
of electromagnetic models.

In addition to acceleration at reconnection layers 
particle acceleration may occur 
    through 
 formation of a spectral cascade of nonlinear  waves in force-free plasma which
transfer energy to progressively larger wave vectors until this energy
is taken up in accelerating a population of relativistic electrons and
positrons.
Formation of turbulent cascade may be initiated, for example, due to development
of dynamic instabilities of the CD 
(Appendix \ref{IKS}).  The generic evolution of this spectrum under
force-free conditions
is complex. However, it appears to be generic that the evolution of the
electromagnetic
field will lead to the formation of expanding surfaces
on which $E\rightarrow B$ \citep[\cf][also Appendix \ref{resicol}]{hh98}. (Recall
that there is no
mathematical limitation on $B^2-E^2$ changing sign under strict
force-free conditions.) In practice, the particles that are
 present are subject to rapid acceleration
through $\vec E\times\vec B$ drift and this is followed  by
$\gamma$-ray emission through synchrotron radiation and inverse Compton
scattering, followed by pair production. This process will
continue until there is enough inertia
and or pressure starts to contribute to the equation of motion so that
relativistic
magnetohydrodynamics becomes necessary to describe the behavior of the
plasma.


In the  case  of polar emission (see below) a possible mechanism of particle acceleration is by
inductive electric fields developing during 
electromagnetic collapse near the axis of the flow \citep{trub02}. This   type of acceleration
resembles laboratory Z-pinches \citep[\eg][]{rud97}. This mechanism may be considered as 
reconnection at an O-point. 

In addition to the acceleration mechanisms which are based on known non-relativistic
schemes, it is feasible that acceleration in relativistic, strongly magnetized plasma
may proceed through mechanisms that do not have non-relativistic or fluid analogues. 
Examples of this type of acceleration include kinetic electromagnetic-type
  instabilities of the shell surface currents, as proposed  by
\cite{su96} and in somewhat different form by \cite{Liang}.
Since kinetic properties of strongly magnetized relativistic plasmas are
only beginning, it is hard to predict acceleration efficiency and particle spectra.
Numerical studies in the coming years will be most important here.

In these case dissipation regions will
 constitute  expanding volumes within which there will be a local
equipartition. These volumes will move with relativistic speeds  in the average
outflow expansion
frame defined by the average contact discontinuity and this will
influence the observed
variability properties.
Under a broad range of initial conditions, we expect that the density
of pairs will increase
and the temperature will fall until pair production balances
annihilation.  Given the tenuous, optically thin environment that is
envisaged, the equilibrium ``temperature''  will be roughly 10-20keV
although a pronounced nonthermal tail should also be present in the
pair distribution function. The gamma ray spectrum may exhibit a
feature at this energy  which will boosted to roughly $\sim1$~MeV. A
more careful calculation is needed to see if the spectral break
sometimes seen at around $\sim200$~keV can be reproduced \citep[see also][]{pw03}.

In this paper we  do not commit to a particular acceleration mechanism, but
give a necessary  qualitative description of the possible
location of acceleration regions in the frame work of our model
and outline possible ways in which magnetic  dissipation  and particle acceleration
may proceed. 
A more detailed  discussion of the acceleration and  emission physics 
is deferred to a later paper.
We envision 
two possible locations of the emission: (i) along the poles, $\theta \leq 1/\Gamma$,
 which we associate with short
bursts  and  (ii) in the 
of the magnetic shell, $\theta  \geq 1/\Gamma$, which we associate with long GRBs.
 We consider these two possibilities in turn.

\subsection{Polar Emission}   

\begin{figure}
\includegraphics[width=0.7\linewidth]{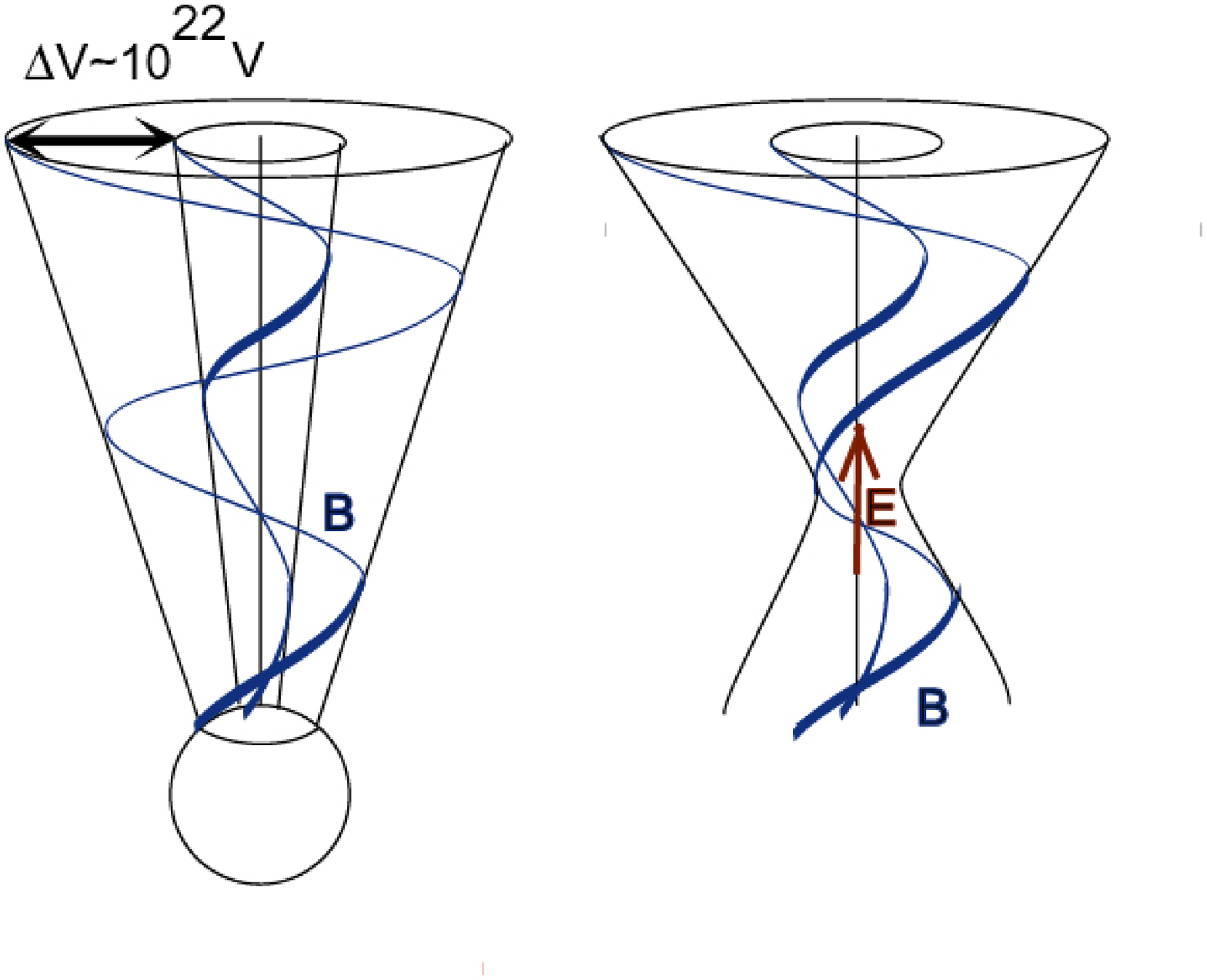}
\hskip -0.5 truein
\includegraphics[width=0.4\linewidth]{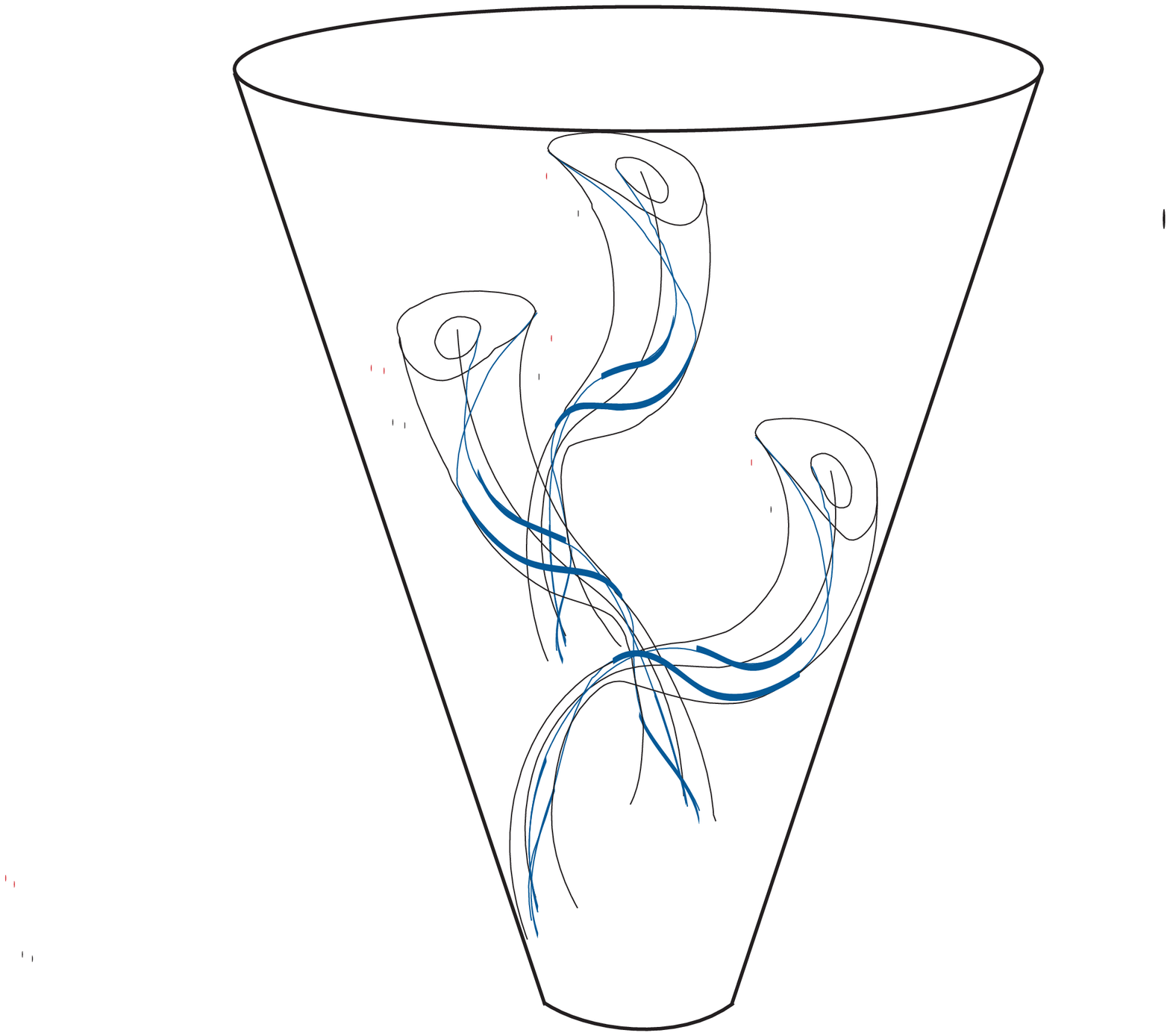}
\caption{
Development of pinching  instabilities leads to core  contraction, filamentation,
 disruption of  currents  and particle acceleration
by DC electric field. }
\label{pinch-coll}
\end{figure}

The dynamics of the flow near the polar regions  is likely to be unstable to 
magnetic pinching, giving  dynamics similar to imploding Z-pinches 
\citep[\eg][]{rud97}
  where 
 explosive X-ray bursts
 as well as 
electron and ion acceleration are often observed. \footnote{The idea of Z-pinch collapse 
has already been suggested in application to GRBs \citep{Trub95}, but
the model lacked important astrophysical ingredients. }
In the proposed model the outgoing wave consists of currents strongly 
concentrated near the axis. Such concentration of 
poloidal currents is likely to  become unstable due to  the 
development of dynamic and resistive instabilities
in the strong current region. 
The fastest growing modes are likely to be pinch, kink and helical modes (see Fig. \ref{pinch-coll}).

At large currents the
 plasmas in Z-pinches and TOKAMAKs
 are known to form thin  filaments  stretching both
along and across the direction of the current \citep{trub02}.
The typical growth rate of filamentation instabilities is 
$\sim \om_p v_d/c$ where $ \om_p$ is the
plasma frequency and $v_d$ is the drift velocity
of the current particles \cite{mol75}.
Thus, as the current is compressed, at some stage it will break
into  filaments.
Secondary filaments will also be compressed and will
 undergo further filamentation.
The development of instabilities will result in  disruption of the current, during which
 large DC electric fields are created near the
O-type point. (The dynamics of current disruption
is considered in more detail in Appendix \ref{Disrupt}).
 This electric field will accelerate run away
particles (both electrons, positrons and ions). Thus, the magnetic energy 
 will finally be radiated as electron  cyclotron emission
 and will also be transported away by 
cosmic rays.
The current disruption at each filament will
produce a burst of emission, giving a complicated
pulse profile for  prompt GRB emission.

Initially, the size of the unstable region is of the order of the core radius $r_D$
(see Eqns. (\ref{rD}-\ref{theta0})).
As the instabilities develop   the magnetic energy is dissipated, which leads to a
loss of magnetic pressure, so that progressively larger scale modes become
unstable (this is similar to an outgoing rarefaction wave). 
A typical angular size  for an unstable current will  be of the order of $1/\Gamma$ 
 since in relativistic flows
lateral (in $\theta$ direction) forces (e.g.,  magnetic hoop stresses)  are strongly suppressed
by a factor $1/\Gamma^2$  \citep[\eg][]{bts99}, 
so that globally (on scales $\theta > 1/\Gamma$) the
 flow  cannot adjust to the change of equilibrium on  a dynamical
time scale --
 the time scale for establishing an equilibrium in
the lateral direction is much larger than the dynamical times scale, by a factor
$\sim \Gamma$.
The fastest evolution of the flow will take place near the axis, within $\theta \sim 1/\Gamma$,
where geometric effects may ``beat'' the relativistic suppression
and produce changes in the field configuration on shorter time scales.

\subsection{Shell Emission}

The expanding shell will be preceded by a surface  (Chapman-Ferraro) current
 emanating from the 
poles and flowing to the equator (or {\it vice versa}). It will be followed,
at a distance of the order of the shell thickness by a reverse 
current. 
There are several possible instabilities associated with this configuration.
(Since the shell decelerates on average, it is stable to the   Kruskal-Schwarzschild 
instability). 

Qualitatively, the  interface between two magnetized media is known
to be unstable in the  case of the  Earth's magnetosphere.
Approximately $10\%$ of the incoming energy flux is dissipated at the day side of the
magnetosphere \citep[\eg][]{cowley82}. In the 
 case of  relativistic flows,  it is feasible
that a similar or even larger fraction of the incoming energy flux may be dissipated.
 Dissipation may be initiated by several types of instabilities.
First, 
the surface Chapman-Ferraro  current becomes unstable due to development of microscopic instabilities
on the scales of tens to hundreds of ion Larmor radii \citep{su96,Liang}. 
As a result, the surface current breaks into  strongly 
non-linear fluctuating current and charge layers. 
These current substructures   become strongly dissipative and accelerate particles. 
Secondly, if the source varies considerably during its activity the 
surface current may become unstable due to impulsive  Kruskal-Schwarzschild instability (see
appendix \ref{IKS}). The impulsive  Kruskal-Schwarzschild instability (IKS)
develops if the  surface separating magnetic field and matter is accelerated by an 
electromagnetic pulse. It is similar to Richtmyer-Meshkov instability in fluid dynamics. 
 At 
the nonlinear stage, IKS instability 
 will lead to formation  of small scale structures that will
become resistive. During  the prompt emission phase, the  presence of resistive 
surface currents may  lead to an absorption of a large
fraction of the  incoming fast mode  energy flux. This  will  result in 
dissipation of magnetic energy and particle
acceleration.

An attractive feature 
 of the shell model is that it predicts
$\gamma$-ray emission over large solid angles   and few orphan afterglows. The 
strength of the burst may also depend on the efficiency of radiation generation, which
may be a strong function of  polar angle, but the  kinetic
total energy (inferred, \eg,   from  early afterglows)  should  remain approximately
constant.
Burst viewed  from a  small angle (as may be inferred from achromatic
breaks in the afterglow emission) should be seen to larger distances
although a large burst to burst dispersion should be expected. 
In addition, if emission is confined to  a narrow region near the surface of the shell,
it will be more variable than in the case of a filled shell, 
since there is no ``radial averaging'' over emitters. 
Simulations
of shell emission are also underway which will clarify some of these points.

\subsection{Variability}

One of the key  observational properties that every model of GRBs should address is the 
 short times scales of   variability, as  short as  $\sim10$~ms. 
One merit of the electromagnetic model  is that it allows the GRB to originate at
a much larger radius, up to $\sim10^{16}$~cm than in the standard, baryonic
 shock model $\sim10^{13}$~cm. 
This  stems from the 
fast that,  in an electromagnetically  dominated medium, the emission 
regions (\eg, large amplitude waves likely to occur in an
intermittent turbulence spectrum) may be moving with 
relativistic velocities in the bulk  frame.
In this section
we consider the  statistical properties of radiation emitted by an ensemble
of the emitters confined to a narrow spherical shell 
and  subject to  relativistic bulk  and random motion.
We show that relativistic internal motion of ''fundamental emitters'' can account
for highly intermittent GRB profile with   smaller bulk 
Lorentz factors or larger radii of emission than is usually inferred.

\begin{figure}
\includegraphics[width=0.9\linewidth]{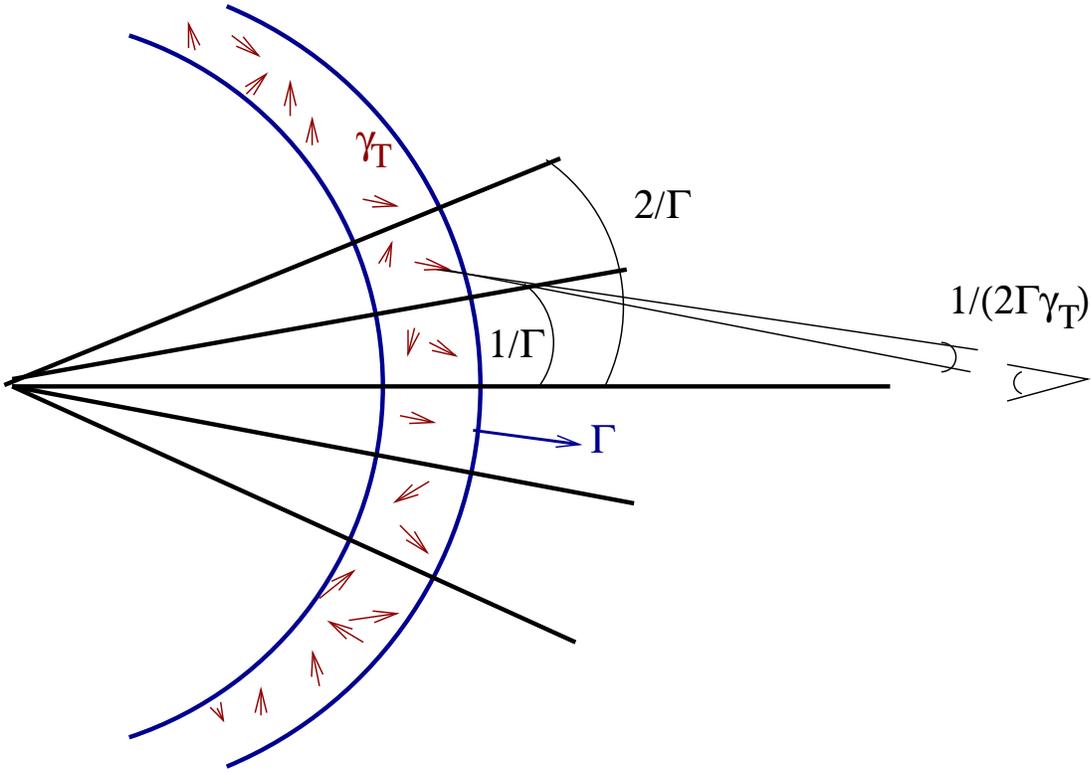}
\caption{
Variability of emission from relativistically moving emitters. 
A shell is moving with relativistic Lorentz factor $\Gamma$. In addition,
primary emitters have ``thermal''
 spread with a typical Lorentz factor $ \gamma_T \leq \Gamma $.
Each isotropic
 emitter produces a pulse of width $ \Delta \theta  \sim 1/( 2 \Gamma \gamma_T)$
when observed  in the lab frame, while only emitters located within the angle
$ 2/\Gamma$ may be seen by an observer.  
}
\label{var}
\end{figure}

Consider a shell of thickness $\Delta R \ll R$ (in the lab frame) moving
relativistically with a bulk Lorentz factor $\Gamma$ (see Fig. \ref{var}). 
Assume that the shell consists of  randomly distributed emitters
which move randomly with respect to the shell rest frame with
 a typical Lorentz factor $\gamma_T$. 
If $\theta'_r$ is the angle (measured in the rest frame of the shell)
 between the radial direction and
 emitter's velocity and $1<\gamma'<\gamma_T$ is the Lorentz
factor of the particle then in the shell frame, emission is beamed into a cone with opening
angle $\Delta \theta_r' \sim  1/ \gamma '$.  
In the observer's frame the emitter's Lorentz factor is
\be
\gamma = \gamma' \Gamma \left( 1 + v' V \cos \theta'_r\right)
\sim  \gamma' \Gamma \left(1+\cos \theta'_r\right)
\ee
where $v'=\sqrt{1-1/\gamma^2} $ and $V=\sqrt{1-1/\Gamma^2}$, and 
 angle between the radial direction and the particle's
velocity in the lab frame is 
\be
\tan \theta_r = { v \sin \theta_r ' \over \Gamma( V+ v \cos \theta_r')}
\approx {\tan \theta_r '/2 \over \Gamma}
\label{s}
\ee
where the approximation assumes $V\sim v \sim 1$.
From Eq. (\ref{s}) it follows that as long as $v \leq V$, the maximum angle 
with respect to the radial direction
that an emitter may have in order to be  seen by an observer  is $\sim 1/\Gamma$, regardless of the
velocity of the emitter.
At the same time, the emission cone in the lab frame is 
$\Delta \theta_r \sim  1/ \gamma  = 1/\gamma_T \Gamma (1+\cos \theta_r')$.
The addition of internal relativistic motion makes the emission beams much narrower
but the visible emitting volume increases only by a factor of 4, when compared with
 cold emitters.
 For relativistic  internal motion, the number
of emitters seen at a given time will decrease as $ 1/\gamma_T^2$.
 At the same time the flux from each emitter
will increase as $\gamma_T^4$ (Two factors of $\gamma_T $
coming from the fact that all the emission is confined within
a narrow beam $1/\gamma_T^2$ and two  factors of $\gamma_T$
 coming  from the relativistic contraction of a pulse).
Thus, the ``thermal''' spread in the motion of emitters 
can drastically change the estimates of the Lorentz factors. 
The  Lorentz factor inferred from variability  is, in fact, a  product
of the bulk and ``thermal'' Lorentz factors. 
Since for a given $\Gamma$
 the number of emitters seen at a given time decreases $\propto 1/\gamma_T^2$, 
modest values $\gamma_T \leq \Gamma$ suffice to produce large variations.

\subsection{Causal structure of electromagnetic  outflows}

Electromagnetic outflows have a different  casual structure from hydrodynamic flows.
Initial evolution of both types of flow is similar: 
 close to the central source both  electromagnetic and  hydrodynamic flows are
  subsonic  and  fully causally connected. Later, after breakout, both types of flows are
 linearly 
 accelerated by magnetic and/or 
 pressure forces $\Gamma \sim r$. Hydrodynamic  flows become causally disconnected
after passing through a sonic point where $\Gamma = \sqrt{3/2}$. 
An electromagnetic flow can  pass through the fast magnetosonic point provided that 
 initially $\sigma_0 > \Gamma^2$,
  and will become  causally disconnected
 over small 
polar angles  $ \Delta \theta \sim 1/\Gamma$.

After the  photosphere, when magnetization
parameter $\sigma$  increases by many orders of magnitude,
causal behavior of the types of flow is drastically different. 
Hydrodynamic flows virtually do not establish a causal contact over angles large than 
$1/\Gamma$.
 Contrary to that,
electromagnetically-dominated flows quickly re-establish causal contact.
Lyutikov \etal (2003) showed that sub-Alfvenic ejecta
  re-establishes a causal contact
over the visible patch of $1/\Gamma$ in just  one dynamical time scale
 (after doubling in radius).
In fact, causal contact may be established over whole expanding
shell  after a time $t_{\rm c} \sim t_s \Gamma^2 $. At this point the shell is  {\it fully
causally-connected }.

Causal contact over large angles of the shell is established
by waves  propagating  almost backward in the shell frame.
Here lies  a principal difference between strongly magnetized  flows  on the one
hand
and weakly and non-magnetized  flows on the other.
Strongly magnetized flows can be subsonic with respect
to the velocity  of the signals in the flow, while weakly magnetized flows
are always supersonic. Thus, in case of weakly magnetized flows
  waves emitted backward in the shell frame are "advected" with the flow and
cannot reach to large angles. Strongly magnetized flows  are subsonic, so that
waves can outrun the flow and reach to large angles. In doing so
they  leave the shell (which has a thickness either $c t_s$ or
$R/\Gamma^2$), but this does not present a problem since the
fast magnetosonic waves become light waves if they propagate into  a low density
medium.

The 
causal structure of electromagnetic
shells plays an important role in maintaining the
coherence of large scale magnetic fields. If a surface of  relativistically
expanding magnetized shell is perturbed at some  radius, it can quickly propagate  information
(\eg magnetic  pressure)
over the visible angle $1/\Gamma$,
 so that the shell can have  quasi-homogeneous  properties   despite
possible inhomogeneities in the circumstellar medium and in ejecta.
In hydrodynamic,  $\sigma \ll 1$,  or hydromagnetic, $\sigma \sim 1$ \citep[\eg][]{sdd01},  models
only under strict homogeneity of the
surrounding medium and of the ejecta will the two causally disconnected  parts of the flow
 have similar properties.

Thus,
the electromagnetic model  provides a solution to the puzzle
of how to launch a blast wave that extends over an angular scale
$>>\Gamma^{-1}$ and where the individual parts are out of causal contact.
In the electromagnetic model, the energy is transferred to the blast wave
by a magnetic shell which is causally connected at the end of
the coasting phase.

\section{Afterglows}
\label{after}

As the magnetic  shell expands, its energy is gradually transfered 
 to the preceding forward shock wave. In a constant density medium
 this transfer occurs
at the end of the coasting phase. 
At later times,
most of
the energy released by the central, spinning, magnetic rotator is  carried
by the shocked interstellar medium.
The structure of the blast wave is well approximated by the self-similar adiabatic
solution \citep{blm76,blm77}, but  with angle-dependent expansion. 
If the total energy released per unit solid angle is $E_\Omega \sim L_\Omega t_s$,
 then \citep{blm76}
\be
\Gamma \sim \left( { 17 E_\Omega \over 2 \rho_{ex} c^5 t^3} \right)^{1/2}
\ee
If, in addition, the energy release is as argued above 
(see Eq. (\ref{Etot})), 
then 
\be
\Gamma \sim  \left( { 17 E_0 \over 2 \rho_{ex} c^5} \right)^{1/2} { t^{-3/2} \over 
\sqrt{ \theta^2+  \theta_0^2}} 
\approx  { \Gamma_0(t) \over \theta}
\mbox{ for $ \theta \gg  \theta_0$.}
\ee

As long as  the blast wave remains relativistic 
 the energy in the magnetic shell decreases slowly (logarithmically),
 so that  the relativistic blast wave stage is coexistent with
the self-similar stage of the magnetic shell 
as long as the shock remains relativistic,
$r<r_{{\rm NR}}\equiv(E_\Omega /\rho c^2)^{1/3} \sim10^{18}$~cm.
 The 
  afterglow phase 
usually becomes unobservably faint after the expansion speed becomes 
mildly relativistic. At that point the forward shock finally detaches
from the magnetic shell and 
 will expand non-relativistically and become
more spherical with time, resembling a normal supernova remnant.

Relativistic particles are accelerated  in the blast wave producing
the observed  afterglow in a manner
which is essentially similar to what is proposed for fluid models.
There are, however, some differences.
First, 
the contact discontinuity itself may be an  important source
of magnetic flux through IKS instability  (appendix \ref{IKS}).
Then the
  afterglow may  result from a mixture
of relativistic particles, derived from the shock front with 
magnetic field derived from the shell,
 much like what seems to happen in regular supernova
remnants.

The 
development of the  IKS instability will 
lead to distortion of the CD.
The perturbations propagate mostly  orthogonally to the
magnetic field.  In the  linear regime, this will lead to ''braiding'' of the  field
lines, so  that large scale ordered field  remains mostly unaffected.
In the nonlinear regime  a  degree of ordering
of the magnetic field may be  preserved as well. 
Later, at the afterglow phase, the  IKS  instability will lead to mixing of the
ejecta with  circumstellar material. This requires that the source
remains active for much longer  than 100 sec. There  are indications, \eg in case
of  GRB030329, 
that this indeed may be the case. 
Since during   the
development of the  IKS  instability 
 field lines will mostly braid keeping large scale field ordered, 
an ordered component
of the field will be introduced to the  shocked circumstellar 
material and may explain the   polarization of afterglows.

In addition to dynamic instabilities, the CD may also be unstable to resistive instabilities
like the  tearing mode. The magnetic field from the  shocked circumstellar material piles up
on the CD, so that however small the field is in the the bulk of the flow, 
it plays an important  part in  a thin
boundary layer near the CD \citep{lyu02}. Thus the CD becomes a rotational discontinuity
and  should be  susceptible to resistive instabilities, similar to the case of 
Earth magnetopause \citep{gkz86}. 
The 
 growth of tearing mode occurs on wave modes propagating orthogonally
to the magnetic field. As  a result ``percolated''  magnetic filaments  form
that connect
outside and inside plasma.

The second difference in the afterglow dynamics between the electromagnetic l 
and fireball models is that the form of the
 forward shock is  a definite function, determined by the current distribution in the shell.
This  is also   the  principal 
difference between relativistic and non-relativistic (Sedov) blast waves.
In the non-relativistic case explosion 
quickly ``forgets'' the details of the central source and dynamics
is determined only by the total energy release, while in the strongly relativistic regime the
forward shock also carries information about the angular energy distribution of the source.
The form of the blast wave would reflect both the form of the driver
(magnetic shell) and subsequent evolution of the shock 
 \citep{kom60,sha79,gkp03}. 
\footnote{Non-spherical shocks and outflows have been extensively studied
in the non-relativistic hydrodynamics \citep{kom60,lp69}
 and in application to the 
propagation of  AGN jets \citep[\eg][]{wii78,norm82}. 
\cite{sha79}
generalized  the work of \cite{kom60,lp69}  
to non-spherical  relativistic shock waves. 
These works are virtually neglected in GRB research.
 As a result incorrect assumptions (\eg, ``gramophone-type'' profiles)
 about the dynamics of the 
non-spherical shocks were made.}
In the strongly relativistic regime, 
the form  of the forward shock is mostly  determined by the 
form of the electromagnetic driver, 
so that the motion of the shock is mostly ballistic \citep{sha79}, 
with {\it little sideways expansion} (so the in the rest frame $v_\theta' \ll c$, 
not $\sim c/\sqrt{3} $
as is commonly assumed);  
 this is confirmed  by recent semi-analytical  \citep{gkp03}  and
fully  relativistic hydrodynamic calculations \cite{cgv03}.

Thus, there is little lateral evolution between the forward shock and the contact
discontinuity. 
 If we continue to use our simple model,
we find that the afterglow expansion varies most rapidly and remains
relativistic for longer times  closer to the symmetry axis.
In particular, if the outgoing waves correspond to the outflow carrying
axial current (section \ref{internal}) the energy carried by the  forward shock 
wave will scale as $E_\Omega \sim 1/\sin^2 \theta$ and the Lorentz
factor of the forward shock $\Gamma_s  \sim \Gamma_0 /\sin \theta$.
(Note that at the afterglow stage the  Lorentz
factor of the forward shock depends only on the energy that the source released per unit
solid angle $E_\Omega$ and {\it different from the 
Lorentz factor of the CD $\Gamma_{CD}$}). 
This type of shock has been named ``structured jet'' (or universal jet) \citep{ros02}, though in our
model there is no proper ``jet'', but simply a  non-spherical outflow. 
As $E_\Omega\propto \sin^{-2}\theta$ the energy contained in each
octave of $\theta$ is roughly constant, 
so that the inferred explosion energy with our simple model
will be roughly independent
of $\theta$ and characteristic of the total energy. This, combined with the
assumption that the total energy of GRBs is related to kinetic energy of a critically rotating
relativistic stellar object, explains a narrow range of inferred GRB energies
\citep{fra01}.
Most importantly, 
the resulting non-spherical  blast wave emits
in all directions. However, the intensity of this emission
is strongest along the poles. 
This means that the most intense bursts and afterglows
in a flux-limited sample will be seen pole-on
and can exhibit achromatic breaks when $\Gamma\sim\theta^{-1}$,
which might be mistaken for jets. The observational appearance
  of such  ``structured jets''
have been investigates by several collaborations \citep{ros02,gkp03,perna03}.
The overall conclusion is  that ``structured jets'' provide the best 
 fit to afterglow emission properties \citep{gkp03}  and 
reproduce well luminosity function of GRBs \citep{perna03} including a possibility
of unified description of  GRBs and X-ray flashes 
(the only X-ray flashes with measured redshift had a total energy comparable to 
classical GRBs \citep{sod03}).

In conclusion, we argue that observational appearance of  GRB afterglows
 depends mostly on two
parameters: (i) explosion energy 
(more precisely, on the
 ratio on the explosion energy to circumstellar density) and, most
importantly, (ii) the
viewing angle that the progenitor's axis is making with the line of sight.
This possibility, that all GRBs are virtually the same but viewed at different angles
resembles  unification scheme of AGNs.

\section{Observational implications}
\label{implic}

{\bf Prompt and afterglow polarization}.
Recent observations by the {\RHESSI} satellite have been interpreted as evidence for
  large polarization of the prompt $\gamma$-ray emission of GRBs \citep{coburn03}. 
This has bee contested by \citep{rutl03}.  If the large polarization is confirmed, 
it could be 
 most naturally explained as 
  a synchrotron radiation from 
a large scale magnetic field. In order to  produce very high polarization
 the coherence scale of the field should be  larger
than the size of the visible emitting region. This  
 can be  achieved  naturally  achieved  
if the flow is magnetically-dominated \citep{lyu03c}.

Several natural correlations between the prompt  GRB polarization and other
parameters follow from the model and  can be tested with
future observations.
The maximum amount of polarization in our model  is related to the spectrum of emitting 
particles, being higher for softer spectra \citep{lyu03c}.
 This  points to a possible correlation between the amount of 
polarization and hardness of the spectrum.

Large scale field structure in the ejecta 
  emission may also be related to  polarization 
of afterglows if fields from the 
 magnetic shell  are mixed in with the shocked
circumstellar material. The fact that the magnetized boundary becomes
``leaky'' and both plasma and magnetic field are transported across it 
 has been amply demonstrated by
decades of space experiments \citep[\eg][]{cowley82}. The transport occurs
 either due to microscopic resistive  instabilities
of the surface current 
(similar to the  so called flux transfer events at the day side
of Earth magnetosphere and instabilities at the heliopause \citep{fahr86}) 
or due to dynamic (\eg RT)  instabilities. In the case of Earth and Solar heliopause
approximately $10\%$ of the incoming plasma is transfered through the CD.
In case of GRB the CD may also be unstable 
 due to dynamical  
impulsive  Kruskal-Schwarzschild instability (appendix \ref{IKS}).
Since both resistive (\eg tearing mode) and dynamics instabilities 
are expected to develop mostly on the waves propagating orthogonally
to the large scale magnetic field, a degree of field ordering will be preserved.
 As a result, a  comparably 
large fractional polarization may be observed in afterglows as well. 
In this case, since the preferred direction of polarization is always
aligned with the flow axis,  {\it the position angle should 
not change through the afterglow}
(if polarization is observed both in prompt and afterglow emission 
the position angle should be the same). 
Also, polarization should be most independent of   the "jet break" moment. 
This is in a stark contrast with  the jet model, 
in which  polarization  is seen only near the "jet break" times and
the position angle is predicted
to experience a flip during the "jet break" \citep{gl99,sari99}
\footnote{Note that these works did not take into account kinematic depolarization \citep[\eg][]{lyu03}
which may give an error of the order of unity.}.
Current polarization data are not completely decisive. In most cases  the  position
angle remains constant (\citealt{goro03,covino03a,covino03b,barth03,ber03}), 
while the 
amount of polarization does not show any correlation  with 
the "jet break". This is consistent with the presence of
large scale ordered magnetic fields in the afterglows.
 But  there are also  exceptions \citealt{rol03,grein03}, when the position 
angle does show some variations.
(A  model of \cite{ros02} of structured jets also predicts 
constant position angle, but since no large scale magnetic field is assumed the
polarization features are still related to the jet break times).

{\bf Structured jet}. As discussed in Section
\ref{after} our model gives a theoretical foundation for  the  ``structured jet''
profile of the external shock. We would like to stress once again that there is no proper
jet in our model, but  a non-spherical outflow, so that a break in the light curve is seen
when $\Gamma \theta_{ob} \sim 1$.

{\bf XRF flashes}.
Another testable prediction of the model is that we should observe
much more numerous X-ray flashes (XRFs), which  may be coming
``from the sides'' of the expanding shell where the
 flow is less energetic and the
Lorentz boosting is weaker. In addition, the total {\it  bolometric } energy 
inferred for XRFs  (presumably from observations of early afterglows before
radiative losses become important) should be comparable to the total bolometric energy
of $\gamma$-ray bursts.
(The total X-ray fluences may be quite different since this would include 
a correction for the  unknown   radiative efficiency as a function of  the polar angle).
 At present, a single  X-ray flash with   
 measured  redshift suggests had an energy similar to those associated with GRBs  
\citep{sod03}. 
Generally, the  distributions of 
parameters of XRFs should continuously match those of GRBs. 

{\bf  Short-long  dichotomy}.
Our model  can also  explain a short versus long  dichotomy in  GRBs. 
We associate  short bursts with the instabilities of polar currents while 
long with instabilities of shell currents. 
Our  analytical solutions  for
 electromagnetic fields diverge on the axis; in reality  the axial current density will have finite
angular 
width.
If an observer is located within this small angle he would see a burst that would be 
 shorter since the Lorentz factors  of the outflow are higher  closer to  the axis. 

{\bf Hard-soft evolution and $E_{peak}- \sqrt{L}$ correlation}.
Magnetically-dominated flows may also explain the observed
 correlations  between various GRB parameters, though the  lack of a testable  particle acceleration
model makes the arguments below only suggestive. 
For example the trend of  GRB spectra to evolve from hard to soft during  a pulse
is  explained as a   synchrotron radiation in an expanding  flow with  magnetic field 
decreasing with radius $B \propto \sqrt{L} /r$ (later in a pulse emission is produced
further out where magnetic field is weaker, so that the peak energy will be lower;
 this is similar to "radius-to-frequency mapping" in radio
pulsars and AGNs). A correlation between peak energy and 
total luminosity, $E_{peak} \sim \sqrt{L}$ \citep{lr02}
 follows from the assumption of a  fixed typical
emission radii and fixed minimum particle energy. 
Both of these correlations are {\it independent} of the bulk Lorentz factor,  but depend on the 
lower energy cut-off  in the 
spectrum of accelerated particles.

{\bf No reverse shock}.
Another consequence of the model is that  
no   emission  from the 
reverse shocks is  expected in electromagnetic models. 
When plasma magnetization becomes of the order of unity 
relativistic MHD shock conditions are
modified: shocks becomes less dissipative; in the 
force-free limit shocks cannot
 exist at all.
Early optical flashes, which are
conventionally associated with reverse shock \citep[\eg][]{mes97,zhang03,kumar03},
should have a different origin. A possible explanation is the formation
of  a radiative precursor that pre-accelerates the medium,
loads it with e+- pairs and produces soft emission \citep{belo02,tm00}

\section{Discussion}

In this paper 
we have explored  the 
 ``electromagnetic hypothesis'' for ultra-relativistic
outflows, namely that they are essentially electromagnetic phenomena
which are driven by the energy released by spinning black holes or neutron
stars and that this electromagnetic behavior continues into the source
region even when the flows become non-relativistic.
In this  model the energy to power the GRBs comes eventually from
 the rotational energy of the progenitor. It is first converted into 
magnetic energy by the dynamo  action of the  unipolar inductor, 
propagated in the form of Poynting flux-dominated flow
and then dissipated at large distances from the sources. 
We have taken an extreme view that the flows become essentially  force-free
and explored the consequences of this assumption.
Real GRB outflows must contain  both baryonic matter
and approach near-equipartition in the emission regions. 

This  model envisions GRB outflows as an electric circuit
in which  the central  source acts as a power-supply generating 
 a very simple current
flow -- along the axis, the surface of the  shell
 and  the equator.  In practice, it is the detailed electrodynamics in the
vicinity of the outgoing light surface that fixes the poloidal
magnetic field and electrical current distributions. We can therefore
change these (still, of course, maintaining the force-free condition)
and solve for a new evolution of the magnetic shell and blast wave. A broader
distribution of currents will generally produce a less
pronounced expansion along the axis and change somewhat the statistics of
the observed afterglows. Alternatively, the central source can 
collimate the flows to even more narrow (cylindrical) expansion. 
In the solution above, we ignored the poloidal
magnetic field and, consequently, the angular momentum. These can
be reinstated perturbatively into the solution. Their influence wanes
with increasing radius.
Other ways to obtain different solutions
include changing the temporal variation of the source from the one satisfying
 self-similar expansion to a more general variation.
 As the individual parts of the
blast wave expand essentially independently, when ultra-relativistic,
there are no new issues of principle to address in solving these problems.
 When the blast wave becomes non-relativistic
 the interior gas will be roughly isobaric and the shell
will become more spherical with time.

 The most striking implications of the electromagnetic
 hypothesis
are that particle acceleration in the sources is due to
electromagnetic turbulence
rather than shocks and that the outflows are cold,
electromagnetically dominated flows, with very few baryons
at least until they become strongly dissipative. 
The major drawback of the current model is the lack of the
 detail model of energy dissipation
and  particle
acceleration in relativistic magnetically-dominated medium. 
This is a difficult problem. The problem with magnetic dissipation
 may be exemplified by the better studied 
way of dissipation of magnetic energy, through reconnection.
Reconnection physics is highly uncertain: it depends crucially on the
kinetic and geometric 
 properties of the plasma, which is very hard to test
observationally. 
Often,
various instabilities (based on 
inertial or cyclotron effects, 
 on ions or electrons) often seem to account  equally well
(with astrophysical accuracy) for the observed phenomena.
 This is an important uncertainty
since the principle  scale of the reconnection region (related, for example,
to  electron or  ion skin depth, Larmor radius or magnetic Debye radius)
is crucial in determining the rate of reconnection  \citep[\eg][]{birk00}. 
 This situation may be contrasted with the shock acceleration schemes,
where  a qualitatively correct result for the spectrum of accelerated particles
can be obtained from simple {\it macroscopic } considerations.

An important implication of the electromagnetic model is that supernova
explosions may be magnetically driven as well \citep{lw71,bk71}. 
A number of research groups are investigating this possibility, mostly through
numerical modelling \citep{myks03,prog03}.

Another uncertainty  in the model is how  baryon-free the flow can become.
At early   stages
 large  fluxes of low energy neutrinos  are expected. They will drive
the material up the field lines
 into the magnetosphere, while gravity will  try to 
pull the matter back into the disk.
We require that the resulting outflow be extremely clear of baryons
($\sim 10^{-9}$ in energy flux);
numerical simulations are required to see if this  is possible
\citep[\eg][]{Thompson01}.

If  
GRBs are electromagnetically-driven they
  should not be associated with strong, high energy
neutrino sources.
A gravitational wave signal is expected in some, though not all, source
models and would be strongly diagnostic if ever detected.

Electromagnetic models are also consistent with being  possible sites 
of UHECR acceleration \citep[]{vie95,wax95,mu95,wda03}
 if the corresponding energy balance of UHECR and
GRBs agree \citep{wax02,vdmg03}
and if assuming that the GZK cut-off is indeed observed
in the spectrum of UHECRs \citep{bw02}.
Primarily,  the total available electric  potential in GRBs, $\sim 10^{22}$ V,
is sufficient for acceleration of UHECRs. 
UHECR energies are mostly likely radiation-limited and
 not acceleration-limited, but, at late stages  of GRB expansion,
synchrotron losses become sufficiently small.

In addition to Fermi acceleration at the  forward  shock,
electromagnetically-dominated flows provide several
 possible acceleration schemes in their own
right. (Note that, in order to explain UHECRs, shock acceleration models
have to assume that turbulent EMF reaches approximate equipartition behind
relativistic shocks \citep[\eg][]{dh01};
 this requirement makes these models akin to electromagnetic models.)
There is a large {\it inductive}
 electric field $E_\theta$ which creates  a large
potential difference (\ref{VI}) between the polar and equatorial flux surfaces.
Thus, in order for a particle to gain the total energy it should traverse from 
pole to equator (or a considerable part of this path). This should be done
on expansion time scale in order to 
avoid adiabatic loss; this is possible at the
end of the relativistic stage.
(Note that,  at late stages, most of the energy is in the forward shock the 
total pole-to-equator electric potential 
drop remains approximately constant as long as the expansion is relativistic
and starts to decrease only at the non-relativistic stage.) 
This crossing of the potential surfaces may be done either kinetically, \eg due to a drift,
or resistively, \eg at O-type point \citep[\eg][]{trub02,vas80}.

In this paper we have discussed the basic principles that may be important
in electromagnetically-dominated outflows. At the present stage the model is 
definitely less developed than the hydrodynamics
 fireball model and a lot needs to be done
to vindicate it,  but the  possible 
fundamental advantages of electromagnetic models 
outlined here  make it, in our opinion,  a very promising approach. 
From a more theoretical
perspective, there is much to be learned about the properties of
force-free electromagnetic fields, especially their dissipation. The
possible relationship of the GRB fluctuation power spectrum to an
underlying turbulence spectrum is especially tantalizing. Undoubtedly,
numerical simulations will be crucial as the problem is essentially
three-dimensional. Force-free electromagnetism is easier to study
than relativistic MHD and may well be a very good approximation
in many of these sources.

If, as we have argued,
 magnetic fields do play an important dynamical role
in GRBs, there is the    between supersonic MHD outflows 
$\sigma \leq \Gamma^2$ \citep[\eg][]{vk03a,fo03}
 and the  subsonic force-free model, $\sigma \geq \Gamma^2$, considered in this paper. 
As we have discussed 
these two alternatives have very different  properties coming from different 
causal structures of the flow. Observationally, it will be hard to distinguish between the 
two since even in the strongly magnetized case we expect that in the emission 
region the fields approach equipartition (since then plasma emits
most efficiently). 

Finally, we give a concise list of  the observational properties of the GRBs which are
explained or foreseen in the electromagnetic model.  GRBs properties should most depend
on the observer angle with respect to the jet axis.
Electromagnetic model  produces ``structured jet''
with  energy $E_\Omega\propto \sin^{-2}\theta$; there is no problem with ``orphan afterglow'' since
GRBs are produced over large solid angle; X-ray flashes are interpreted
as GRBs seen ``from the side'', but their total energetics should be comparable
to proper GRBs; overall energy of GRBs is expected to have a small scatter if  it is related
to a critically rotating relativistic object;
  the model  can  qualitatively
reproduce hard-to-soft  spectral evolution as a synchrotron  emission in
ever decreasing magnetic field $B \propto \sqrt{L} /r $ ($L$ is luminosity, $r$ is emission
radius),
 akin  to "radius-to-frequency mapping" in radio
pulsars and AGNs; similarly, the    correlation   $E_{peak} \sim \sqrt{L}$
is also a natural consequence. Finally, high polarization of prompt
emission  may also be  produced:
it should  correlate with the spectral index; mixing 
between circumstellar material and ejecta will
result in  the same  position angles  of the prompt emission
 and afterglow which should  be
constant over time; fractional
polarization should
be  mostly independent of the ``jet break'' time,
but may show variations due to turbulent mixing.

\section{Application to other sources}

Many of the principles described in this work may be 
applied to other astrophysical 
sources like pulsars, (micro)quasars and  AGNs \citep{bla02}.
Despite the apparent differences 
(inner boundary condition for neutron stars and black holes, 
time (non-)stationarity, optical thickness)
it is  plausible that all these sources 
 produce ultra-relativistic magnetically-dominated   outflows
with low baryon density.
Energy, transported primarily by magnetic fields, 
is dissipated far away from the source due to  the development of MHD
current instabilities. Particles are  accelerated in localized current
sheets by DC electric fields and/or electromagnetic turbulence
producing bright knots (in AGNs) and a variety of bright
spots in pulsar jet,  best observed in the Crab. {\it In situ}
acceleration is required in both AGNs and pulsars since the
life-time of emitting particles is shorter than the  dynamical
time. Qualitatively, magnetically-dominated jets are expected to be more
stable than their fluid counterparts \citep{bes93}.
In the case of AGN,  recent observation of   TeV emission from blazers with
a very short time scale variability \citep{ken02}
favors magnetically-dominated jets \citep[\eg][]{Bolog} since the  two leading models of jet 
composition - leptonic and ionic -
both have difficulties in explaining the fast variability (problem for
ion-dominated jets, since cyclotron times are very long for ions)
 and total energy content
(since leptons suffer strong radiative losses at the source due to radiation
drag).
The flow evolution in all these systems may proceed in a similar wave.
Jets are launched in the vicinity of a central object (a  BH-disk system or 
a neutron star) where the flow is subsonic and in a dynamical balance.
As the flow propagates out jets become laterally imbalanced, just as with GRBs,
 the lateral dynamics is suppressed by relativistic kinematics, so that
the jet cannot adjust to the change of equilibrium on  dynamical
time scale.
After  a jet has propagated $\Gamma^2 $ dynamical times, the
lateral dynamics starts to become important.  A particularly strong evolution  
will  take
place in the core of the jet, where  magnetic hoop stresses may lead
to the collapse of the flow toward the axis. 
For example, AGN jets may originate close to the central black hole with
a Schwarzschild radius $r_s \sim 10^{13}$ cm with
a typical Lorentz factor   $\Gamma \sim 10-30$.
Then the dynamical evolution and the onset of emission occurs on a scale
$ \sim \Gamma^2 r_s  \sim 10^{15}-10^{16}$ cm.
In pulsars acceleration occurs on a scale of light cylinder,
$r_s \sim 10^{10}$ cm,with a Lorentz factor $ \Gamma\sim 100-1000$,
then the dynamical instabilities occur at $\sim 10^{14} - 10^{16}$ cm.

\begin{acknowledgements} 
Over the past three and a half years that this work has been in progress we 
benefited from numerous  discussions with our colleagues. In particular, RB
thanks Jerry Ostriker for early discussions of electromagnetic
models of gamma ray bursts and ML thanks Chris Thompson for discussion and comments on the
manuscript.
This research has been supported by NSERC grant  RGPIN 238487-01 and
  NASA grant 5-2837 and 5-12032.
\end{acknowledgements}

\appendix

\section{Dynamics of magnetized pair-loaded  flows (minifireballs) }
\label{warm}

In this appendix we consider the far field dynamics of a spherically symmetric 
magnetized wind. 
The  dynamics of such warm magnetized wind
  is    controlled by three parameters:
energy $L$, 
 mass flux $\dot{M} $  and the electro-motive force ${\cal E}$ produced by the central
source.
 The  central source loses energy in two 
forms: mechanical $L_M$ and  electromagnetic $L_{EM}$ luminosities.
\be
L = L_M + L_{EM}
\ee
We wish to understand how the parameters of a fully relativistic
flow   (velocity $\beta$, 
pressure $p$,
magnetization $\sigma$) evolve for an arbitrary ratio of both $L_{EM} / L_M$ 
and $p/\rho$ ($\rho$ is the rest-frame mass density).

The 
asymptotic evolution of the flow
is determined by  conserved quantities which 
may be chosen as the  total luminosity $L$,  the mass flux $\dot{M}$
and the  EMF  ${\cal E}$.
 Thus, the central source
works both as thruster and as a dynamo.
We follow the flow evolution starting from some small inner radius 
where we assume that a flow with given $L, \, \dot{M} $ and ${\cal E}$
(in a form of toroidal magnetic field)
is generated. 
Thus, 
we avoid the  important question of how the flow 
is launched.
Another important assumption that we make is that  the expansion is radial.

The formal treatment of the problem
 starts with the 
set of RMHD  equations which 
can be written in terms of conservation laws (\ref{x3}) in which we assume that
fluid is polytropic with adiabatic  index $\Gamma_{a}$: $w =\rho+ {\Gamma_{a} \over \Gamma_{a} -1} p$.
(We assume that the adiabatic index $\Gamma_{a}$
 is constant for algebraic simplicity.)

Writing  out eqns. (\ref{x3}) in coordinate form and 
assuming a stationary, radial,
 spherically symmetric outflow with toroidal magnetic field,
we find
\ba 
&&
{1\over r^2} \partial_r \left[ r^2 ( w+b^2) \beta \Gamma^2 \right] =0
\label{x4} \\ &&
{1\over r^2} \partial_r 
\left[r^2\left((w+b^2) \beta^2 \Gamma^2 + (p+b^2/2)\right) \right] - 
{2 p \over r}=0
\label{x5} \\ &&
{1\over r} \partial_r  \left[r b \beta \Gamma \right] =0
\label{x51} \\ &&
{1\over r^2 } \partial_r  \left[r^2 \rho  \beta \Gamma \right] =0
\label{x6}
\ea
The above relations can be simplified if by defining
\be 
L = 4 \pi r^2 \beta \,{\Gamma }^2\,\left( b^2 + {\Gamma_{a} \over \Gamma_{a} -1} 
 \,p + \rho  \right) , \,\,\,
\dot{M} =  4 \pi r^2 \beta \,\Gamma \,\rho  , \,\,\,
{\cal E}= 2 \sqrt{\pi} r\, \beta \,\Gamma \, b
\ee
where
 ${\cal E}$ is the electromotive force.

It is convenient to introduce two other parameters: the 
magnetization parameter $\sigma$ as the ratio of the rest-frame
magnetic and particle energy-density
 and a fast magnetosonic  wave  phase velocity
$\beta_f$ 
\be 
\sigma={ b^2 \over w} = {{\cal E}^2  \over L \beta - {\cal E}^2} , \,\,\,
\beta_f^2=  
{\sigma \over 1+ \sigma} + {\Gamma_{a} p \over (1+ \sigma) w}=
\left( \Gamma_{a}-1 \right) \left( 1 - { \Gamma \dot{M} \over L}  \right)+
(2-\Gamma_{a}) { {\cal E}^2  \over L \beta}
\label{pp}
\ee
Using the  three conserved quantities $L, \, \dot{M}$ and $ {\cal E}$
the evolution equation becomes
\be
{ 1 \over 2  \beta^2 \Gamma  }
\partial_ {r }  \Gamma= { (\Gamma_{a}-1)
\left( \beta L - \beta \Gamma \dot{M}  - {\cal E}^2 \right) \over
 r\, \left(\beta L  ( \beta^2 +1 -\Gamma_{a}) + (\Gamma_{a}-1) \beta \Gamma  \dot{M}  
-(2-\Gamma_{a}) {\cal E}^2 \right) }
\label{gt}
\ee
Eliminating 
${\cal E}$ in favor of $\Gamma_f$ we 
get a particularly transparent form for the evolution of Lorentz
factor
\be
{  \left( {\Gamma }^2 - {{{\Gamma }_f}}^2 \right) \over \beta^2  \Gamma^3}
\partial_r \Gamma=
{ 2  p \Gamma_{a} \over ( w - \Gamma_{a} p) r } 
\label{dg}
\ee
Equation (\ref{dg}) is   nozzle-type   flow
(e.g. \citep{LLIV}); in astrophysical context it is best known
for Parker's solutions of the solar wind \citep{par60}. 
The lhs of  eq. (\ref{dg}) contains a familiar 
critical  point at the sonic transition $\Gamma= \Gamma_f$.
The positively defined first term on the rhs  describes the  evolution
of Lorentz factors due to pressure effects. 
In the case of purely radial expansion the magnetic gradient
forces are exactly balanced by the hoop stresses, so that
magnetic field does not contribute to acceleration.
From Eq. (\ref{dg}) it follows that 
super-fast-magnetosonic flows accelerate while
sub-fast-magnetosonic flows
 decelerate. 
It also it follows that 
terminal Lorentz factor of the flow  is determined by  the condition
$\partial_r \Gamma =0$ which implies  that either 
$p=0$ or $\beta =0$.
Condition $\beta_{\infty} =0$ can be reached only for subsonic flows with ${\cal E}=0$;
(a steady state perfect MHD radial 
magnetized flow with ${\cal E} \neq 0$ cannot slowdown to a halt at 
infinity).
Neglecting the  $\beta_{\infty} =0$ solution, the terminal
velocity of magnetized flow is determined only by the condition $p=0$:
\be
L=\Gamma_{\infty}  \dot{M} + { {\cal E}^2 \over \beta _{\infty}}
\label{L1}
\ee

For each set of parameters ($L$, $ \dot{M}$ and ${\cal E}$) there
are  generally two solutions for the terminal four-velocity corresponding to the
supersonic and subsonic solutions.
Th only exception is the unmagnetized branch, $ {\cal E}_ 0$. 
In the absence of magnetization
 the terminal (supersonic) Lorentz
factor is uniquely  determined by $ \Gamma_{\infty} = L/ \dot{M}$.
 In GRB applications we are mostly interested in the supersonic solutions
(the subsonic solutions are important for pulsar winds, \cite[\eg][]{KC84}).

For supersonic solutions, the unmagnetized branch with $ {\cal E}_0$
is qualitatively different from magnetized branch.
In the absence of magnetization,
 the terminal (supersonic) Lorentz
factor is determined by $ \Gamma_{\infty} = L/ \dot{M}$ 
(in the subsonic regime 
$\beta \sim r^{-2}$  and $p, \, \rho =$ const as with the  breeze solution
of the solar wind).
For finite 
  magnetization  there are two solution: a larger one corresponding
to super-fast-magnetosonic flow and  a smaller one corresponding to a
sub-fast-magnetosonic flow.
In particular,
for  large  $L / \dot{M} \gg 1 $ the supersonic terminal proper velocity
reaches 
\be
\Gamma_\infty \beta_\infty \sim  { L - {\cal E}^2 \over  \dot{M} }
\ee
Thus, for a given ratio $ L  / \dot{M}$ the terminal Lorentz
factor decreases with increasing $ {\cal E}$ - magnetic field 
provides an effective inertial  loading of the flow.

For non-zero magnetization the terminal 
velocity cannot be determined uniquely from a given $\dot{M}$ and ${\cal E}$ - 
this problem is the results of our neglect of  details of  
acceleration \citep{Michel69,gj70}.
Yet, there is a trick, initially suggested by \citep{Michel69}, which
allows one to select from a family of solutions 
a "minimum torque" solution (in the  terminology of \cite{Michel69}). 
A more careful examination \citep{gj70,KFO} 
proves that indeed,
for strongly magnetized flows, when the pressure contribution
to acceleration is not important, the minimum torque solution is the
correct one.

In case of non-zero magnetization, 
for a given ratio ${\cal E}$  solutions exist only for
$ L  / \dot{M}$ larger than some critical value
 $L / \dot{M}=(1 - ( {\cal E}^2/L)^{2/3})^{-3/2}$,
corresponding to  $\beta_\infty =  ( {\cal E}^2/L)^{1/3}$
(alternatively, $\beta_\infty = 1/\sqrt{ 1+ (\dot{M}/ {\cal E}^2)^{2/3}}$,
reached at ${\cal E}^2   \left( 1+ (\dot{M}/ {\cal E}^2)^{2/3} \right)^{3/2}$).
This is the Michel solution, which we can reformulate now stating that
for fixed $\dot{M}$  and ${\cal E}$ the  minimum energy  loss is reached
at $\beta_{min}$.
It is expected that, as the flow become stronger magnetized, its terminal evolution
asymptotes to the Michel solution.

The above relations may be expressed in term of  terminal
 magnetization parameter $\sigma_\infty$  and the terminal velocity
of the supersonic  wind
\be  
\Gamma_\infty = { L \over \dot{M}_0 (1+ \sigma_\infty)},
, \,\,\,
\sigma_\infty = { {\cal E}^2 \over \Gamma_\infty \beta_\infty \dot{M}_0}.
\label{LK}
\ee
Assumption  $\beta_\infty = \beta_{min}$, 
then gives
\be
\left( 1 + ( \beta_\infty \Gamma_\infty \sigma_\infty)^{2/3} \right)^{3/2} = 
{ \Gamma_\infty \sigma_\infty \over \beta_\infty^2}
\label{W}
\ee
which in the 
 strongly relativistic limit  gives the Michel solution
\be
 \Gamma_\infty = \sqrt{\sigma_\infty}
\ee

We  can also   relate
the terminal magnetization $\sigma_\infty$ to the magnetization at the
source -  more specifically to magnetization at the sonic point
$\sigma_f$.
Using (\ref{LK})
 we find that at 
any point in the flow the magnetization parameter is  given by
\be
\sigma =
 { \beta_\infty  \sigma_\infty \over 
\beta (1+\sigma_\infty) - \beta_\infty \sigma_\infty}
\label{O}
\ee
To relate the magnetization at infinity to the 
magnetization at the sonic point we need solve for the
velocity at the sonic point $\beta_f$.
Using (\ref{pp}) and (\ref{O}) we find an equation for $\beta_f$
\be
\beta_f^2 = (\Gamma_{a}-1 ) \left(1- {\Gamma_f \over \Gamma_\infty (1+\sigma_\infty) } \right)
 + (2-\Gamma_{a}) { \sigma_\infty \over 1+\sigma_\infty} 
{\beta_\infty\over \beta_f} 
\ee
In the 
 absence of mass flux, $ \dot{M}=0$, $\Gamma_\infty = \infty$,
\be
\beta_f = 
\left\{
\begin{array}{ll}
\sqrt{\Gamma_{a}-1}  + { (2-\Gamma_{a})  \sigma_\infty \over 2  (\Gamma_{a}-1 ) }&
\mbox{ if $ \sigma_\infty \ll 1$} \\
1-{ 2-\Gamma_{a} \over (4- \Gamma_{a}) \sigma_\infty}
& \mbox{ if $ \sigma_\infty \gg  1$}
\end{array} \right.
\ee
Thus, for strongly magnetized flow, $\sigma_\infty \gg 1$,
 $\Gamma_f = \sqrt{2 \sigma_\infty}$ for $\Gamma_{a}=4/3$.
Using above expressions for the sonic velocity we find a relationship
between the magnetization at the sonic point and 
  infinity (for supersonic flows):
\be
\sigma_f =  
\left\{
\begin{array}{ll}
{\sigma_\infty \over \Gamma_{a}-1 } &
\mbox{ if  $ \sigma_\infty \ll 1$} \\
{ (4- \Gamma_{a}) \sigma_\infty \over 2} & \mbox{ if $ \sigma_\infty \gg  1$}
\end{array} \right.
\ee
Thus, we always have $\sigma_\infty < \sigma_f$, but they
remain of the same order of magnitude:
the magnetization of the flow changes only slightly as the
flow propagates away from the launching point to infinity. 
The reason for constant $\sigma$ in the supersonic regime is that
both in the case $p \gg \rho $ (linear acceleration stage) and $p \ll \rho $
 both the plasma and the magnetic field
energy densities in the  flow change with the same radial dependence ($\sim r^{-4}$
and $\sim r^{-2}$ correspondingly).

\begin{figure}
\plotone{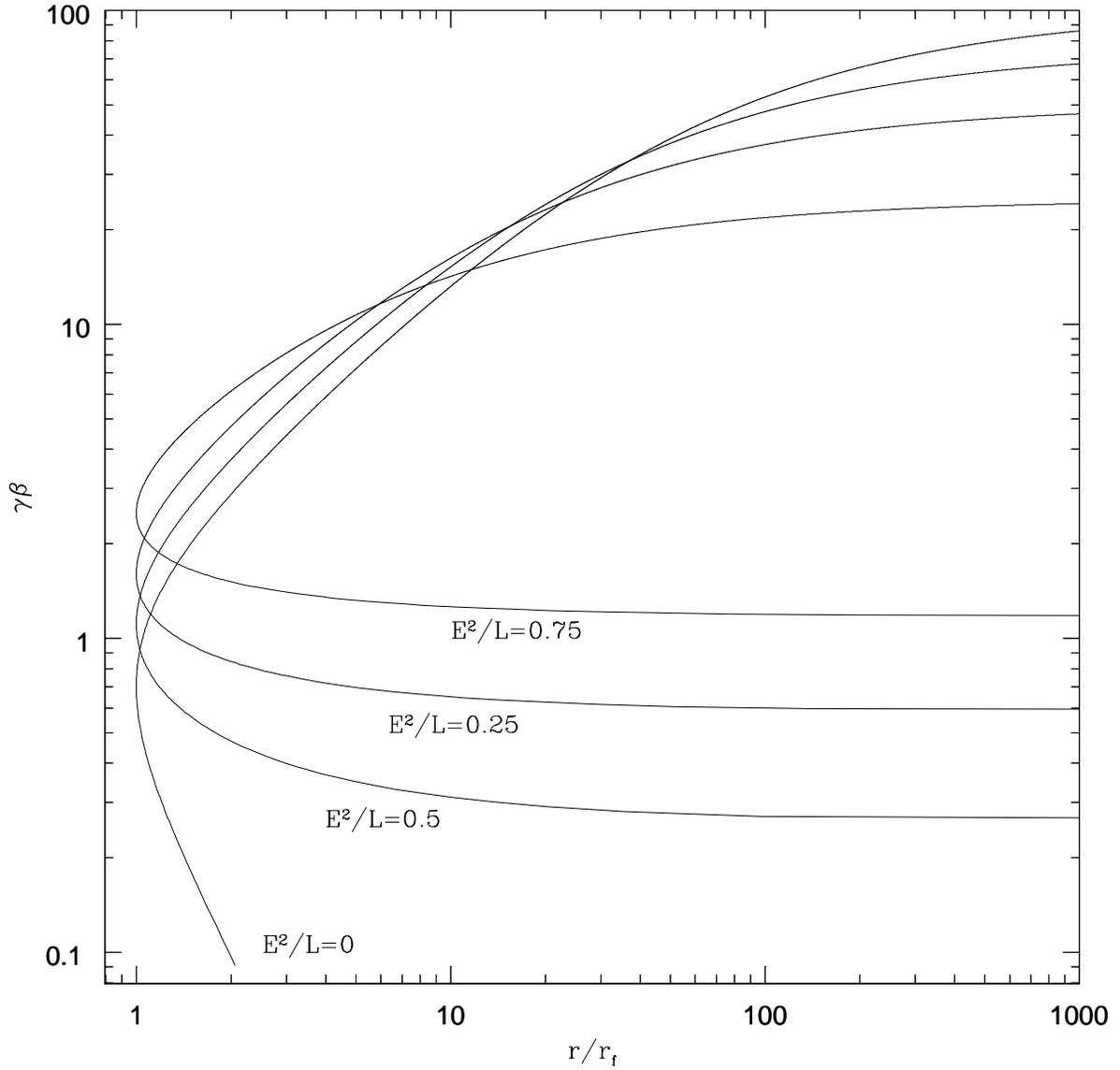}
\caption{  Evolution of strongly magnetized optically thick  flows.
Four-velocities 
 of flows
 are given 
as functions of  $r/r_f$ for $L/\dot{M}=100$ and 
 different values of the parameter ${ {\cal E}^2 \over L}$.
 Flows start at $ r= r_f$ with $\beta = \beta_f$; supersonic
flows first  accelerate as $ \beta \gamma \sim r$, reaching a 
terminal value given by the larger root of  eq. (\ref{L}), 
while subsonic decelerate initially as $\beta \sim r^{-2}$  
reaching asymptotic value given by the smaller root of eq. (\ref{L}).
}
\label{rofu1}
\end{figure}

\begin{figure}
\plotone{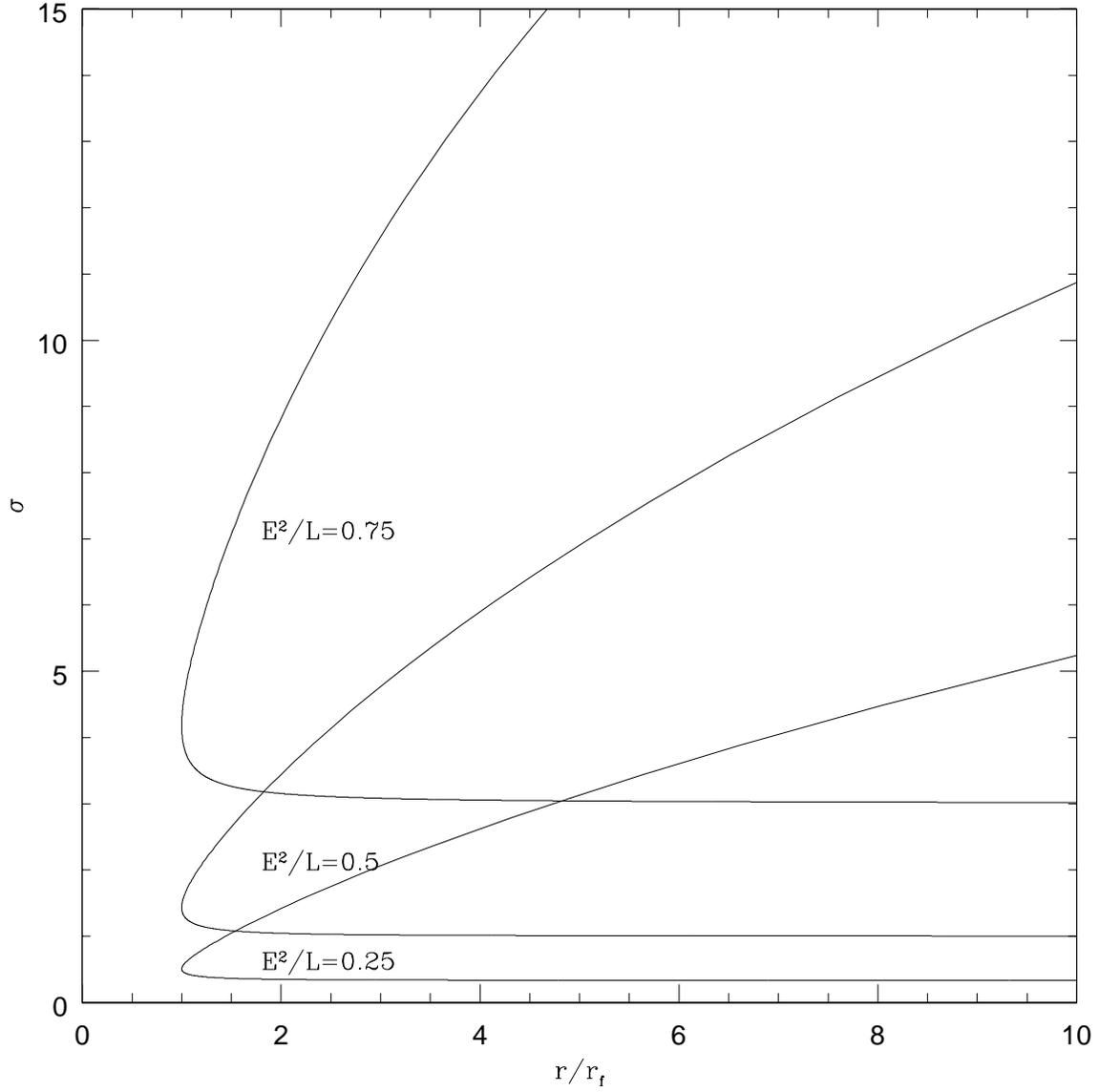}
\caption{
Magnetization parameter $\sigma$.
For supersonic flows (lower branch)
 the magnetization remains constant, reaching
$\sigma_\infty = \left( 1 - { {\cal E}^2 \over L} \right)^{-1} $
as   $r \rightarrow \infty $.
Subsonic flows become strongly magnetized as they expand (upper branch); the 
  magnetization parameter increases  $\sigma \sim r^{2/3}$.
}
\label{sigma}
\end{figure}

Generally 
there are two branches of  solutions:
supersonic   and subsonic.
In the  limit  $\Gamma_\infty \rightarrow \infty$ (this is equivalent to neglecting the mass loss rate of the central
source in comparison with
the energy loss rate),
  the nozzle equation
(\ref{gt})
simplifies
\be
{\Gamma^2 \over \beta} \partial _{\ln r} =
{ (\Gamma_{a}-1) \left( \beta - {\cal E}^2/L \right)
\over
\beta  \left( \beta^2 - \Gamma_{a} +1 \right) - (2-\Gamma_{a}){\cal E}^2/L  }
\label{GE}
\ee
 which can be integrated to give
\be
r\propto
  \left(  \beta  \Gamma \right)^{(2-\Gamma_{a})/(2(\Gamma_{a}-1))} 
\left(  \beta  - {\cal E}^2/L  \right)^{-1/(2(\Gamma_{a}-1))}
=
 { \beta  \Gamma \over (\beta - {\cal E}^2/L )^{3/2}}
\, \mbox{for $\Gamma_{a}=4/3$}.
\ee
Thus, 
the supersonic 
flow is accelerated by pressure effects as long as $p\gg \rho$, reaching
a  coasting phase  with $\Gamma \sim \Gamma_\infty$ when $p \leq \rho$.
while 
 subsonic flows reach a minimum velocity  given by 
\be
\beta = { \sigma_\infty \over 1+ \sigma_\infty }.
\label{sw}
\ee
For arbitrary flow parameters the evolution equations are integrated numerically
(Fig. \ref{rofu1}).
Given the evolution of the flow and the relation for local $\sigma$ 
(\ref{O}) we can find the 
evolution of the magnetization parameter (Fig. \ref{sigma}).

\section{Shocks driven by magnetic explosion}
\label{shocked}

As the magnetic shell expands into an ambient gas
a shock  forms ahead of it.
The shock would quickly go into self-similar regime described by the 
Blandford-McKee solution (\cite{blm76}).
In this appendix we relate the  two self-similar
 solutions - inside the magnetic shell and unmagnetized
Blandford-McKee solution for the flow between the forward shock and  the CD.
It is straightforward to generalize these results to include external
magnetic field using the solutions for the self-similar structure of
magnetized blast waves derived by Lyutikov (2001).
We also neglect the lateral dynamics of the shocked material and 
assume that  between
the blast wave and the CD (which are separated by $\sim R/\Gamma^2$)
  the plasma   moves radially. This is justified as long as the Lorentz
factor changes on angular scales larger that $ 1/\Gamma^2$.

For a self-similar blast wave the contact discontinuity is located at a fixed
$\chi$ (The self-similar variable $\chi$ in this section
is chosen in such a  way that $\chi =1$ on the forward shock).
Differentiating Eq. (\ref{chi}) with respect to time and using
$ r'(t) = 1-1/(2 \gamma^2) = 1-1/(y g)$ we find
\be
\chi_{CD} =
\frac{{\left( 1 + 2\,\left( 1 + m \right) \,y \right) }^2}
  {g\,y\,\left( 1 + 2\,{\left( 1 + m \right) }^2\,y \right) }
\approx {2 \over g}
\ee
(c.f. \cite{blm76} Eq. (39)).

We have to balance the pressures on both sides of the CD.
The momentum flux of the shocked material in the frame of the CD is
\be
\left( 4 \gamma^2 -1 \right)  p \beta
\approx  4/3 \Gamma^4 g(\chi_{CD}) f(\chi_{CD}) w_{\rm ext}
\label{WQW}
\ee
where $w_{\rm ext}=n_{\rm ext} m_i $ is the enthalpy ahead of the shock.
and $g (\chi_{CD})$ and $f(\chi_{CD}) $
and the \cite{blm76} values on the CD.

Using Eq. (\ref{WQW}) we find
\be
w_{\rm 1} = \kappa w_{\rm ext}, 
\hskip .3 truein \kappa = {8 \over 3} g(\chi_{CD}) f(\chi_{CD}) 
\label{H}
\ee
for the enthalpy on the CD.
Relation (\ref{H}) determines the boundary condition on the CD
and, thus, the normalization for the  solution inside the CD.

\begin{figure}[h]
\includegraphics[width=0.9\linewidth]{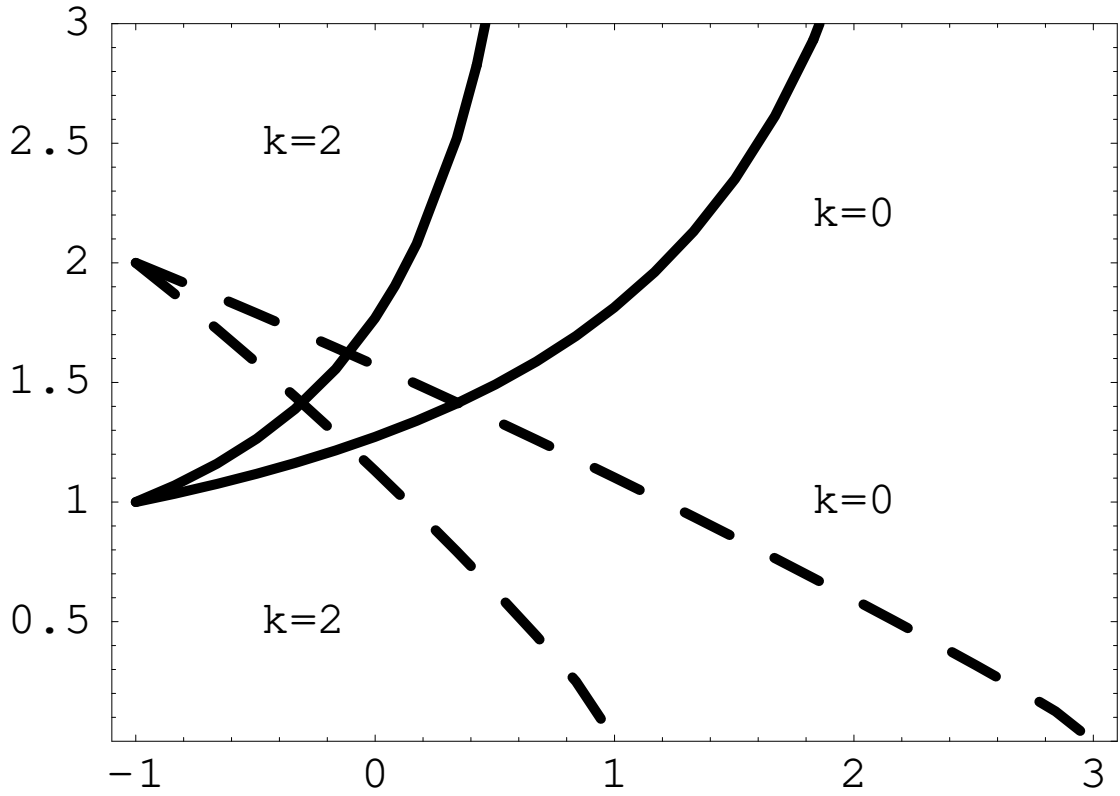}
\caption{Lorentz factor (dashed lines)
 and location of the contact discontinuity (solid lines) of the
B\&M solutions as a function
of $m$ for two choices of parameters: $k=0$ and $k=2$. At the limiting value
$m_{max}=  k+1$ (point explosion cases) the contact discontinuity moves to
$\chi = \infty$.
At the extreme case, for $m=-1$,
 the contact discontinuity merges with the forward shock at $\chi=1$.
}
\label{cd}
\end{figure}

The relations for the $g(\chi_{CD})$ and $f(\chi_{CD})$ are easily found from
\cite{blm76}:
\ba
&&
g(\chi_{CD}) =
   {\left( \frac{a - 2\,b}{a + 2\,b} \right) }^{\frac{16 + 28\,m - 3\,k\,m + m^2}{4\,b}}\,
  {\left( \frac{ 2( 12 - 3\,k - m)}{3(3 - k - m)} \right) }^{\frac{2 + m}{2}}
\nn &&
f(\chi_{CD}) =
{\left( \frac{a + 2\,b}{a - 2\,b} \right) }^
   {\frac{64 + 3\,k^2 + 2\,k\,\left( -16 + m \right)  - 24\,m - m^2}{4\,b}}\,
  {\left( \frac{2(12 - 3\,k - m)}{3(3 - k - m)} \right) }^{\frac{-4 + k + m}{2}}
\ea
where
\be
a = 32 - 9\,k - 5\,m, \hskip .2 truein
b^2 = 4\,\left( 1 + m \right)  +
    \frac{{\left( 12 - 3\,k + m \right) }^2}{4}
\ee 
where $k$ is the power law of the density variation $\rho \sim r^{-k}$ 
(see Fig. \ref{cd}).
These relations allow us to sew the two solutions - inside the CD and
 between the CD and forward shock.

\section{Self-similar expansion of magnetic shell: Relativistic MHD
approach}
\label{BMR}

In this appendix we illustrate the similarities of the RMHD and 
RFF approach re-deriving the  self-structure  of the magnetic shell
from RMHD equations. The self-similar approach is useful when the 
thickness of the magnetic shell is much smaller than it radius. 
For a power-law dependence of Lorentz factor of the shell on time,
$\Gamma \propto t^{-m/2}$ the 
 the fields are concentrated near the CD for  $1<m<3$.
In this case    we can
neglect the divergence of the characteristics and 
apply the hydrodynamic  Blandford-McKee approach.

Writing  out Eqns (\ref{x3}) in coordinate form and
assuming an azimuthally  symmetric outflow with toroidal magnetic field,
the conservation of energy and momentum, induction equation 
 and mass conservation  give
\ba 
&&
\partial_t \left[ ( w+ b^2) \gamma^2 -(p+b^2/2)\right]+
{1\over r^2} \partial_r \left[ r^2 ( w+b^2) \beta \gamma^2 \right] 
+{1\over r \sin \theta} \partial_{\theta} 
\left[  \sin \theta  ( w+ b^2) \gamma^2 \lambda \right]
  =0
\label{x41} \\ &&
\partial_t \left[ ( w+ b^2) \gamma^2 \beta \right]+
{1\over r^2} \partial_r 
\left[r^2\left((w+b^2) \beta^2 \gamma^2 + (p+b^2/2)\right) \right] -
{ 2  p + ( w+ b^2) \gamma^2 \lambda^2  \over r}
\nn  &&
+ {1\over r \sin \theta} \partial_{\theta} 
\left[  \sin \theta  ( w+ b^2) \gamma^2 \lambda \beta  \right]
 =0
\label{x51-1} \\ &&
\partial_t \left[ ( w+ b^2) \gamma^2 \lambda  \right]
+ {1\over r^2} \partial_r \left[ r^2 ( w+b^2) \gamma^2  \lambda \beta  \right]
+  { ( w+b^2) \gamma^2  \lambda \beta  \over r} -
\cot \theta { p - b^2/2 \over r} +
\nn  &&
{1\over r \sin \theta} \partial_{\theta} 
\left[  \sin \theta 
\left(( w+ b^2) \gamma^2 \lambda^2+ ( p+b^2/2 ) \right) \right]
=0
\label{x501} \\ &&
\partial_t \left[  b \gamma \right]+
{1\over r} \partial_r  \left[r b \beta \gamma \right] +
{1\over r} \partial_{\theta} \left[ b \gamma \lambda  \right] =0
\label{x511} \\ &&
\partial_t \left[ \rho \gamma \right]+
{1\over r^2 } \partial_r  \left[r^2 \rho  \beta \gamma \right]
+ {1\over r  \sin \theta} \partial_{\theta}  \left[\rho   \lambda  \gamma \right]
 =0
\label{x61}
\ea
where $\lambda = v_\theta$.

Our goal is to find relativistic,
self-similar solutions to Eqns (\ref{x41}-\ref{x61})
in the limit  $w, \rho, p \ll b^2$ and $\lambda \ll 1$.
We assume that 
matter inertia and pressure  can be neglect and set
$
w=p=0
$. The dynamical equations, which are equivalent to force-free equations (\ref{twomax},
\ref{FF}),
become 
\ba
&&
\partial_t \left[  b^2 \gamma^2 -b^2/2\right]+
{1\over r^2} \partial_r \left[ r^2 b^2 \beta \gamma^2 \right]
+{1\over r \sin \theta} \partial_{\theta}
\left[  \sin \theta   b^2 \gamma^2 \lambda \right]
  =0
\nn &&
\partial_t \left[  b^2 \gamma^2 \beta \right]+
{1\over r^2} \partial_r
\left[r^2\left(b^2 \beta^2 \gamma^2 + b^2/2\right) \right] -
{  b^2 \gamma^2 \lambda^2  \over r}
+ {1\over r \sin \theta} \partial_{\theta}
\left[  \sin \theta   b^2 \gamma^2 \lambda \beta  \right]
 =0
\nn &&
\partial_t \left[  b^2 \gamma^2 \lambda  \right]
+ {1\over r^2} \partial_r \left[ r^2 b^2 \gamma^2  \lambda \beta  \right]
+  { b^2 \gamma^2  \lambda \beta  \over r} +
\cot \theta {  b^2 \over 2 r} +
{1\over r \sin \theta} \partial_{\theta}
\left[  \sin \theta
\left( b^2 \gamma^2 \lambda^2+ b^2/2 ) \right) \right]
=0
\nn &&
\partial_t \left[  b \gamma \right]+
{1\over r} \partial_r  \left[r b \beta \gamma \right] +
{1\over r} \partial_{\theta} \left[ b \gamma \lambda  \right] =0
\label{x62}
\ea

The strongly relativistic solutions that we are interested 
in involve expansion
of all quantities assuming  large $\gamma$-factors.
In addition, assuming that 
 $ \lambda \gamma \ll 1$ and balancing the powers of $\lambda$ in eq.
 (\ref{x501}) we conclude that in order for the self-similar solutions
involving radial and longitudinal expansion to exist, we need
 $\lambda \propto 1/\gamma^2$.


Following  \cite{blm76}
we chose the self-similar variable
\be
\chi = 1+2(m+1) \xi =[1+2(m+1) \Gamma^2](1-r/t)
\label{chi}
\ee
where $\xi = (1-r/R)  \Gamma^2$, 
$R= t \left( 1-1/(2(m+1) \Gamma^2)\right)$ is the radius
 of the contact discontinuity and 
we assumed that the Lorentz factor scales with radius as
$\Gamma^2 \propto t^{-m}$.
We limit ourselves to the strongly relativistic case 
expanding all relations  to the first order in $1/\Gamma^2$.

Treating $(\chi,y)$, where $y= \Gamma^2$, as new independent variables we find
\ba&&
\partial_ t = - m y \partial_{y} +  ((m + 1)( 2 y - \chi) +1) \partial_{\chi}
\nn
&&
\partial_r = -  (  1 +  2(m + 1) y)  \partial_{\chi}
\nn &&
\beta = 1-{1 \over 2 y g}
\nn &&
r=  t \left( 1-{\chi \over 1+2(m+1) y}\right)
\ea

The boundary conditions on the CD require that 
 Lorentz factor of the magnetic field lines
equals the Lorentz factor of the CD itself,
while pressures on both sides of the CD should be equal.
This  and the scaling $\lambda \sim 1/\Gamma^2$,
allow the following parameterization:
\ba
&&
\gamma^2= \Gamma^2 g( \chi) 
\nn &&
 b= \sqrt{ 2 w_{ 1}  } \Gamma h( \chi) 
\nn &&
\lambda = l( \chi) / \Gamma^2
\ea
with $g(1) =h(1)  = 1 $ and $\Gamma\equiv  \Gamma(\theta)$.

The  equations for the self-similar 
variables
  are \footnote{General relations for radial motion
($l=0$) including effects of
nonzero pressure are derived in Lyutikov (2002)}.
\ba
&& 
 { \partial \ln g \over  \partial  \ln \chi}=  { 1- m \over 1+m}
\label{151} \\ &&
 { \partial \ln h  \over  \partial  \ln \chi}=  { 1- m \over 2( 1+m)}
\label{152} 
\\ && 
 (1+m) (1-\chi g) { \partial  l \over  \partial \chi}
  = -  {1 \over \sin \theta y^{1/2} }
{ \partial \sin \theta y^{1/2} \over \partial \theta}
+
\left( m-1- (m+1) \chi g  \right)l
\label{15}
\ea
Taking into account  boundary conditions,
the solutions are 
\ba &&
g = {\chi }^{\frac{1 - m}{1 + m}},
 \\ &&
h =  {\chi }^{\frac{1 - m}{2\,\left( 1 + m \right) }}
 \label{h} 
\ea
(compare with \ref{BB}).

One particular case is  $l \equiv 0$  -
 purely radial motion  with parameters depending on the 
polar angle. In this case from (\ref{15}) it follows
\be
y = \Gamma^2 =  {\Gamma_0^2 \over \sin^2 \theta }
\label{y0}
\ee
 where $\Gamma_0= \Gamma^2(\pi/2)$.
This solution corresponds to the fully balanced case (\ref{bala}).

If the magnetic shell expands into the progenitors wind
with density varying as a power law of radius, $\rho _{\rm 1} \propto
r^{-k}$ the relations (\ref{15}) may be easily generalized.
For the laterally balanced solution ($l=0$) we find
\ba
&& 
{1\over g} { \partial \ln g \over  \partial \chi}= 
\frac{2(1-m) -k }{2 (m+1)}
\nn &&
{1\over g} { \partial \ln h  \over  \partial \chi}= 
\frac{2(1-m) -k }{4 (m+1)}
\ea
with  solutions
\ba
&&
g = \chi^{ { 2(1-m) -k \over 2 (1+m)}}
\nn &&
h= \chi^{ { 2(1-m) -k \over 4 (1+m)}}
\ea
It is required for consistency that $k<4$ and $ m> -1$.

\section{Instability of the contact surface: 
 Impulsive  Kruskal-Schwarzschild}
\label{IKS}

In this appendix we consider the dynamical instability of the CD due to  a
fluctuating luminosity of the central source. 
Conventionally, the 
Kruskal-Schwarzschild  (KS) instability refers to an instability of a plasma
supported against gravity by magnetic field. If   plasma
is ''below'' the magnetic field (so that the effective gravity is directed
from the magnetic  to the plasma phase) then the configuration is stable. 
This is similar to the Rayleigh-Taylor (RT) stability/instability.

A stable  contact discontinuity separating two fluids
may  become unstable when a  shock passes through it. In fluid dynamics
this is known as  Richtmyer-Meshkov (RM) instability.
It is {\it independent}
of the whether the contact surface is RT  stable or
unstable. Physically, small ripples of the CD distort
the flow of the shocked material and  create vorticity which destroys the
CD. Mathematically, the RM instability is treated as an impulsive acceleration
of the CD
during the passage of the shock.
Similarly, we expect that the 
KS-stable CD will be unstable under influence of an impulsive
electromagnetic perturbation propagating in the magnetically dominated medium.
This instability may be called impulsive Kruskal-Schwarzschild instability, 
(or, equivalently,  magnetic Richtmyer-Meshkov instability). 

The 
mathematical analysis of the  Richtmyer-Meshkov instability is complicated, yet
a simple  original derivation of  Richtmyer (1960) gives
a good estimate of the growth rates in a wide range of cases \citep[see also][]{Inog99}.
Below we follow the logic of Richtmyer (1960) in estimating the 
growth rate of impulsive KS instability. 
A more detailed analysis will be reported elsewhere.

Consider a pressure balanced 
 CD separating regions of magnetic field and plasma, so that
$B^2/8 \pi = p$. 
If the  perturbation of the  CD  with   a wave number $k$ is  propagating 
orthogonally to the direction for the magnetic field the behavior
of the displacement $\xi$ is determined by
\be
{ d^2 \xi \over dt^2} - g k \xi =0
\label{KS1}
\ee
In the case $g > 0$, perturbations grow exponentially: this is the   KS instability.
In the  case $g  < 0$, this describes oscillatory perturbations
propagating orthogonally to magnetic field (akin to surface gravity waves).

Assume next  that the magnetic pressure suddenly increased by $\delta B$.
This may be due to (fast) magnetosonic wave pulse propagating towards the boundary. 
This will launch a shock  wave in the plasma, so that the 
CD will acquire some velocity $\Delta v$. 
In this case the effective gravity on the CD  is 
$ g = \Delta v \delta(t)$. Integrating Eq. (\ref{KS1}) once we find
\be
{ d  \xi \over dt} =  \Delta v  k   \xi_0
\ee
where $\xi_0$ is  the initial perturbation.
Thus perturbations will grow. Growth is initially  
linear in time (as oppose to exponential in
case of SM instability). 
This  is the  impulsive  Kruskal-Schwarzschild (IKS below)  instability. 
It's growth rate 
is 
\be
\gamma_{IKS} \sim \Delta v k
\label{gammaIKS}
\ee

We also note, that IKS, like 
RM,  is not a classical instability: it does not involve ''self-amplification'', so that
at the linear stage perturbations grow 
linearly, not exponentially, with time. The reason is that the 
initial pulse generates velocity fields which evolve dynamically and distorts
the CD by inertial motion, so that the typical  velocity  of a  perturbation  
will decrease with time. 

Next we consider the  IKS instability in GRBs.
The pressure balance in the frame of CD gives (\ref{b1})
\be
{B^2 \over 8 \pi \Gamma^2} = p' = \kappa \rho \Gamma^2
\ee
where 
$p'$ is the pressure of the shocked material and 
$\kappa$ is a constant  relating the  pressure
on the CD to the external density. In case of self-similar motion the
value of $\kappa$ can be easily found from the Blandford-McKee solution (appendix
\ref{shocked}). 

Magnetic field $B(t')$ at   time $t'$ is related to the retarded time
luminosity of the source $t'_{ret}$:
\be 
B(t') = {1 \over r} \sqrt{L (t'_{ret}) \over c}
\ee
As we have discussed in Section \ref{relatexpan}, for constant luminosity
these equations determine self-similar evolution of the magnetic shell. 
Since Lorentz factor decrease with time the  effective gravity in the frame
of the CD is directed from the magnetic shell (''light fluid'') to 
shocked plasma (''heavy fluid''), so that  the system is KS stable.

Next assume that at a retarded time $t'_{0,ret}$  the  luminosity of the source
 increases by $\delta L$ so that
\be
L=L_0+  \delta L\Theta(t'_{ret} - t'_{0,ret})
\ee
($\Theta$ is Heaviside function).
When the fast magnetosonic waves propagating information about the change in the
 source luminosity   reach the CD, it will launch a forward shock
(or a sound wave) in the 
already shocked circumstellar medium. As a results   the velocity of the CD
will change. 
The changes in the velocity may be related to the change in luminosity
\be
{ \delta L \over L} = 2 { \delta B  \over B} =
 4 { \delta \Gamma \over \Gamma} =
{4 \delta v  \Gamma^2}/c = 4 \delta v'/c
\label{delta}
\ee
where $\delta v'$ is a change in the velocity of the CD in its frame and 
 we  assumed that all the  changes are small, $\leq 1$.

Note, that the electromagnetic pulse  associated with
the increase of luminosity should not necessarily launch a shock. All  that is needed
is a sharp velocity change of the CD. In the hydrodynamic case such a sharp
velocity change can be  only due to a shock (any subsonic flow would communicate information
pre-accelerating the CD). In the case of force-free fields,
a sharp velocity change may be due to an  electromagnetic  pulse propagating with the speed of
light.

We can now estimate the growth rate of IKS instability in GRBs.
From (\ref{gammaIKS})  and  (\ref{delta})  we find the growth rate in the frame of the CD:
\be
\gamma_{IKS}' \sim {1 \over 4} { \delta L \over L} c k
\ee
Qualitatively, for $\delta L/L \sim 1$
\be
\gamma_{IKS}' \sim { c k \over 4}
\ee
Thus,  the IKS instability grows on time scales of the order of sound crossing time.

At the largest scale, $r' \sim r/\Gamma$, 
the instability will be suppressed by the
spherical expansion of the flow. This will occur when 
$\gamma_{IKS} \sim 3 \dot{r'}/r' \sim 3 / t'
$ 
(c.f. the  Hubble flow). 
In the observer's frame the 
largest unstable mode will have a scale somewhat  smaller that  the ``horizon''
 $\leq  r/\Gamma$.

In this appendix we briefly discussed the instability of the CD due to source
non-stationarity, assuming that the source changes its luminosity on the scales
much shorter that the dynamic scales of the outflow. This is indeed expected
since the source may change on a scale of milliseconds. On the other
hand, if the source  luminosity changes on a  scale comparable 
to the outflow time scale, the impulsive KS instability becomes a conventional
 KS (in)stability. The relation between the two is very much similar
to the relation between RT and RM instabilities and is quantified by the
Froude number
$
Fr = {\Delta  v} \sqrt{k/g}
$. For  RT and KS instabilities $Fr \sim 1$,  while for RM and IKS $Fr \gg 1 $.

\section{Disruption of polar current}
\label{Disrupt}

In this Appendix we consider  disruption of the axial current  
due to  the development of dynamic instabilities near the polar axis.
As we have argued, the  lateral (in  the $\theta$ direction) dynamics is suppressed
in relativistically expanding flows. As a results,
at large distances from the source, when the Lorentz factor has decreased considerably,
the flow will be dynamically  unbalanced. In this case the   development of dynamic instabilities
 may proceed in an explosive fashion, 
so that the axial current is disrupted in  finite time. 
Development of explosive instabilities   is a 
 characteristic feature of  hydrodynamically
unstable media \citep[\eg][]{ps93,trub96}.
Plasma systems, like  solar flares
and TOKAMAKs disruption, often show   explosive  behavior
\citep[\eg][]{ca97}. In these cases 
 amplitude of perturbations $A$ tends to infinity,
$A \sim (t_0-t)^\alpha$, on  \Alfven
 time scale.

To illustrate  explosive
current disruption that may occur in the polar regions of GRBs
 we consider hydromagnetic collapse of a current carrying
pinch \citep[see also][]{lib99}. For simplicity we assume that collapse is non-relativistic and  that the
 plasma obeys the  ideal MHD equations. Since the size of the current carrying region $r_D $ 
(Eq \ref{rD}) is much 
larger than the light cylinder radius $r_s$, we neglect the stabilizing effects of the
poloidal magnetic field. The core is assumed 
to be  cylindrically collimated and  there is no velocity shear in the plasma rest frame. 
The core plasma is  hot with temperature
$T$, non-force-free and   described by
 adiabatic index $\Gamma_a$. 
Under these assumptions the governing equations are
(all the quantities are measured in the plasma rest frame)
\ba &&
\partial_t \rho + {1 \over \varpi } \partial_\varpi( \varpi\rho u) =0,
\nn &&
\partial_t B_{\phi} + \partial_\varpi(  u B_{\phi}) =0,
\nn &&
\partial_t T + u \partial_\varpi T + (\gamma-1) { T \over \varpi}  \partial_\varpi( \varpi u)=0,
\nn &&
\rho  \left( \partial_t u + u \partial_\varpi u \right) + \partial_\varpi P +
 { B_{\phi} \over 4 \pi \varpi } \partial_\varpi( \varpi B_{\phi}  ),
=0
\label{h1}
\ea
where $u$ is the (cylindrical) radial velocity.

We will look  for self-similar solutions of the system (\ref{h1}) 
where all the variables depend on the self-similar coordinate
\be
\xi = { \varpi \over R_c(t)},
\ee
where $ R_c(t)$ is the radius of the pinch at time $t$.  Introducing 
scale parameter
\be
\alpha (t) = {  R_c(t)  \over  R_{c,0} },
\ee
where $ R_{c,0} \geq r_D $ is the initial radius of the core, we parameterize the variables
\be 
u = \dot{R_c} \xi U(\xi),
\hskip .3 truein 
\rho = \rho_0 \alpha ^{2 \chi} \Lambda(\xi),
\hskip .3 truein
B_{\phi} =  B_{\phi,0} \alpha ^{\mu} H_{\phi} (\xi),
\hskip .3 truein
T=  T_0 \alpha ^{- 2 \lambda} \Theta(\xi).
\ee
Assuming a special form of the velocity profile $U(\xi)=1$ we find
\be
\xi=-1, \, \mu=-1, \,\lambda = \gamma-1
\ee
with the following equations for the functions
$\Lambda, \,  H_{phi}$ and $\Theta$:
\ba &&
{  H_{\phi} \left( H_{\phi} \xi \right)'_\xi \over \xi^2  \Lambda} ={\rm C_1}
\nn &&
{ \left(  \Theta \Lambda  \right)'_\xi \over \xi  \Lambda}=- {\rm C_2}
\ea
where ${\rm C_1}$ and ${\rm C_2}$ are  constants which can  be set to unity.
The radial dynamics of the pinch is then governed by
\be
\alpha^{\prime \prime } (t)  =
  {\beta  
\over \alpha(t)^{2  \gamma -1} }  - { 1 \over  \alpha(t)},
\label{kk}
\ee
where $\beta = 4 \pi T_0 \rho_0 / B_0^2$ is the plasma pressure  parameter
and time is measured in \Alfven times, $\tau_A = B_{\phi,0} /
2 \sqrt{ \pi \rho_0} R_0$.
The first term on the rhs of Eq. (\ref{kk}) is responsible for the pressure support,  
 while the second term is due to the
magnetic field pinching.

In the spirit of our electro-magnetic approach we assume that the 
pressure forces cannot halt a collapse and set  $\beta=0$.
Choosing constant density, $ \Lambda = {\rm const}$, we find that
\be
 H_{\phi}  \propto \xi
\ee
The radial dynamics of the pinch is then governed by
\be
\alpha^{\prime \prime } (t)  =
 - { 1 \over  \alpha(t)}
\label{ss}
\ee
Since the second derivative is  negative, the scale factor 
 $\alpha$ will become zero in finite time.
This can be seen by direct implicit integration of
eq. (\ref{ss})
\be
t= \sqrt{\pi \over 2} Erf( \sqrt{\ln 1/\alpha})
\ee
where $ Erf$ is the error function.
For $ \alpha \rightarrow 0$ this gives
\be
\sqrt{\pi \over 2} -t = { \alpha \over \sqrt{ \pi \ln 1/\alpha } }
\ee
Thus, the radius of a pinch becomes zero at $t=\sqrt{\pi \over 2} $
\Alfven times.

Hydrodynamical description of the pinch collapse is applicable only for initial
stages of the collapse.
As the pinch contracts its radius will become smaller than $r_D$. At this point the
 MHD approximation
becomes invalid: there is not enough particles in the core to support the  required
current.  As a result  large inductive electric field $E_z \geq B_\phi$
 will develop. This will lead to particle acceleration and production of synchrotron
radiation.

\section{Resistive collimation}
\label{resicol}

As we have discussed in Section \ref{Collim} late collimation of
relativistic outflows is hard to achieve due to 
effective relativistic suppression of lateral dynamics. In this section we show that
collimation of relativistic outflows may be more efficient
if we allow for dissipation of magnetic energy close to the axis and  if the
dissipated energy  can decouple from the plasma flow, \eg as radiation.
If initially dissipation occurs near the axis, \eg to enhanced resistivity brought
about by current concentration, the resulting loss of magnetic pressure, which
resists the inflow of plasma towards the axis, will propagate as a
rarefaction wave away from the axis. After the propagation of the
rarefaction waves plasma acquires lateral velocity  which will leads to the
pile-up of magnetic field near the axis and to faster radial expansion
(toothpaste tube effect).

As a simple model problem consider the structure of a core part of the flow in the
flow rest frame. For didactic purposes we assume that the central core
is cylindrically collimated and that there is no velocity shear in the plasma rest frame. 
In this case, the electric field vanishes and 
 there is only the magnetic field due to a line  current $I'$: $b=2I'/\varpi'$ ($\varpi$ is the 
cylindrical radius in the plasma rest frame).

\begin{figure}
\plottwo{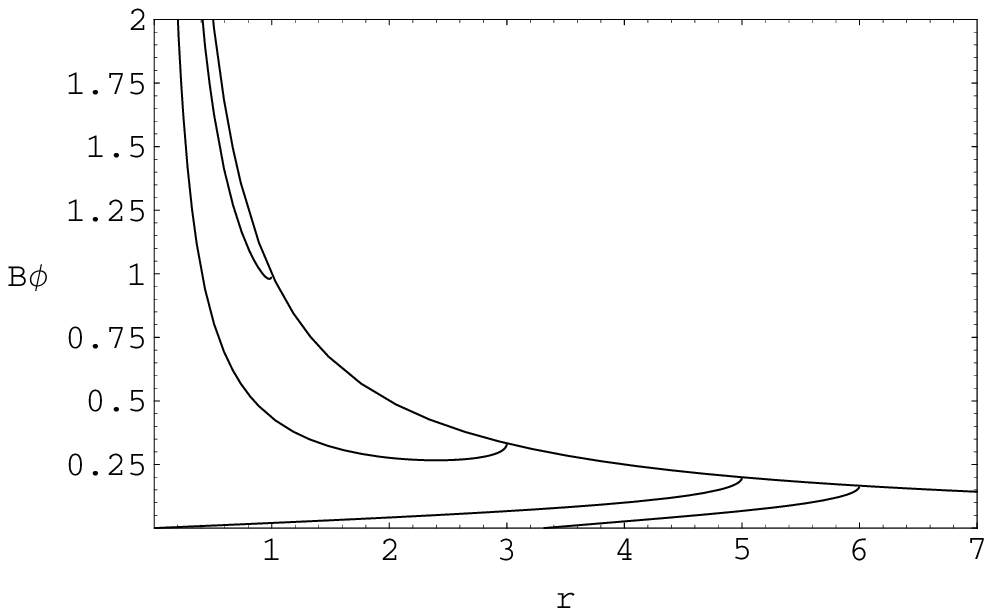}{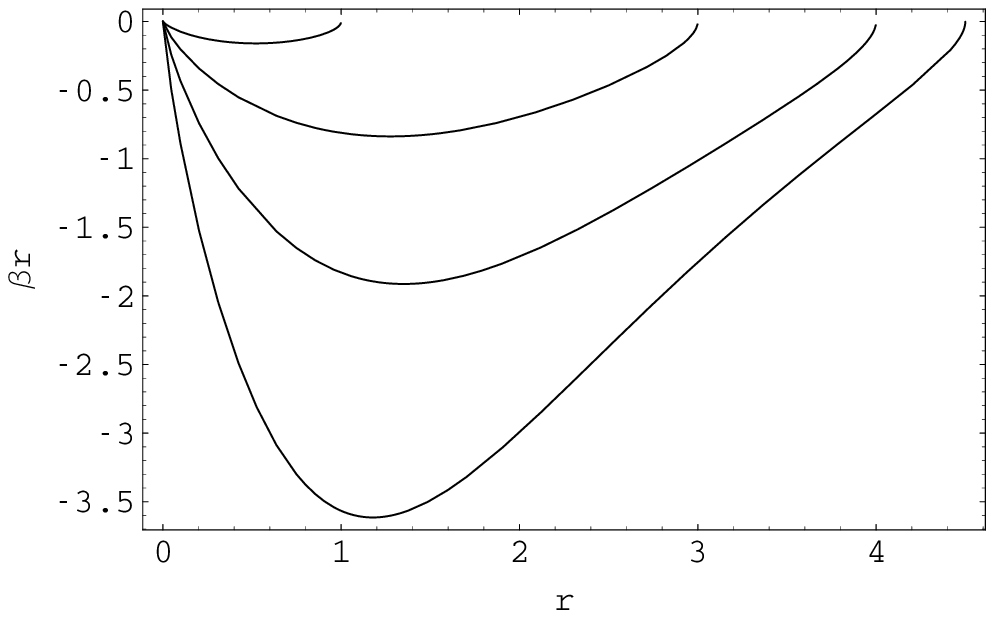}
\caption{(a) Evolution of the toroidal magnetic field  in the cylindrically collimated core
due to dissipation of  axial current 
(Eq. \ref{Bf}). At $t=0$  the axial  current becomes dissipative
with linearly decreasing total current, $\tau_0=5$,  $t=0,1,3,5,6$. 
Electromagnetic rarefaction  wave propagates away from the axis. (b) Radial 
electromagnetic velocity for $t=1,3,4,4.5$. When $\beta_r$ exceeds unity (in absolute value)
force-free approximation breaks down. }
\label{Bphi}
\end{figure}

Consider cylindrical line current 
 RFF pinch in which at moment $t=0$ resistivity is turned on. 
 Since
there are  no distributed charges, no
distributed currents, equations
of resistive RFF then become
\ba&&
\partial_t \B = - \curl \E
\nn &&
\partial_t \E=  \curl \B
\label{eq1}
\ea
Eq. (\ref{eq1}) then becomes an equation for cylindrical waves emitted by 
the line current.
\be
{\partial_\varpi ( \varpi\partial_\varpi A_z ) \over \varpi} - \partial_t^2 A_z = 2 I(t) {\delta(\varpi) \over \varpi}
\ee
For a given dependence $I(t)$ its solution can be found by standard
Green function method:
\be
A_z= \int_0^{t-\varpi} d \tau { 2 I(\tau) \over \sqrt{ (t-\tau)^2 - \varpi^2} }
\ee
Analytical solution exits, for example, for 
$I(t)=I_0 \Theta(-t)+ I_0 (1-t/\tau_0)\Theta(t) $
where $\tau_0$ is  typical decay time for the axial current.
We find
\be
A_z = 2 I_0 \left( \ln \varpi- { \sqrt{t^2-\varpi^2} + t \ln \left( (t- \sqrt{t^2-\varpi^2})/ \varpi \right) 
 \over \tau_0} \right)
\ee
for $t>r$, and $A_z = 2 I_0 \ln r$ for $t<r$.
The fields are then given 
\ba && 
B_\phi = {2I_0 \over \varpi} \left( 1- {\sqrt{ t^2 -\varpi^2} \over \tau_0} \right)
\nn &&
E_z = { 2 I_0  \over \tau_0} \ln \left( (t- \sqrt{t^2-\varpi^2})/\varpi \right) 
\label{Bf}
\ea
(see Fig. \ref{Bphi}).

We can introduce a 
 radial velocity with which electromagnetic energy is advected towards the resistor
on the axis
\be
\beta_\varpi = {E_z \over B_\phi} = 
{ \varpi\ln \left( (t- \sqrt{t^2-\varpi^2})/\varpi \right) \over \tau_0 - \sqrt{ t^2 -\varpi^2} }
\label{betavarpi}
\ee
 Eq. (\ref{betavarpi}) has important implications. 
For $\sqrt{ t^2 -\varpi^2} \sim \tau_0$ electromagnetic velocity becomes
of the order of the velocity of light and formally becomes larger than
the speed of light (see Fig. \ref{Bphi}).
This illustrates a property of relativistic force-free plasma: there is 
no dynamical constraint that would preclude electric fields becoming larger
than magnetic. Of course, as $E \rightarrow B$ the applicability of 
RFF will break down and one need to use full relativistic MHD equations.
One possibility then is that the plasma inertia will always keep the plasma from developing
unphysical solutions. Alternatively, RFF dynamics will lead to ''wave breaking'' - creation
of dissipative regions where particles will be accelerates \cite[\cf][]{lb00}.


\begin{figure}[h]
\includegraphics[width=0.9\linewidth]{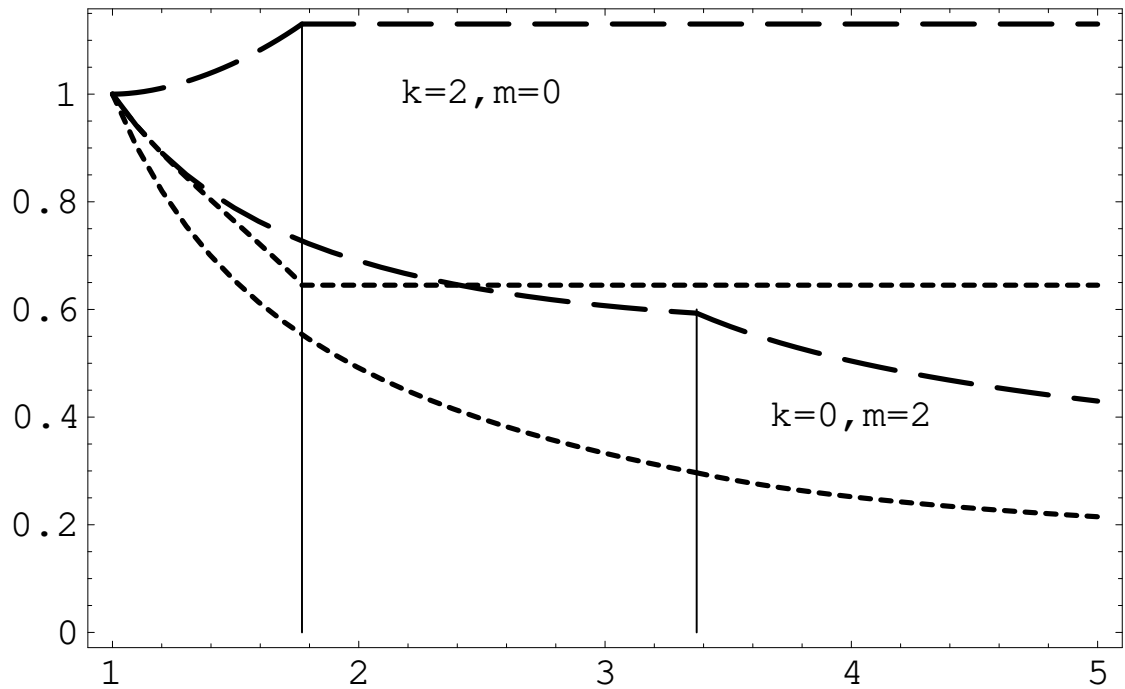}
\caption{Lorentz factor (long  dashed lines) and  pressure (short dashed lines)
 as a function
of $\chi$ for two choices of parameters: $\{m=2,k=0\}$ and $\{m=0,k=2\}$.
Locations of the contact discontinuities are denoted by solid vertical lines.
}
\label{cd1}
\end{figure}

\end{document}